\documentclass{article}
\pdfoutput=1

\usepackage{arxiv}
\usepackage{natbib}

\usepackage[utf8]{inputenc} % allow utf-8 input
\usepackage[T1]{fontenc}    % use 8-bit T1 fonts
\usepackage{hyperref}       % hyperlinks
\usepackage{url}            % simple URL typesetting
\usepackage{booktabs}       % professional-quality tables
\usepackage{amsfonts}       % blackboard math symbols
\usepackage{nicefrac}       % compact symbols for 1/2, etc.
\usepackage{microtype}      % microtypography
\usepackage{lipsum}
\usepackage{graphicx}
\graphicspath{ {./images/} }
\usepackage{amsmath}
\usepackage{amssymb}
\usepackage{bm}
\usepackage{subfigure}
\usepackage{multicol}
\usepackage{multirow}
\usepackage{xr}
\makeatletter
\newcommand{\E}{\mathbb{E}}

\newcommand{\iidsim}{\overset{iid}{\sim}}
\newcommand{\indsim}{\overset{ind}{\sim}}
\newcommand{\Norm}{\text{N}} %Normal distribution
\newcommand{\Be}{\text{Beta}} %Beta distribution
\newcommand{\Bin}{\text{Bin}} %Binomial distribution
\newcommand{\NegB}{\text{NB}} %Negative-Binomial distribution
\newcommand{\Cat}{\text{Cat}} %Categorical distribution
\newcommand{\Gam}{\text{Gam}} %Gamma distribution
\newcommand{\DP}{\text{DP}} %Dirichlet process
\newcommand{\Dir}{\text{Dir}} %Dirichlet 
\newcommand{\IG}{\text{IG}} %Inverse gamma distribution
\newcommand{\1}{\textbf{1}} %Indicator

\newcommand{\bY}{\mathbf{Y}}

\newcommand{\bbv}{\mathbf{b}}

\newcommand{\bZ}{\mathbf{Z}}
\newcommand{\bp}{\mathbf{p}}
\newcommand{\btheta}{\bm{\theta}}

\newcommand{\bmu}{\bm{\mu}}
\newcommand{\bphi}{\bm{\phi}}

\title{Shared Differential Clustering across Single-cell RNA Sequencing Datasets with the Hierarchical Dirichlet Process}

\author{
 Jinlu Liu \\
  School of Mathematics\\
  University of Edinburgh\\
  \texttt{Jinlu.Liu@ed.ac.uk} \\
   \And
 Sara Wade \\
  School of Mathematics\\
  University of Edinburgh\\
  \texttt{Sara.Wade@ed.ac.uk} \\
  \And
  Natalia Bochkina \\
  School of Mathematics\\
  University of Edinburgh\\
  \texttt{N.bochkina@ed.ac.uk} \\
}

\begin{document}
\maketitle

\begin{abstract}
Single-cell RNA sequencing (scRNA-seq) is a powerful technology that allows researchers to understand gene expression patterns at the single-cell level and uncover the heterogeneous nature of cells. Clustering is an important tool in scRNA-seq analysis to discover groups of cells with similar gene expression patterns and identify potential cell types. Integration of multiple scRNA-seq datasets is a pressing challenge, and in this direction, a novel model is developed to extend clustering methods to appropriately combine inference across multiple datasets.
The model simultaneously addresses normalization to deal with the inherent noise and uncertainty in scRNA-seq, infers cell types, and integrates multiple datasets for shared clustering in principled manner through a hierarchical Bayesian framework.
A Gibbs sampler is developed that copes with the high-dimensionality of scRNA-seq through consensus clustering.
This work is motivated by experimental data from embryonic cells, with the aim of understanding the role of PAX6 in prenatal development, and more specifically how cell-subtypes and their proportions change when knocking out this factor.
\end{abstract}


\section{Introduction} \label{s:intro}

Technological developments in the detection of genetic sequences, such as single-cell RNA-sequencing (scRNA-seq), have enabled scientists to measure gene expression on a single-cell level. As opposed to bulk sequencing experiments, which measure average expression levels across the bulk cell population, scRNA-seq enables investigation into the heterogeneity of the cells in the population. Clustering is an important tool in scRNA-seq analysis, which is used to disentangle this heterogeneity and discover groups of cells with similar gene expression profiles, yielding potential cell types. 
Experiments routinely collect multiple scRNA-seq datasets across samples (e.g. individuals,  experimental conditions, disease subtypes, time points, etc.) and methods to effectively integrate these datasets are required (as highlighted in the grand challenges of single-cell data science \citep{lahnemann2020eleven}).
In this work, we aim to extend clustering methods to appropriately integrate multiple datasets in order to identify potential cell types and understand how their proportions differ across datasets in a principled manner, as well as identify potential unique and/or rare patterns that may be present in only a subset of the datasets.

Our work is motivated by experimental data collected to shed light on the development and fates of embryonic cells and the importance of the transcription factor PAX6 in this process \citep{mi2013}. More generally, a single cell develops into an estimated 30 trillion cells in humans, and there is great interest in using single-cell technology to understand this process. The transcription factor PAX6 plays an important role in early development and is believed to control the expression of receptors that allow cells to respond correctly to signals from other cells during development \citep{isabel2014}. To empirically study the effect of PAX6, the experimental data was collected at day E13.5 from mouse embryos under control and mutant conditions in which PAX6 is deleted. By employing integrated clustering methods for combined inference across the control and mutant datasets, we can discover potential cell types, which may be shared or unique to a condition, and utilize statistical tools to examine how their proportions change when knocking out this factor, providing insight into the role of PAX6 in cellular development.

However, challenges arise in scRNA-seq data analysis due to the increased uncertainty and noise in measurements when moving from bulk to single-cell RNA-seq \citep{lahnemann2020eleven}. Specifically, only a small fraction of the total RNA present can be recorded, and thus, the data are very sparse, with zero values representing either true zero counts or missing values, also called dropouts. In addition, the fraction of transcripts recovered, also called the capture efficiency, varies across cells, causing high variability in expression levels and dropout rates. Moreover, batch effects may be present across experimental conditions, e.g. the control and mutant group, leading to further variability in capture efficiencies.

There has been a large amount of research in the field of normalization of observed gene counts \citep{vallejos2017normalizing} to account for dropouts, imputation, over-dispersion and batch effects. Often methods originally developed for bulk RNA-seq are employed; however as single cells are highly heterogeneous, the assumptions of these methods are typically not met, potentially leading to adverse consequences in downstream analysis \citep{vallejos2017normalizing}. More recently, methods tailored to scRNA-seq have been introduced \citep[e.g.][]{Greg,kharchenko2014bayesian,l2016pooling}, in which normalization is carried out in a preprocessing step by dividing the raw counts by estimated cell-specific scaling factors (i.e. capture efficiencies). Subsequently, the data are simply log-transformed,  after adding an offset to avoid the log of zero, in order to apply standard statistical tools.  Alternatively, capture efficiencies can be jointly estimated with other parameters of interest, such as gene expression, through approaches that account for the count nature of the data directly, for example, through a negative-binomial  model which also allows for over-dispersion \citep{basics, risso2018, Tang}.

In order to cluster cells and identify potential cell types, most approaches first apply such normalization strategies to the data in a preprocessing step, followed by clustering in a downstream analysis. In addition, many methods also employ some form of dimension reduction, typically via principal component analysis (PCA) \citep{ascend, rahul2015, Lin} or t-distributed stochastic neighbour embedding (t-SNE) \citep{monocle}. Subsequently, a variety of clustering methods have been employed, such as hierarchical clustering \citep{ascend, Lin, flowsom, sc3}, k-means \citep{pcareduce, safe}, density-based methods \citep{martin1996, lan2016}, or  model-based clustering \citep{de2008clustering, TSCAN, chandra2023escaping}. For recent reviews of clustering methods for scRNA-seq data, we refer the reader to \cite{vladimir2019} and \cite{raphael2020}.

However, separating the workflow into the steps of normalization, dimension reduction and clustering can adversely affect the analysis, resulting in improper clustering and characterization of cell subtypes \citep{prabhakaran2016dirichlet, vallejos2017}. More recent proposals  integrate normalization, parameter estimation, and clustering in a combined model-based framework; not only does this allow for simultaneous recovery of clusters, inference of cell subtypes and normalization of the data based on cells with similar expression patterns, but it also  provides measures of uncertainty through the model-based approach. Proposals include 1) \citet{prabhakaran2016dirichlet} who employ a Dirichlet process (DP) mixture of log-normal distributions and demonstrate the superiority of their approach compared with global normalization followed by clustering; 2) \citet{sun2018} and \citet{duan2019} who consider a DP mixture of multinomial distributions; and 3) \citet{wu2019nonparametric}  who combine the nested-hierarchical DP \citep{rodriguez2008} with a zero-inflated Poisson-log-normal distribution to cluster both subjects and cells in a nested fashion.

In this direction, we construct a novel Bayesian model which combines normalization, parameter estimation, and clustering, and integrates multiple datasets through a hierarchical framework for shared clustering across datasets. In particular, we build on the \textit{bayNorm} model \citep{Tang}, which directly accounts for the count nature and overdispersion of the data through a negative-binomial model and addresses normalization and imputation by estimating the capture efficiencies through an empirical Bayes approach. Specifically, we extend bayNorm to incorporate clustering through mixture models and integrate multiple datasets for shared clustering with a hierarchical framework based on the hierarchical Dirichlet process \citep[HDP,][]{teh2006hierarchical}. This allows us to identify potential cell types and infer varying cell-type proportions across datasets in a principled fashion.
Moreover, cells are clustered based on both mean expression and dispersion, allowing us to directly account for the mean-variance relationship, which provides robust estimates, particularly for sparse data and/or small clusters \citep{basics3}.

An important aspect of clustering scRNA-seq data is the discovery and identification of genes that distinguish one cluster from the others, often referred to as \textit{marker genes} \citep{raphael2020}. Most methods identify marker genes after clustering, by performing some statistical tests, e.g. \citet{rahul2015} identify marker genes by applying the Wilcoxon rank-sum test to the expression values  and \citet{minzhe2015} use a mixture of the rank-sum test and Welch's t-test depending on the sample size. Motivated by  \citet{basics2}, we go beyond simple comparison of mean expression levels; specifically,  marker genes that characterize differences between cell subpopulations are detected based on the log-fold change (LFC) of expression values and dispersions across subpopulations. Other methods identify  marker genes simultaneously within the clustering process \citep[e.g.][]{amit2015,andre2016,jesse2018}.

In summary, we develop a novel Bayesian model (Section 2) that simultaneously normalizes the data, infers cell subtypes with unique mean expression and dispersion patterns, and integrates multiple datasets for shared clustering. A posterior inference scheme that copes with the high-dimensionality of scRNA-seq through consensus clustering \citep{coleman2022consensus}, as well as marker gene detection and posterior predictive checks are developed and described in Section 3. The effectiveness and robustness of the proposed model and probabilistic tools for detection of marker genes are demonstrated on simulated datasets in Section 4, and results on the motivating experimental data to examine how cell-type proportions change with PAX6 is deleted are presented in Section 5.

\section{The Model} \label{s:model}

To introduce notation, we observe multiple scRNA-seq datasets over the same genes, and the raw RNA counts for each dataset $d$ are collected in the matrix  $\bY_d$ for $d=1,\ldots D$. Each $\bY_d$ has elements $y_{c,g,d}$, with rows  representing cells $c=1,\ldots, C_d$ and columns representing genes $g=1,\ldots, G$. The number of genes $G$ is common across the datasets, while the number of cells $C_d$ is dataset-specific. 

\subsection{The \textit{bayNorm} Model}\label{sec:baynorm}

In this work, we build on \textit{bayNorm} \citep{Tang},  an integrated modelling approach to address simultaneously normalisation (correcting for variability in capture efficiencies), imputation (accounting for dropouts), and batch effects. 
Specifically, \textit{bayNorm}  assumes a binomial likelihood for the observed  raw counts, given the unobserved true latent counts, denoted by $y^0_{c,g,d}$, and the cell-specific capture efficiencies, denoted by $\beta_{c,d}$:
\begin{align*}
y_{c,g,d} \mid  y^0_{c,g,d}, \beta_{c,d} \indsim \Bin( y^0_{c,g,d}, \beta_{c,d}).
\end{align*}
The binomial distribution is a simple model for the transcript capture in scRNA-seq data, assuming the the observed count is obtained through independent Bernoulli experiments determining whether each of the true transcripts is captured, with a constant cell-specific probability (or capture efficiency).
The latent counts are modelled with a negative-binomial distribution:
\begin{align}
 y^0_{c,g,d} \mid \mu_{g}, \phi_{g} \indsim \NegB( \mu_{g,d}, \phi_{g,d} ),  \label{eq:latentcounts}
\end{align}
with gene-specific mean expression level $\mu_{g,d}$ and dispersion parameter $\phi_{g,d}$. The negative-binomial, which can be represented as a Poisson-gamma mixture, is required when modelling RNA counts to capture the burstiness and excess variability observed, compared with a Poisson model. The latent counts can be marginalized to obtain the model:
\begin{align}
 y_{c,g,d} \mid \mu_{g,d}, \phi_{g,d}, \beta_{c,d} \indsim \NegB( \mu_{g,d} \beta_{c,d}, \phi_{g,d} ).
 \label{eq:ymod}
\end{align}
We remark that potential identifiability issues are apparent in the marginalized model in eq.~\eqref{eq:ymod}; specifically, if all capture efficiencies are multiplied by a common factor and all mean expressions are divided by that same factor, the model is unchanged. 
To address this, an informative approach is used to estimate the capture efficiencies, which assumes they are proportional to cell-specific scaling factors times the estimated global mean capture efficiency across all cells for each experiment \citep[See Supplementary Note 1,][]{Tang}. We also note that batch effects are mitigated by allowing both the capture efficiencies and mean expression and dispersion to be batch-specific.

An important aspect of \textit{bayNorm} is that it allows imputation of the latent counts by computing the posterior of $y^0_{c,g,d}$, which is shown to be a shifted negative-binomial distribution:
\begin{align*}
y_{c,g,d}^0 = y_{c,g,d} + \zeta_{c,g,d},
\end{align*} 
where $\zeta_{c,g,d}$ represents the \textit{lost} count, which has a negative-binomial distribution with mean $ \mu_{g,d} (1- \beta_{c,d}) (y_{c,g,d} + \phi_{g,d}) / (\mu_{g,d}\beta_{c,d}+ \phi_{g,d})  $ and size $ y_{c,g,d} + \phi_{g,d}$. This accounts for dropouts through imputation of $y^0_{c,g,d}$, in contrast to other normalization schemes \citep[e.g.][]{Greg,kharchenko2014bayesian,l2016pooling}, which simply rescale the raw data by the estimated cell-specific scaling factors (i.e. zero counts remain zero after rescaling).

\subsection{The NormHDP model}
We develop a novel model (NormHDP) that extends \textit{bayNorm} to incorporate shared clustering across multiple datasets through a hierarchical Bayesian framework. Specifically, we allow for \textit{cell-specific} mean expression $\mu_{c,g,d}$  and dispersion  $\phi_{c,g,d}$ in eq. \eqref{eq:latentcounts} and assume that they are generated from unknown, discrete, data-specific distributions $P_d$ for $d=1,\ldots D$, which are modelled with a hierarchical Dirichlet process \citep[HDP, ][]{teh2006hierarchical}:
\begin{align*}
y_{c,g,d} \mid \mu_{c,g,d}, \phi_{c,g,d}, \beta_{c,d} &\indsim \NegB( \mu_{c,g,d} \beta_{c,d}, \phi_{c,g,d} ),\\
(\bmu_{c,d}, \bphi_{c,d}) |P_d \indsim P_d, \quad
P_d |P &\iidsim \DP(\alpha P) \quad \text{and} \quad
P \sim \DP(\alpha_0 P_0),
\end{align*}
where $\bmu_{c,d}= (\mu_{c,1,d},\ldots, \mu_{c,G,d})^T \in \mathbb{R}_+^G$ and $\bphi_{c,d}= (\phi_{c,1,d},\ldots, \phi_{c,G,d})^T \in \mathbb{R}_+^G$ collect the mean expression and dispersion parameters for the $c$-th cell in dataset $d$ across all genes; $\alpha>0$ and $\alpha_0>0$ are the concentration parameters and $P_0$ is the base measure of the HDP; and DP denotes the Dirichlet process \citep{ferguson1973} discussed below.

\subsubsection{The Hierarchical Dirichlet Process}

The HDP \citep{teh2006hierarchical} defines a distribution over a set of exchangeable random probability measures and is a widely adopted nonparametric prior due to its large support, interpretable parameters, and tractability. Realisations of the HDP are discrete with probability one, and an explicit construction is provided by the stick-breaking representation:
\begin{align*}
P_d= \sum_{j=1}^\infty p_{j,d} \delta_{\btheta_j^*},
\end{align*}
with $\btheta_j^*=(\bmu_j^*,\bphi_j^*)$ and $\btheta_j^* \iidsim P_0$,
\begin{align*}
p_{j,d}= v_{j,d} \prod_{j'<j} (1-v_{j',d}), & \quad \quad v_{j,d}|(p_1,\ldots,p_j) \sim \Be\left(\alpha p_j, \alpha \left(1- \sum_{j'=1}^j p_j \right) \right),\\
p_{j}= v_{j} \prod_{j'<j} (1-v_{j'}) & \quad \text{and} \quad v_j \sim \Be(1,\alpha_0).
\end{align*}
Notice that the probability measures $P_d$ share a common set of atoms $\btheta_j^*$, representing different cell subtypes with unique expression levels  $\bmu_j^*$ and dispersion $\bphi_j^*$, but have cell-subtype proportions $p_{j,d}$ that differ across datasets. The discrete nature will induce ties in the cell-specific values of $\btheta_{c,d} =(\bmu_{c,d},\bphi_{c,d})$ with positive probability, and thus a random clustering of the cells is obtained, where two cells belong to the same cluster if they share the same expression level  $\bmu_j^*$ and dispersion $\bphi_j^*$. Moreover, clusters can be shared across multiple datasets. In fact, the law of this random clustering can be analytically obtained and described by the hierarchical Chinese restaurant franchise \citep{teh2006hierarchical}. The HDP is a nonparametric prior that avoids pre-specifying a finite number of cell subtypes and instead assumes the number of cell subtypes in any finite sample is data-driven and grows with the number of cells.

The HDP can also be constructed as the limit of a finite-dimensional HDP defined as
\begin{align}
P_d^J= \sum_{j=1}^J p^J_{j,d} \delta_{\btheta_j^*},
\label{eq:fdHDP1}
\end{align}
with $\btheta_j^* \iidsim P_0$,
\begin{align}
\begin{split}
(p^J_{1,d},\ldots, p^J_{J,d}) |(p^J_{1},\ldots, p^J_{J}) \sim \Dir( \alpha p^J_{1},\ldots, \alpha p^J_{J})
\end{split}
\end{align}
and
\begin{align}
(p^J_{1},\ldots, p^J_{J}) \sim \Dir\left( \frac{\alpha_0}{J},\ldots, \frac{\alpha_0}{J}\right).
\label{eq:fdHDP2}
\end{align}
Here, $J$ represents the truncation level of the finite-dimensional approximation. As shown in \cite{teh2006hierarchical} and from the results of \cite{kingman1975}, it follows that
$ P_d^J\Rightarrow P_d.$

\begin{figure}[!h]
\centering
\makebox{\includegraphics[width=1\textwidth]{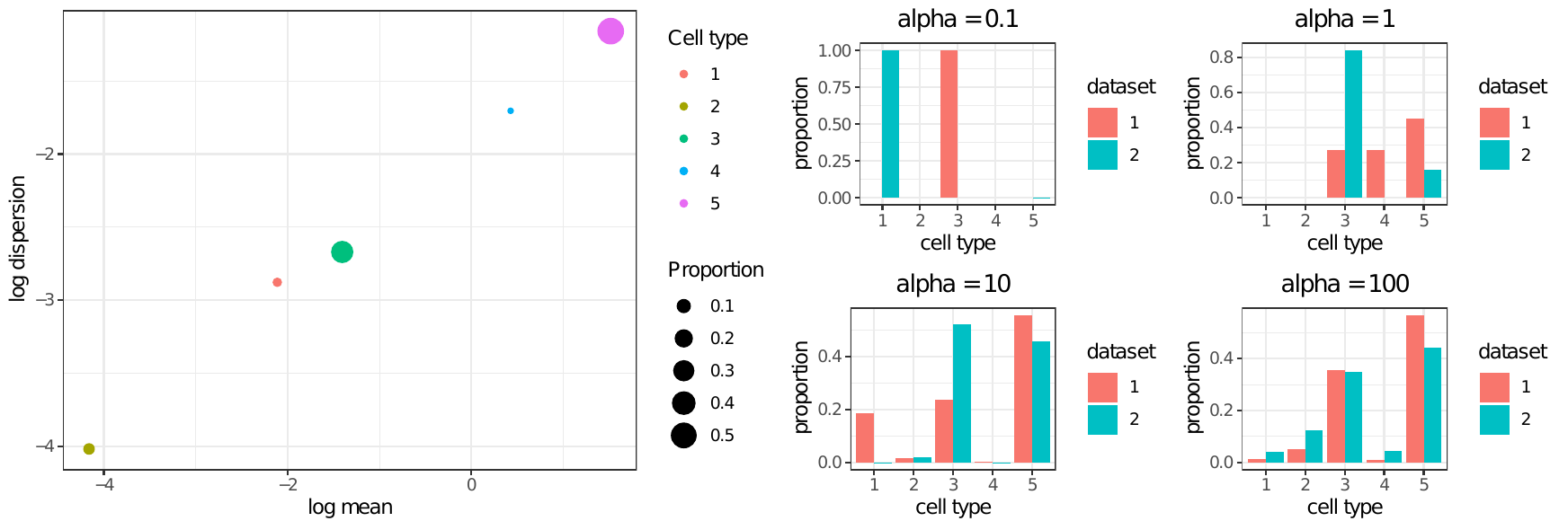}}
\caption{Simulation from the finite-dimensional HDP model, with $G=1$ genes and  $J=5$ cell subtypes. Left: illustration of the mean and dispersion for each cell subtype, with size reflecting the cell-subtype proportion in the base measure. Right:  cell-subtype proportions across two datasets for varying values of $\alpha$. For a large value of $\alpha=100$, the cell-subtype proportions for both datasets are similar to those of the base measure. As $\alpha$ decreases, the cell-subtype proportions become more distinct across datasets and degenerate to a single cell subtype per dataset as $\alpha \rightarrow 0$.}
\label{fig:sim}
\end{figure}

Figure \ref{fig:sim} illustrates a simulation from the model in a simple setting with only one gene and a truncation at  $J=5$ cell subtypes. On the left, the mean and dispersion for each cell subtype is depicted, with size reflecting the cell-subtype proportion $p_j$ of the base measure. The concentration parameter of the base measure is $\alpha_0 =5$; as $\alpha_0 \rightarrow \infty$, the Dirichlet prior degenerates to a point mass at $(1/J,\ldots, 1/J)$, while as $\alpha_0 \rightarrow 0$, the Dirichlet prior places all mass on the vertices of the simplex. Thus, large values of $\alpha_0$ favor equal cell-subtype proportions in the base measure and small values of $\alpha_0$ result in a large proportion for a single cell subtype. On the right of Figure \ref{fig:sim}, the cell-subtype proportions across two datasets are depicted for varying values of the HDP concentration parameter $\alpha$, which controls the similarity of the data-specific cell-subtype proportions $p_{j,d}$ to the overall cell-subtype proportions  $p_j$ of the base measure. For a large value of $\alpha=100$, the cell-subtype proportions for both datasets are similar to those of the base measure. As $\alpha$ decreases, the cell-subtype proportions become more distinct across datasets, and each dataset will contain only a single cell subtype in the limit as $\alpha \rightarrow 0$. Based on initial discussions with neuroscientists, such clustering structure reflects the anticipated clustering in the motivating data; specifically, we anticipate similar cell-subtypes, with some difference across control and mutant conditions, and possibly the presence of small unique cell-subtypes.

\subsubsection{Prior Specification}\label{sec:prior}

The NormHDP model is completed with prior specification for the other parameters, namely, the base measure $P_0$ of the HDP, the capture efficiencies $\beta_{c,d}$ and additional hyperpriors (e.g. for $\alpha, \alpha_0$). 

\paragraph{The base measure $P_0$} The prior for the atoms $\btheta_j^*$, which characterize the different cell subtypes, is determined by the base measure $P_0$ of the HDP. One main difference with other proposals in literature (e.g. \cite{wu2019nonparametric}) is that our model allows both the mean expression and dispersion to be cell subtype-specific, i.e. the atoms are $\btheta_j^*= (\bmu_j^*,\bphi_j^*)$.
The motivation for this is two-fold; it allows 1) more general patterns to characterize differences between cell subpopulations and 2) inclusion of prior dependence between expression levels and dispersions to account for the mean-variance relationship.
Recent studies have demonstrated the utility of exploring more general patterns, beyond focusing solely on differential expression \citep{korthauer2016}. For example, \cite{basics} and \cite{basics2} develop tools to assess differential variability, which has led to novel biological insights \citep{martinez2017aging}.  Moreover, a strong relationship is typically observed between mean expression and variability \citep{brennecke2013accounting}, suggesting that marker genes which are differentially expressed across subpopulations tend also to be differentially dispersed. Also, including prior dependence between expression levels and dispersions has shown to be important for sparse data and/or small sample sizes \citep{basics3}; for our motivating dataset, we show that gene counts are truly sparse (Web Appendix \ref{web_appendix_PAX6}). Based on preliminary analysis of our motivating dataset (Web Appendix \ref{web_appendix_PAX6}), a parametric linear dependence appears sensible (in contrast to the  nonlinear dependence in \citet{basics3}). Thus, the base measure is assumed to have the form:
\begin{align*}
    P_0(d\bmu^*) = \prod_{g=1}^G \text{log-N}(\mu_g^*|m_u, a_u^2)
\end{align*}
and
\begin{align}
    P_0(d\bphi^*| \bmu^*) = \prod_{g=1}^G \text{log-N} (\phi_g^*| b_0 +b_1 \log(\mu_g^*), a_\phi^2). \label{eq:priorlin}
\end{align}
To enhance flexibility and assess robustness, we also consider a simple extension of eq.~\eqref{eq:priorlin} based on a quadratic prior relationship between the mean-dispersion parameters.

\paragraph{Hyperparameters of $P_0$} For the hyperparameters of the mean-dispersion model $\bbv = (b_0,b_1)$ and $a_\phi^2$, we set $\bbv|a_\phi^2 \sim \Norm(\mathbf{m}_b, a_\phi^2 \mathbf{V}_b)$ and $a_\phi^2 \sim \IG(\nu_1, \nu_2)$, with default values of $\mathbf{m}_b= \mathbf{0}$ , $\mathbf{V}_b = \mathbf{I}$, $\nu_1=2$, and $\nu_2=1$ \citep{basics3}. We also consider an empirical prior by setting the values of $\mathbf{m}_b$, $v_1$ and $v_2$ based on the estimated linear relationship using the \textit{bayNorm} estimates of the mean and dispersion parameters. The parameters $m_u$ and $a_u^2$ and can also be set to default values of $m_u =0$ and $a_u^2 = 0.5$ \citep{basics3} or set empirically based on the mean and range of the mean expression estimates from \textit{bayNorm}.

\paragraph{Capture efficiencies} \citet{vallejos2017normalizing} acknowledge the inherent randomness in the capture efficiencies; if cells were processed twice, the related scaling factors would vary. Thus, instead of using fixed estimates as in \textit{bayNorm}, we model the capture efficiencies as $\beta_{c,d} \iidsim \Be(a^{\beta}_{d}, b^{\beta}_{d})$ to account for their randomness. However, as highlighted in Section \ref{sec:baynorm}, identifiability issues exist. Interestingly, parallels can be made with other domains, namely, the bias issues in scRNA-seq due to dropouts are akin to under-reporting of events for economic, health, and social indicators \citep{lopes2022bias}. In this field, compound models are considered, which involve modelling the latent true event count (similar to the latent count $y^0_{c,g,d}$) and associated reporting probabilities (related to the capture efficiencies). To mitigate identifiability issues, introducing prior information on the reporting probabilities is necessary. One approach is to make use of a validation dataset on the reporting process \citep{whittemore1991poisson,stamey2006bayesian,dvorzak2016sparse}, however this is rarely available. Alternatively, informative priors based expert knowledge have been successful \citep{moreno1998estimating,schmertmann2018bayesian}, as well as a more recent hierarchical approach with an informative prior on only the mean reporting probability \citep{stoner2019hierarchical}. Following this framework, we employ informative priors by setting $a^{\beta}_{d}$ and $b^{\beta}_{d}$ empirically based on the mean and variance of the $\widehat{\beta}_{c,d}$ obtained from \textit{bayNorm}. As suggested by one reviewer, the capture efficiencies also play an interesting role in inducing dependence across the genes, and a further discussion on this is provided in Web Appendix \ref{sec:gene_dep}.

\paragraph{Concentration parameters} The HDP concentration parameters influence the overall number of cell subtypes and the amount of information borrowed across datasets. Thus, we infer and account for their uncertainty through the hyperpriors: $\alpha \sim \Gam(1,1)$ and $\alpha_0 \sim \Gam(1,1)$.

\section{Posterior Inference}\label{sec:post}

Due to the infinite number of parameters, inference schemes for HDP models often rely on an approximation based on a finite truncation to $J<\infty$ components. Truncations based on the stick-breaking construction \citep{ishwaran2001gibbs} are widely used but can suffer from an inflated proportion $p_J$ for the last atom if the truncation level is not sufficiently large. Alternatively, the finite-dimensional construction in eq.~\eqref{eq:fdHDP1} can be used, which provides a good approximation of the HDP for sufficiently large $J$ (see \cite{ishwaran2002exact} for thorough study of the finite-dimensional Dirichlet approximation to the DP); moreover, it has the nice feature of exchangeability of the proportions $p_j$. 
We further note that finite-dimensional Dirichlet approximations belong to the class of over-fitted or sparse mixtures, which have shown to be consistent for the true number of clusters, when this number is finite but unknown \citep{judith2011}. In contrast, DP mixtures can lead to posterior inconsistency for the number of clusters, if the true number of clusters (in an infinite sample) is finite. However, this consistency requires correct specification of local likelihood \citep{Jeffrey2}.

\begin{figure}[!h]
\centering
\makebox{\includegraphics[width=0.8\textwidth]{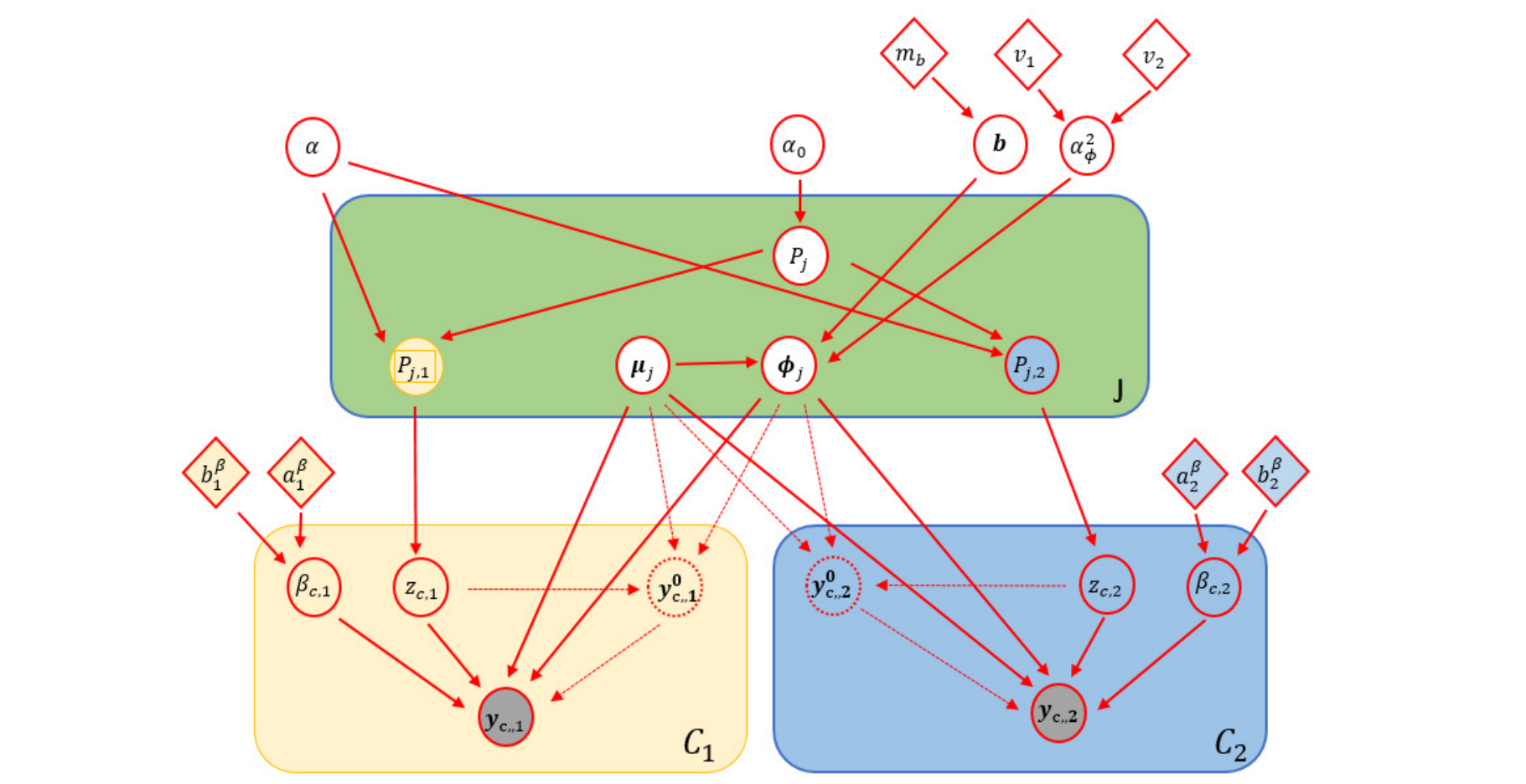}}
\caption{Graphical model of NormHDP with two datasets. Observed gene counts are shown in grey circles and latent genes counts are shown in circles with dashed outlines. Shapes with white fills are for global parameters, and shapes with yellow and blue fills are specific to dataset 1 and 2, respectively. Circles denote parameters of interest, and diamonds denote fixed hyperparameters.}
\label{fig:demo_1}
\end{figure}

Thus, focusing on the finite-dimensional approximation of the HDP in eqs. \eqref{eq:fdHDP1}-\eqref{eq:fdHDP2} and introducing the latent allocation variables $z_{c,d}$, where  $z_{c,d}=j$ if $\btheta_{c,d}=\btheta_j^*$, we define the augmented model:
\begin{align*}
 y_{c,g,d} \mid z_{c,d}=j, \mu_{j,g}^*, \phi_{j,g}^*, \beta_{c,d} &\indsim \NegB( \mu_{j,g}^* \beta_{c,d}, \phi_{j,g}^* ),\\
 z_{c,d}| (p^J_{1,d},\ldots, p^J_{J,d}) &\indsim \Cat(p^J_{1,d},\ldots, p^J_{J,d}),\\
 (p^J_{1,d},\ldots, p^J_{J,d}) |(p^J_{1},\ldots, p^J_{J}) &\sim \Dir( \alpha p^J_{1},\ldots, \alpha p^J_{J}),\\
(p^J_{1},\ldots, p^J_{J}) &\sim \Dir\left( \frac{\alpha_0}{J},\ldots, \frac{\alpha_0}{J}\right),\\
\mu^*_{j,g} \iidsim \text{log-N}(m_u, a_u^2), &\quad
\phi^*_{j,g}| \mu^*_{j,g} \indsim  \text{log-N} ( b_0 +b_1 \log(\mu_{j,g}^*), a_\phi^2),\\
\beta_{c,d} &\iidsim \Be(a^{\beta}_{d}, b^{\beta}_{d})
\end{align*}
and hyperpriors: $\alpha \sim \Gam(1,1)$; $\alpha_0 \sim \Gam(1,1)$; and $(\bbv,a_\phi^2) \sim \text{NIG}(\mathbf{m}_b, \mathbf{V}_b,\nu_1, \nu_2)$. 
A graphical representation of the NormHDP model is shown in Figure \ref{fig:demo_1}.

A Markov chain Monte Carlo (MCMC) algorithm is developed for full posterior inference. The algorithm is a Gibbs sampler which produces asymptotically exact samples from the posterior by iteratively sampling the parameters in blocks. For the allocation variables $(z_{c,d})$, mean-dispersion hyperparameters ($\bbv, \alpha_\phi^2$) and dataset-specific component probabilities ($\bp_d =(p^J_{1,d},\ldots, p^J_{J,d})$), the full conditional distributions correspond to standard distributions and can be sampled from directly. For the remaining variables, samples are obtained via adaptive Metropolis-Hastings \citep{griffin2013advances}.  Full implementation details are provided in Web Appendix \ref{Web Appendix A}. We note that each iteration of the Gibbs sampling algorithm has a computational complexity of $ \mathcal{O}((C_1 + \dots + C_D) JG)$.

\begin{figure}[!h]
\centering
\makebox{\includegraphics[width=0.7\textwidth]{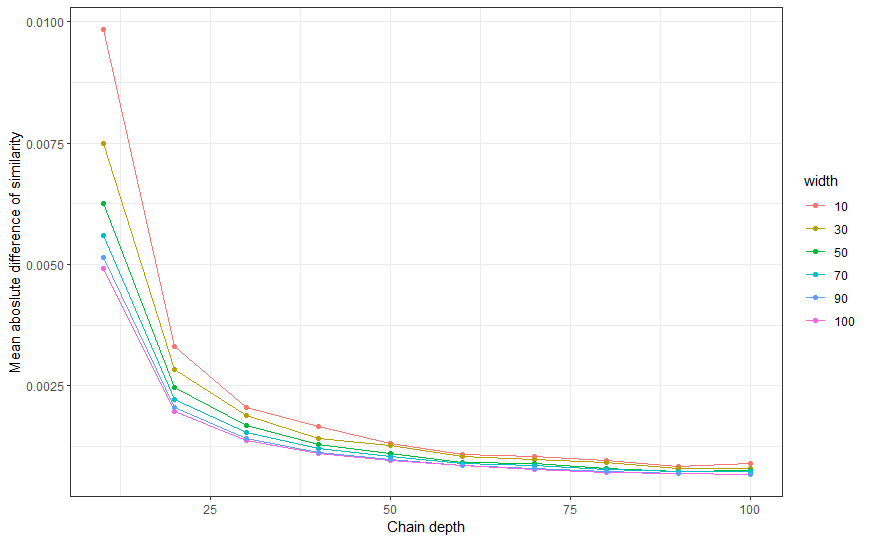}}
\caption{The mean absolute difference between the sequential consensus matrices for the experimental scRNA-seq data, for different chain widths of ${10,30,50,70,80,100}$ and chain depths of ${10, 20, \dots, 100}$. The mean absolute difference levels off before $100$ iterations.}
\label{fig:concensus_cluster_plot}
\end{figure}

However, while in theory, running a single long MCMC chain provides convergence to the posterior of interest, in practice, this produces unsatisfactory results, due to the computational issues that arise in clustering high-dimensional data \citep{celeux2019computational}, such as sRNA-seq. Namely, in high-dimensions, MCMC chains are highly sensitive to initialization and tend to get trapped very quickly in local modes, with little to no movement in the cluster allocations after the burn-in period. To overcome such issues and reduce computational cost, \citet{coleman2022consensus} develop a general scheme to explore the posterior based on an ensemble of Bayesian clustering results. The basic idea is to run a large number of chains (referred to as the width) and a small number of iterations (called the depth ). Following \citet{coleman2022consensus}, the width and depth are selected by monitoring the mean absolute difference in the posterior similarity matrix (see Section \ref{sec:clustering}), comparing successive iterations for fixed width to determine when the chains have reached the local modes and successive widths for fixed depth to understand if the uncertainty in the clustering has stabilized. Figure \ref{fig:concensus_cluster_plot} provides an example for the data in Section \ref{sec:real}, highlighting the utility of this approach to improve exploration of the posterior. After combining the MCMC draws across the multiple chains, in the following, we use $T$ to denote the total number of MCMC draws and use the superscript notation $z_{c,d}^{(t)}$ to denote the $t$-th sample. The MCMC is re-run conditional on the clustering estimate from consensus clustering to make further inference on the patterns within each cluster.

\subsection{Latent Counts}

To estimate the normalized count of gene $g$ in cell $c$, we report the posterior mean of the latent count $y_{c,g,d}^{0}$, along with posterior summaries to characterize uncertainty.
Following \cite{Tang}, it can be shown the posterior of the latent count given the allocation variables, capture efficiencies and unique parameters has a shifted negative-binomial distribution:
\begin{align*}
y_{c,g,d}^0 = y_{c,g,d} + \zeta_{c,g,d},
\end{align*} 
where $\zeta_{c,g,d}$ represents the \textit{lost} count, which has a negative-binomial distribution with mean $ \mu^*_{z_{c,d},g} (1- \beta_{c,d}) (y_{c,g,d} + \phi^*_{z_{c,d},g}) / (\mu^*_{z_{c,d},g}\beta_{c,d}+ \phi^*_{z_{c,d},g})  $ and size $ y_{c,g,d} + \phi^*_{z_{c,d},g}$.
Thus, the posterior mean of the latent count given the allocation variables, capture efficiencies and unique parameters can be written as:
\begin{align*}
\E[y_{c,g,d}^0 | y_{c,g,d},  z_{c,d} = j,\beta_{c,d}, \bmu^*_{j}, \bphi^*_{j}] = y_{c,g,d} \frac{ \mu^*_{j,g} + \phi^*_{j,g}}{ \mu^*_{j,g}\beta_{c,d}+ \phi^*_{j,g} } +\mu^*_{j,g}\frac{ \phi^*_{j,g}(1- \beta_{c,d})}{ \mu^*_{j,g}\beta_{c,d}+ \phi^*_{j,g} } ,
\end{align*}
and the posterior mean of the latent counts can be approximated by the MCMC average:
\begin{align*}
\E[y_{c,g,d}^0 | \bY] \approx  \frac{1}{T} \sum_{t=1}^T \E[y_{c,g,d}^0 | y_{c,g,d},  z_{c,d}^{(t)} = j,\beta_{c,d}^{(t)}, \bmu^{*\,(t)}_{j}, \bphi^{*\,(t)}_{j}],
\end{align*}
where $ \bY = ( \bY_1,\ldots, \bY_D)$. 
We can also examine the full posterior of the latent counts and compute credible intervals by imputing multiple values of $y_{c,g,d}^{0}$ from the shifted negative-binomial distribution at each MCMC draw.

\subsection{Clustering}\label{sec:clustering}
To summarize the posterior of the allocation variables and uncertainty in the clustering structure, we construct the posterior similarity matrix (PSM) to measure the similarity between individual cells, both within  and across datasets. In particular, each element of the posterior similarity matrix, $\text{PSM}_{c,c'}$, represents the posterior probability that cells $c$ and $c'$ are clustered together, which is approximated by
\begin{align*}
\text{PSM}_{c,c'}  \approx \frac{1}{T}\sum_{t=1}^K \text{I}(z_{c}^{(t)} = z_{c'}^{(t)}).
\end{align*}
Cells are ordered in blocks corresponding to the different datasets; thus, the diagonal blocks represent the posterior similarity matrix within each dataset and the off-diagonal blocks represent the posterior similarity matrix across datasets. Within each block, cells are sorted based on hierarchical clustering to improve visualization. Based on the posterior similarity matrix, we obtain a point estimate of the clustering structure by minimizing the posterior expected variation of information \citep{Sara}. To subsequently, analyze the patterns and uncertainty within each cluster for this optimal clustering, an additional MCMC draws values of all parameters with the allocation variables fixed at the optimal clustering (see Web Appendix \ref{Web Appendix A}).

\subsection{Detecting Marker Genes} \label{sec:markers}

Motivated by \citet{basics2}, probabilistic tools are developed based on the LFC to detect marker genes that distinguish between different cell types. \citet{basics2} focus on comparing two cell populations; thus, a simple extension is proposed which takes all clusters into consideration. Going beyond comparison of mean expression levels, we aim to detect marker genes both in terms of differential mean expression (DE) and differential dispersion (DD). For example, DD allows identification of genes whose expression may be less stable in one cell subtype.

We define two types of marker genes; \textit{global} marker genes differ between at least two clusters, whereas  \textit{local} marker genes for a given cluster differ compared with all other clusters. 

\begin{figure}[!h]
\centering
\subfigure[Density of hypothetical latent counts]{\includegraphics[width=0.8\textwidth]{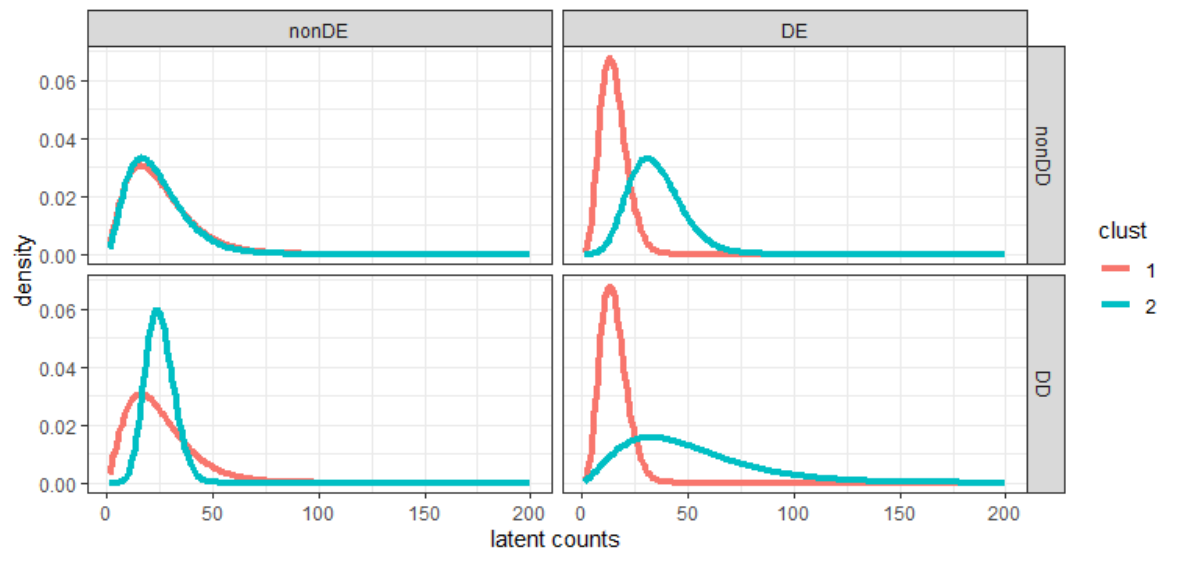}} 
\subfigure[Different types of global marker genes]{\includegraphics[width = 0.9\textwidth]{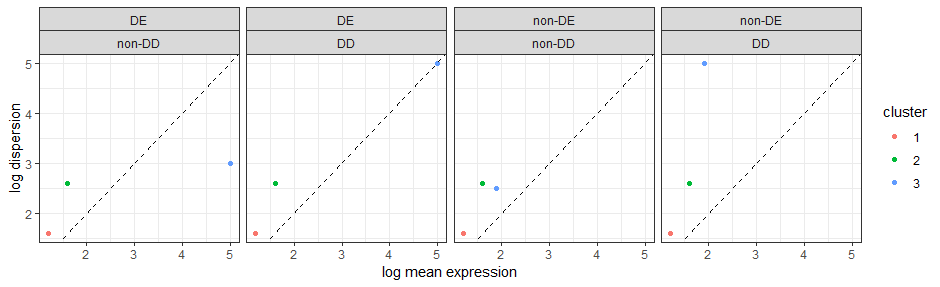}}
\caption{Top: Density of hypothetical latent counts across the four combinations of non-DE/DE and non-DD/DD genes.  For illustration, we have assumed a total of 2 clusters. Bottom: a demonstration of four different types of global marker genes with 3 clusters.}
\label{fig:marker_genes_demo}
\end{figure}

Given the clustering allocation  $\bZ=(z_{c,d})_{c=1,d=1}^{C_d,D}$ of all cells across all datasets, we first focus on comparing the mean expression and dispersion across two clusters $j$ and $j'$. We highlight that different dispersion parameters quantify changes in heterogeneity across cell subpopulations, while also accounting for the well-known mean-variance relationship in count data \citep{basics2}. Figure \ref{fig:marker_genes_demo} depicts four scenarios for a single gene: on the top right, cells are differently expressed across the cell types with similar heterogeneity; on the bottom left, overall expression levels are similar but less stable with varying heterogeneity; and on the bottom right, cell types differ in both overall expression and heterogeneity. The mean-variance relationship induces apriori correlation in the chance of DE and DD, i.e. genes that are DE tend also to be DD, but Figure \ref{fig:marker_genes_demo} demonstrates how the other cases also occur.

To measure these changes, the posterior probability associated with the LFC decision rule is computed for each pair of clusters $j$ and $j'$ and each gene $g$.  Specifically, let
 $P_g (j,j')$ and $L_g(j,j')$ be the posterior tail probabilities that the absolute LFC of the mean expression and dispersion between clusters $j$ and $j'$ is greater than the threshold $\tau_0$ and $\omega_0$, respectively;
\begin{align*}
P_g (j,j') = \text{Pr}\left( \left| \log \left(\frac{\mu^*_{j,g}}{\mu^*_{j',g}} \right) \right| > \tau_0 \mid \bZ, \bY \right)
\end{align*}
and
\begin{align*}
    L_g (j,j') = \text{Pr}\left( \left| \log \left(\frac{\phi^*_{j,g}}{\phi^*_{j',g}} \right) \right| > \omega_0 \mid \bZ, \bY \right).
\end{align*}
In light of the correlation between the unique parameters, we also developed tools to compare the LFC in the residual overdisperion \citep{basics3} (method outlined in \ref{subsec:residual_overdispersion}).

\subsubsection{\textit{Global} Marker Genes}
\textit{Global} marker genes are identified by considering the maximum of the posterior tail probabilities across all pairs of clusters:
\begin{align*}
P_g^* = \max_{(j,j')} P_g (j,j') \quad \text{and} \quad L_g^* = \max_{(j,j')} L_g (j,j').
\end{align*}
Genes with high values of $P_g^*$ or $L_g^*$ have a high posterior probability that the LFC in the mean expression or dispersion is greater than a threshold across at least two clusters.
Formally genes are classified as DE if the maximum probability, $P_g^*$, is greater than the threshold value $\alpha_M$, and genes are classified as DD if the maximum probability, $L_g^*$, is greater than the threshold value $\alpha_D$. By default, these threshold values ($\alpha_M, \alpha_D$) are set to control the expected false discovery rate (EFDR)  to 5 percent \citep{basics2}. In our context, these are given by:
\begin{align*}
    \text{EFDR}_{\alpha_M} (\tau_0) = \frac{\sum_{g=1}^G \left(1-P_g^*(\tau_0) \right) \text{I}(P_g^*(\tau_0) > \alpha_M)}{\sum_{g=1}^G (1-P_g^*(\tau_0))}
\end{align*}
and
\begin{align*}
    \text{EFDR}_{\alpha_D} (\omega_0) &= \frac{\sum_{g=1}^G \left(1-L_g^*(\omega_0) \right) \text{I}(L_g^*(\omega_0) > \alpha_D)}{\sum_{g=1}^G (1-L_g^*(\omega_0))}.
\end{align*}
Global marker genes can be computed by conditioning on the clustering estimate to detect important genes that distinguish between the identified cell subpopulations.  Alternatively, uncertainty in the clustering structure can also be incorporated by integrating the maximum probability $P_g^*$ or $L_g^*$ with respect to the posterior of $\bZ$, which can be approximated by averaging $P_g^*$ or $L_g^*$ across the MCMC samples. 

\subsubsection{\textit{Local} Marker Genes}
\begin{figure}[!h]
\centering
\makebox{\includegraphics[width=0.45\textwidth]{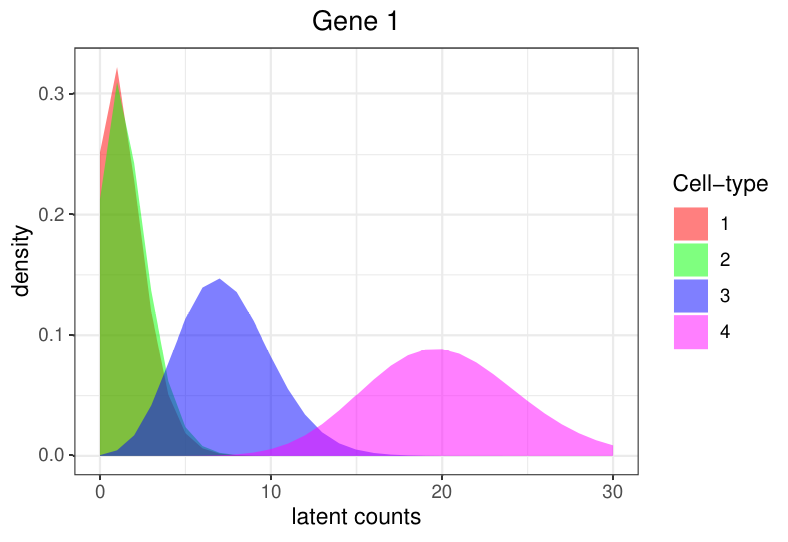}
\includegraphics[width=0.45\textwidth]{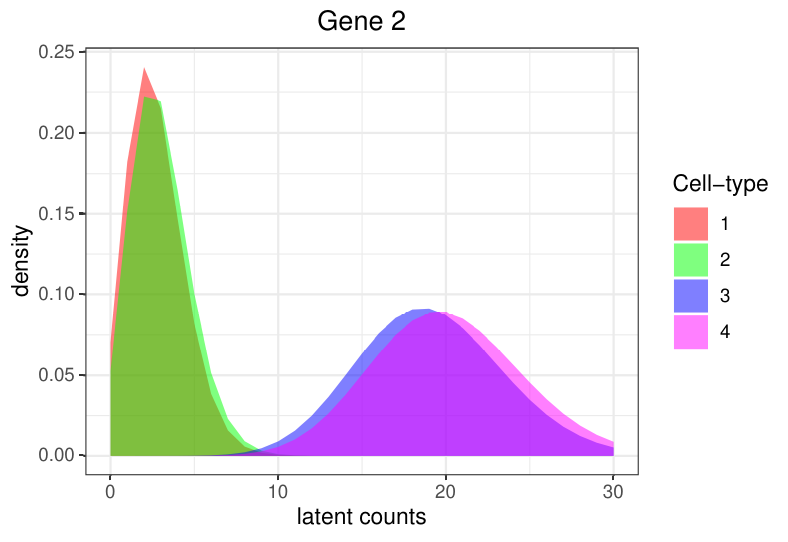}}
\caption{Density of hypothetical latent counts across four cell subtypes for two genes. Both genes are global markers, while the gene on the left is a local marker for cell-subtypes 3 and 4 and the gene on the right is  \textit{not} a local marker for any cell subtype.}
\label{fig:demo_2}
\end{figure}

While global marker genes distinguish between at least two cell subtypes, one might also be interested in identifying \textit{local} marker genes, or cluster-specific marker genes, with unique expression or dispersion for a specified cell-subtype $j$ in comparison with all others. In this case, the minimum of the posterior tail probabilities is computed:
\begin{align*}
P_{g,j}^* = \min_{j' \neq j} P_g (j,j') \quad \text{and} \quad L_{g,j}^* = \min_{j' \neq j} L_g (j,j').
\end{align*}
For cluster $j$, genes with high values of $P_{g,j}^*$ or $L_{g,j}^*$ have a high posterior probability that the LFC in the mean expression or dispersion is greater than a threshold between cluster $j$ and any other cluster. If the minimum posterior tail probability is greater than a threshold (calibrated through EFDR), the gene is detected as locally DE or DD for the specified cluster. Figure~\ref{fig:demo_2} illustrates the hypothetical latent count density for two genes and four cell subtypes. Both genes are global markers, but the gene on the left is a local marker for cell-subtypes 3 and 4 only, while the gene on the right is not a local marker for any cell subtype.

\subsection{Detecting Differences in Datasets} \label{sec:differences_in_pj}
From NormHDP, it is possible to investigate if differences exist between the dataset-specific component probabilities.
Specifically, given the clustering estimate, for each cluster $j$ and for each pair of datasets $d$ and $d'$, we compare their component probabilities, $p_{j,d}$ and $p_{j,d'}$, by computing the posterior probability  $$\pi(p_{j,d}>p_{j,d'} \mid \mathcal{D}) \approx \frac{1}{T} \sum_{t=1}^T \1(p_{j,d}^{(t)} > p_{j,d'}^{(t)}).$$
In the scRNA case, if the posterior probability is less than $0.05$, we classify the cluster as being \textit{over-represented} in the mutant group. If the posterior probability is greater than $0.95$, we classify the cluster as being \textit{under-represented} in the mutant group. The remaining clusters are the stable clusters.

\subsection{Posterior Predictive Checks}\label{sec:ppp}

Posterior predictive checks are used to assess the fit of the model together with the inferred parameters to the observed data. In particular, we generate replicated datasets from the posterior predictive distribution (approximated based on the MCMC draws) and compare key statistics between the observed and replicated data, such as the mean and standard deviation of the log counts (shifted by 1) and the dropout probabilities. Following \citet{alex2007} in the context of gene expression data, we employ mixed posterior predictive checks, where the posterior is used to simulate a subset of parameters and the prior  is used to simulate the remaining variables.  In particular, 
we generate replicated datasets after first simulating dispersions ($\bphi$) from their log-normal prior given the posterior samples of $\bbv$, $\alpha_\phi^2$ and $\bmu$. Pseudo-code for generating replicated datasets is given in Web Appendix \ref{subsec:residual_overdispersion}.

\section{Simulation Study}\label{sec:sim}

We consider three simulated scenarios to examine different aspects of our model. In Simulation 1, data is simulated based on the proposed model, and we investigate the ability of the model to recover the true parameters and clustering. In Simulation 2, we study robustness of the model under misspecification of the true mean-variance relationship. In Simulation 3, we assume that only a fraction of genes distinguish between clusters to demonstrate the effectiveness of the proposed probabilistic tools for detecting global marker genes. For each of the simulated scenarios, we run $100$ parallel chains, each with $100$ iterations to obtain the clustering estimate; %, and make inference on all remaining parameters based on his fixed clustering. 
in all runs, the algorithm is able to find the correct clustering in less than four iterations. The subsequent MCMC with fixed clustering is run for $T=8,000$ iterations, following a burn-in of $5,000$ and using thinning of $5$; traceplots (shown in  Web Appendix \ref{sec:sim1_and_2_results}) suggest convergence.

\subsection{Simulation 1 and 2}

\begin{figure}[!h]
\centering
\subfigure{\includegraphics[width=0.26\textwidth]{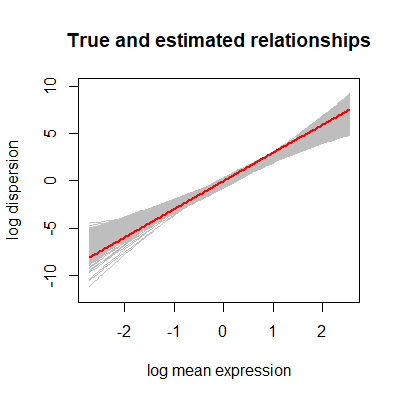}}
\subfigure{\includegraphics[width=0.3\textwidth]{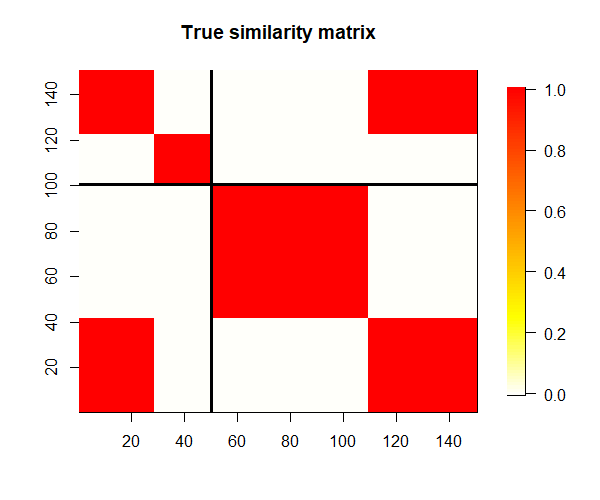}}
\subfigure{\includegraphics[width=0.3\textwidth]{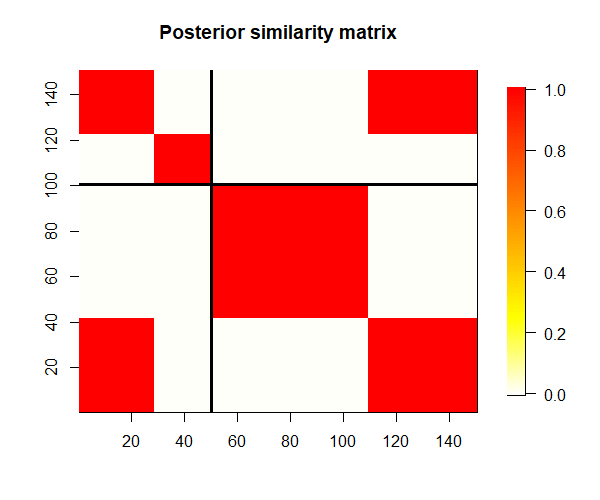}}
\subfigure{\includegraphics[width=0.26\textwidth]{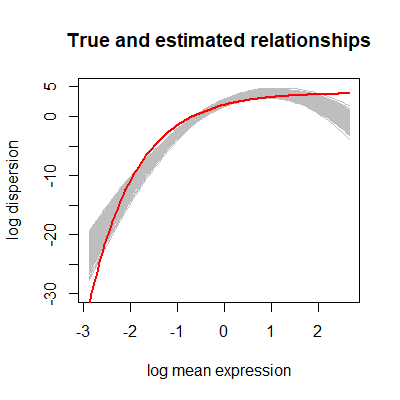}}
\subfigure{\includegraphics[width=0.3\textwidth]{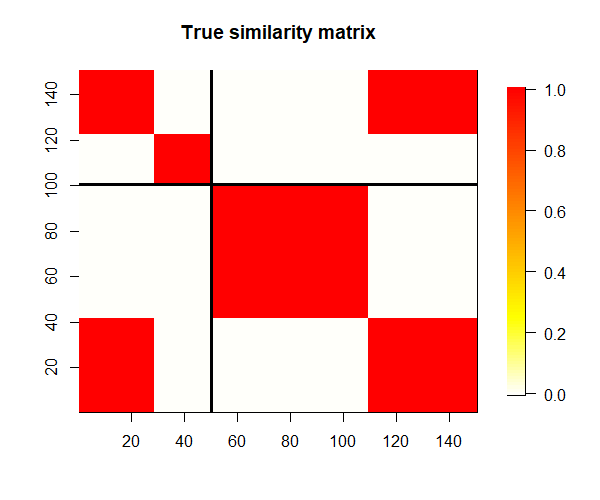}}
\subfigure{\includegraphics[width=0.3\textwidth]{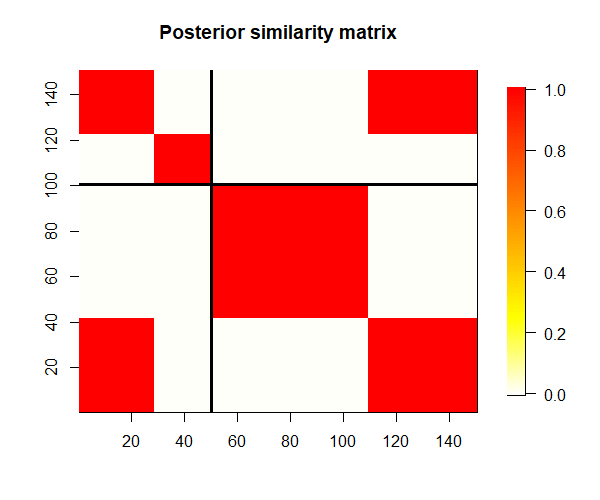}}
\caption{Summary of Simulation 1 (top row) and 2 (bottom row). Left: the true (red) and posterior (grey)  relationship between mean-dispersion parameters on the log-scale. Middle: true similarity matrix. Right: posterior similarity matrix.}
\label{fig:summary_copy}
\end{figure}

In Simulation 1, we assume the true relationship between the mean-dispersion parameters is linear on the log-scale, while in Simulation 2, we assume it is non-linear and non-quadratic, but monotonically increasing on the log-scale (Figure \ref{fig:summary_copy}). For both, we simulate data with $C_1= 50$ cells in dataset 1 and $C_2= 100$ cells in dataset 2, with $G= 50$ genes. We assume there are 3 clusters, with true cell proportions $(p_{1,1},p_{2,1}, p_{3,1}) =  (0.6,0.4,0)$ for dataset 1 and $(p_{1,2},p_{2,2}, p_{3,2}) = (0.4,0,0.6)$ for dataset 2; simulation details are provided in Web Appendix \ref{sec:sim2} and \ref{sec:sim3}.

Note that to avoid simulating datasets with empty cells and genes that are not expressed, we generate the true capture efficiencies with a mean of 0.70 which is much higher than the default value of 0.06 for droplet based protocol \citep{klein2015droplet}. The capture efficiencies in \textit{bayNorm} are estimated as proportional to cell-specific scaling factors with the global mean set to this default value \citep[Supplementary Note 1 of][and also detailed in  Web Appendix \ref{sec:baynorm_beta}]{Tang}; thus, the empirical prior based on the \textit{bayNorm} capture efficiencies estimates has prior mean equal to the default value.  Due to the identifiability issues discussed in Section \ref{s:model}, recovery of the true capture efficiencies is difficult if the true mean is far from the prior. Hence, for this task, we employ more informative priors by setting the global mean capture efficiency in \textit{bayNorm} to 0.70. In Web Appendix \ref{sec:sim3_results}, we demonstrate good recovery of the true mean capture efficiencies under our  Bayesian approach even when the prior mean is slightly misspecified, which is instead problematic for \textit{bayNorm}.

\subsubsection{Results}

\begin{table}[!h]
\centering
\caption{Comparison of the different clustering solutions based on the VI and ARI to measure distance between the true and estimated clustering for Simulations 1 and 2.} 
\bigskip
\scalebox{0.8}{\begin{tabular}{llllllll}
       & \multicolumn{2}{l}{General} & \multicolumn{2}{l}{Empirical} & \multirow{2}{*}{Seurat} & \multirow{2}{*}{CIDR} & \multirow{2}{*}{TSCAN} \\ \cline{2-5}
       & Linear      & Quadratic      & Linear       & Quadratic       &                         &                       &                        \\ \cline{1-8}
Simulation 1 - VI & 0.00        & 0.00           & 0.00         & 0.00            & 0.31                    &       1.30                &           0.58             \\ \cline{1-8}
Simulation 2 - VI & 0.00        & 0.00           & 0.00         & 0.00            & 0.31                    &       1.38                &           0.75             \\ \cline{1-8}
Simulation 1 - ARI &  1.00      &     1.00     & 1.00        & 1.00            & 0.87                    &       0.28                &           0.72                \\ \cline{1-8}
Simulation 2 - ARI & 1.00        & 1.00           & 1.00         & 1.00            & 0.87                    &       0.26         & 0.63
\end{tabular}}
\label{Table: Compare_VI}
\bigskip
\end{table}

For both simulations, we investigate prior sensitivity by comparing general priors based on standard hyperparameter values with empirical priors based on hyperparameters specified using initial \textit{bayNorm} estimates (as described in Section \ref{sec:prior}). In addition, to enhance flexibility, we consider both linear and quadratic relationships in the prior model for the mean-variance relationship. Results presented focus on the quadratic model with empirical priors (the remaining results are shown in Web Appendix \ref{sec:sim1_and_2_results}). But, we highlight that recovery of the true clustering  is robust to the choice of priors and model specification  (Table \ref{Table: Compare_VI});  for Simulation 2, however, the quadratic model is better in recovering the true mean-dispersion relationship in comparison to the linear model.
In Table \ref{Table: Compare_VI}, we also compare the clustering solutions based on our proposed model (and under the different settings) to three competing methods for clustering scRNA-seq data: 1) Seurat \citep{rahul2015}, 2) CIDR  \citep{Lin}, and 3)TSCAN \citep{TSCAN}. The VI and adjusted Rand index (ARI) are used to compare the clustering solutions with the truth, where a small value of VI and large value of ARI  indicate good performance. In both simulations, NormHDP performs the best, followed by Seurat, TSCAN and CIDR.

\begin{figure}[h]
    \centering
    \subfigure{\includegraphics[width=0.8\textwidth]{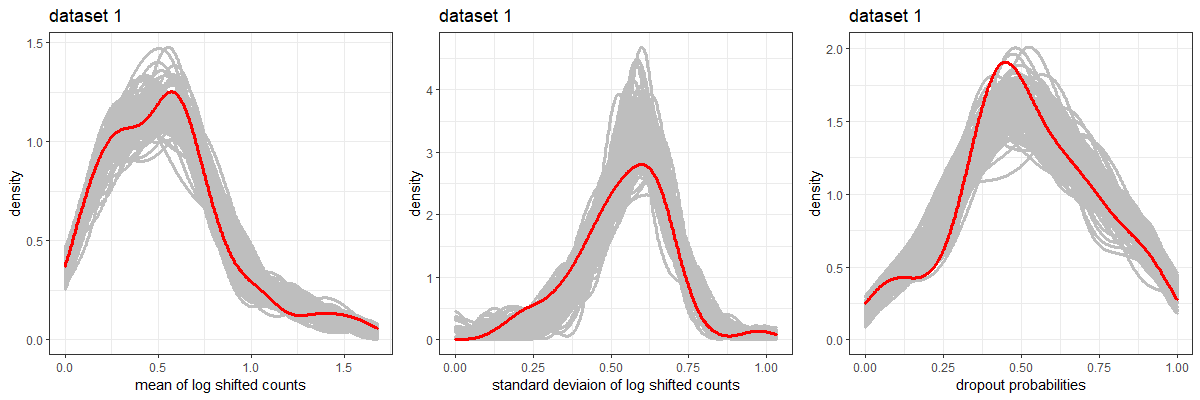}}
    \subfigure{\includegraphics[width=0.8\textwidth]{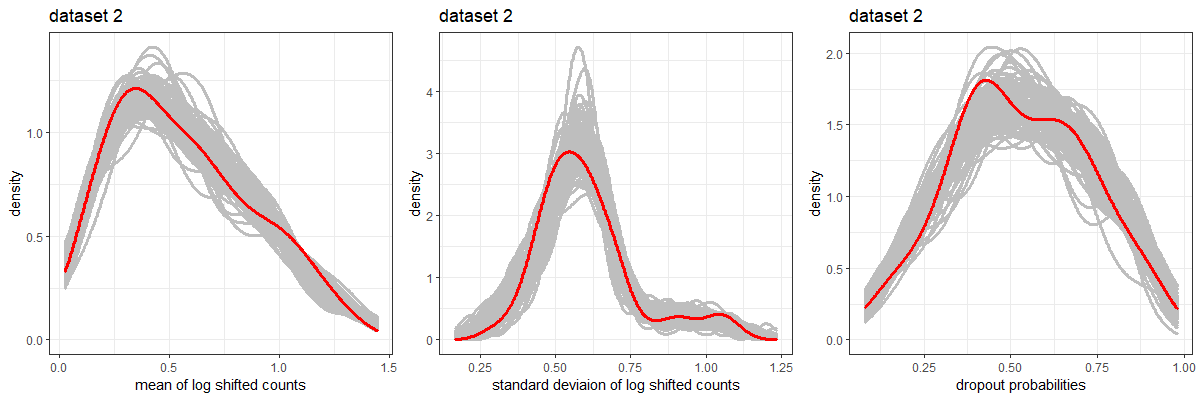}}
    \caption{Posterior predictive checks for Simulation 1. Grey and red lines are the KDEs of the replicated and true simulated datasets, respectively. Left to right: the density of mean of log shifted counts, standard deviation of log shifted counts and dropout probabilities.}
    \label{fig:case1_summary_appendix}
\end{figure}

\begin{figure}[h]
    \centering
    \subfigure{\includegraphics[width=0.8\textwidth]{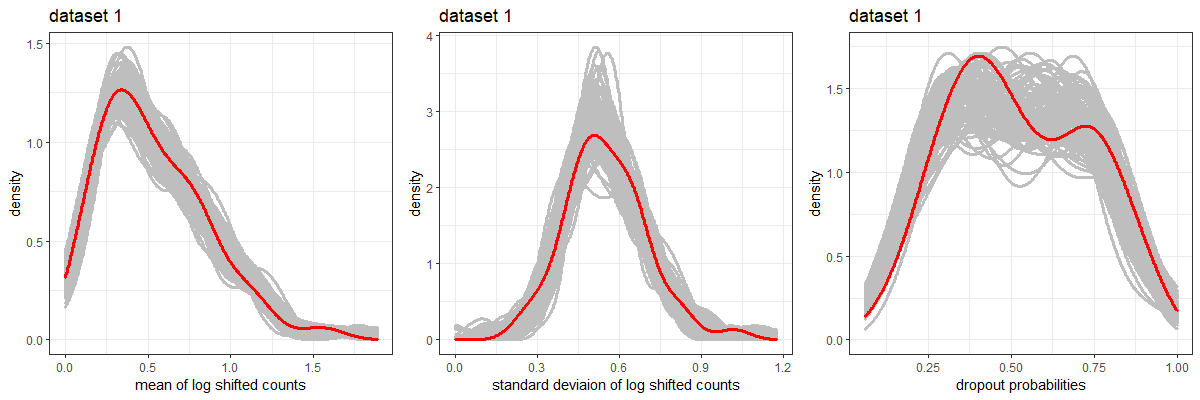}}
    \subfigure{\includegraphics[width=0.8\textwidth]{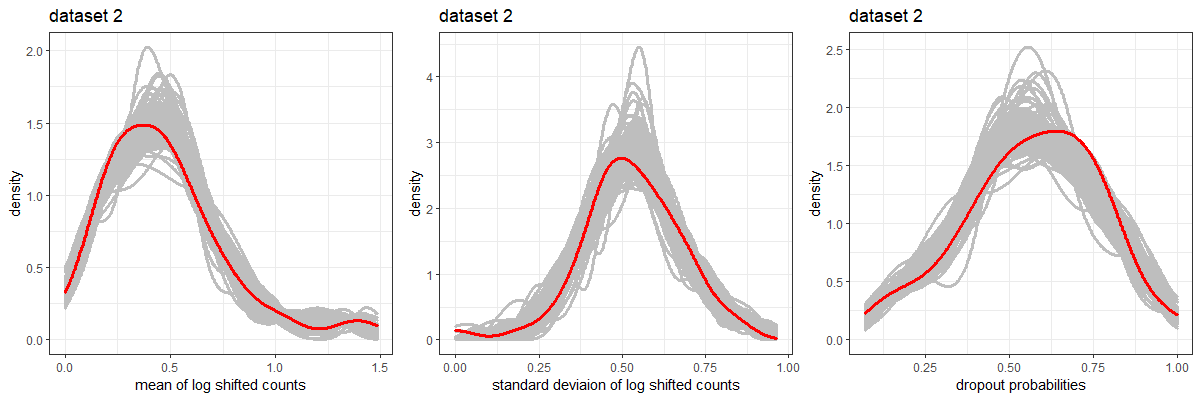}}
    \caption{Posterior predictive checks for Simulation 2. Grey and red lines are the KDEs of the replicated and true simulated datasets, respectively. Left to right: the density of mean of log shifted counts, standard deviation of log shifted counts and dropout probabilities.}
    \label{fig:case2_summary_appendix}
\end{figure}

Posterior predictive checks are presented in Figure \ref{fig:case1_summary_appendix} and \ref{fig:case2_summary_appendix} for Simulation 1 and 2, respectively. For the simulated and replicated datasets, we use kernel density estimation (KDE) to estimate densities of key statistics, namely the 1) mean of log shifted counts, 2) the standard deviation of log shifted counts and 3) the dropout probabilities of each gene. The KDE of the simulated dataset is similar to the KDEs of the replicated datasets, highlighting the sensible fit of the proposed model.

\subsection{Simulation 3}

\begin{figure}[!h]
    \centering
    \subfigure{\includegraphics[width=0.4\textwidth]{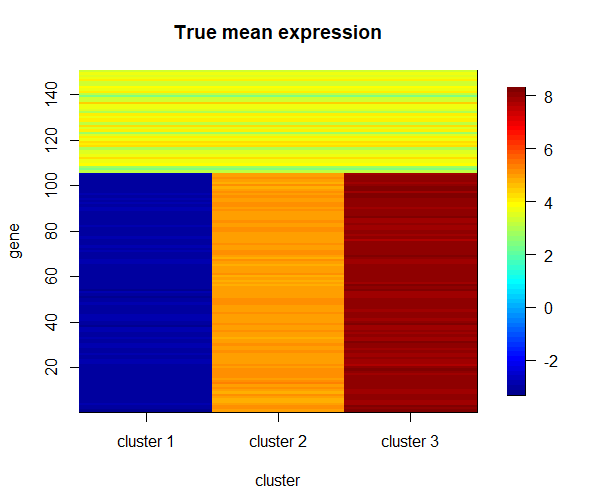}}
    \subfigure{\includegraphics[width=0.4\textwidth]{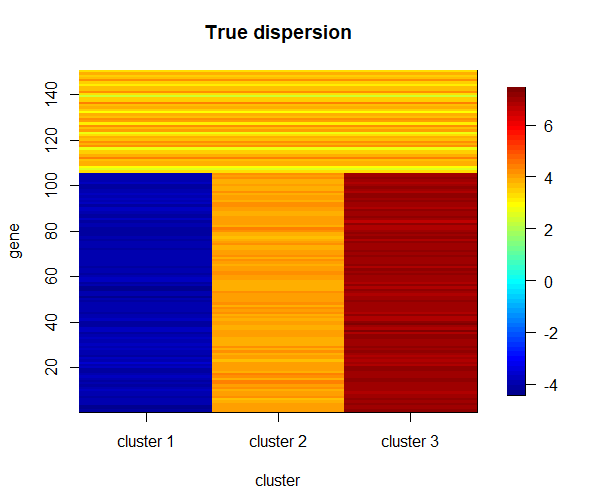}}
    \subfigure{\includegraphics[width=0.4\textwidth]{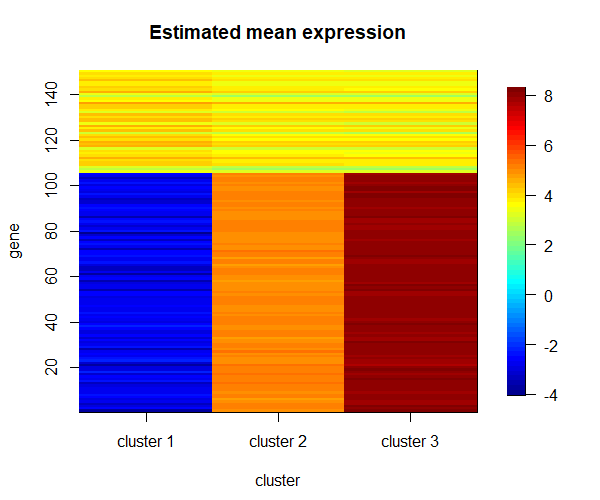}}
    \subfigure{\includegraphics[width=0.4\textwidth]{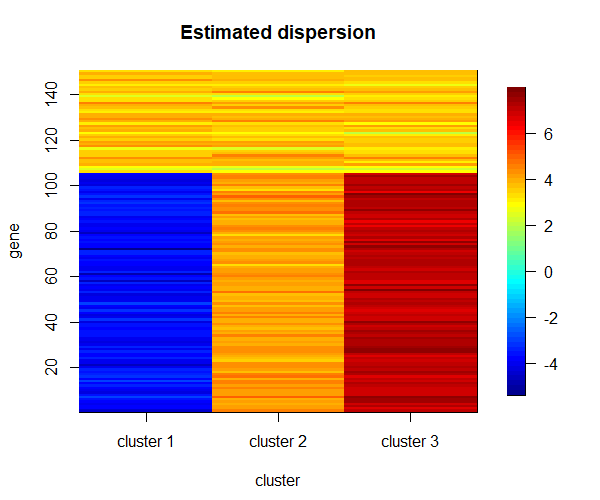}}
    \caption{Simulation 3: heat-maps for true and estimated mean expressions and dispersions.}
    \label{fig:case3_summary}
\end{figure}

For Simulation 3, we generate 20 replicated sets of data, with $C_1 =300$ and $C_2 = 400$ cells, for dataset 1 and 2, respectively, with $G=150$ genes. We assume a total of 3 cell subtypes, with proportions $(p_{1,1},p_{2,1}, p_{3,1}) =  (0.8,0.2, 0)$  and $(p_{1,2},p_{2,2}, p_{3,2}) = (0.8,0,0.2)$ for dataset 1 and 2, respectively. In addition, we assume that the first 70 percent of the genes are both DE and DD, and remaining genes are both non-DE and non-DD. Unique parameters and allocation variables used to simulate each set of data are identical across replicates. We also assume that the mean expressions $(\mu_{j,g})$ for DE genes follows a log-normal distribution with log-mean $m_j$ for each cluster $j$ and $m_{1:J} = (-3,5,8)$. The true relationship between mean and dispersions is assumed to be linear on the log-scale. Simulation details are provided in Web Appendix \ref{sec:sim3}.
In this case, we focus on the linear model with empirical priors. 
Figure \ref{fig:case3_summary} shows that NormHDP recovers well the true mean and dispersion parameters (result based on single set of data). In addition, heat-maps for the true and estimated latent counts show similar patterns (Web Appendix \ref{sec:sim3_results}).

\begin{figure}[!h]
    \centering
    \subfigure{\includegraphics[width=0.45\textwidth]{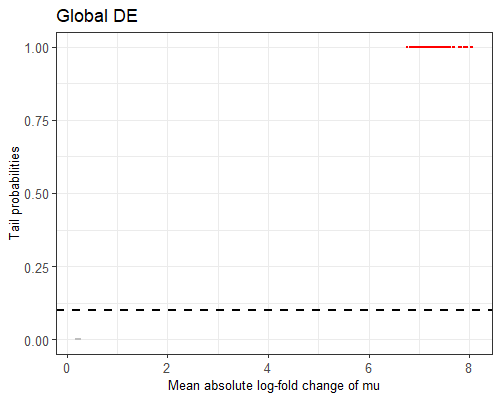}}
    \subfigure{\includegraphics[width=0.45\textwidth]{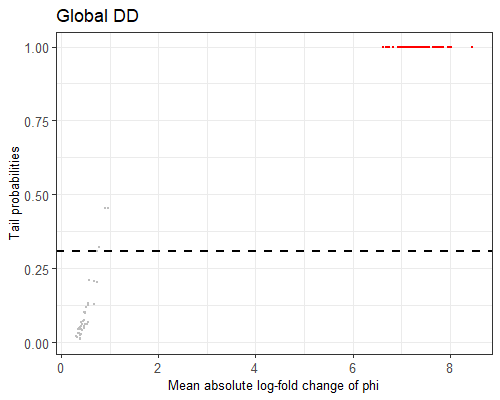}}
    \caption{Simulation 3: relationship between the mean absolute LFC and tail probability for Simulation 3. The horizontal dashed lines represent the threshold for classifying DE and DD based on the tail probabilities. True DE or DD genes are represented in red and true non-DE or non-DD genes coloured grey.}
    \label{fig:DE_DD_test}
\end{figure}

Using the global marker gene detection method proposed in Section \ref{sec:markers}, the range of the false discovery rate (FDR), across the 20 replicated sets of data, corresponding to the mean expressions is $(0, 0.019)$ and dispersions is $(0,0.045)$, hence NormHDP is sufficient in detecting the true global marker genes. Relationships between mean absolute LFCs and tail probabilities for a single set of data are shown in Figure \ref{fig:DE_DD_test}, in which case the FDR is $0$ and $0.028$ for the mean expression and dispersion, respectively. The thresholds $\tau_0$ and $\omega_0$ for computing the LFC are set to $1$ for both DD and DE. 

\section{Experimental Data on Embryonic Cell Development}\label{sec:real}

The experimental scRNA-seq datasets \citep{tan2022pax6} analyzed in this paper were collected and prepared by Dr. Tan Kai Boon and the research group lead by Prof. D. Price and Prof. J. Mason at the Centre for  Discovery Brain Sciences, University of Edinburgh. The study aims to shed light on the importance of the transcription factor PAX6 in the development and fates of embryonic cells. Tamoxifen administration was carried out at day E9.5, with mouse embryos sacrificed and dissected at day E13.5. Thus, the resulting scRNA-seq data, collected at day E13.5, is obtained under control (HET) and mutant (HOM) conditions in which PAX6 has been deleted. Standard pre-processing for scRNA-seq is carried out, following procedures in \citep{seurat}, to remove non-informative genes and cells from the raw datasets to improve model performance and avoid misinterpretation; pre-processing details are included in Web Appendix \ref{sec:filter}, which involves selecting cells and genes based on quality control metrics \citep{Satija}. After pre-processing, the HET and HOM datasets contain $C_1 = 3,096$ and $C_2=5,282$ cells, respectively, both with $G=2,529$ genes. To investigate the role of PAX6 and differences when PAX6 is not present, the proposed NormHDP model is employed for integrative clustering across the control and mutant datasets. This allows for identification of cell subtypes, that can be shared across datasets, and detection of differences in cell-subtype proportions  when PAX6 is knocked out, providing an understanding of how PAX6 influences the presence/absence of cell subtypes.

For inference, we run consensus clustering with $100$ parallel chains and $100$ iterations in each chain, which as shown in Figure \ref{fig:concensus_cluster_plot}, is sufficient to explore the posterior clustering structure. After obtaining a point estimate of the clustering by considering MCMC draws from all chains, we run an additional MCMC chain to infer the remain parameters with  $T = 8,000$ iterations, burn-in of $5000$, and thinning of $5$. For robustness to non-linearity, we focus on the quadratic model for the mean-variance relationship with empirical priors and fix the truncation level to $J = 30$. Traceplots demonstrating mixing and convergence are shown in  Web Appendix \ref{sec:real_results}. In all chains, less than $30$ components are occupied, thus $J=30$ provides a sufficient level of truncation.

\begin{figure}[!t]
    \centering
    \subfigure{\includegraphics[width=0.45\textwidth]{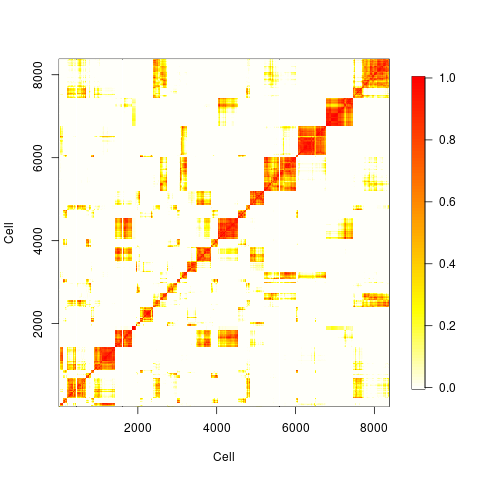}}
    \subfigure{\includegraphics[width=0.45\textwidth]{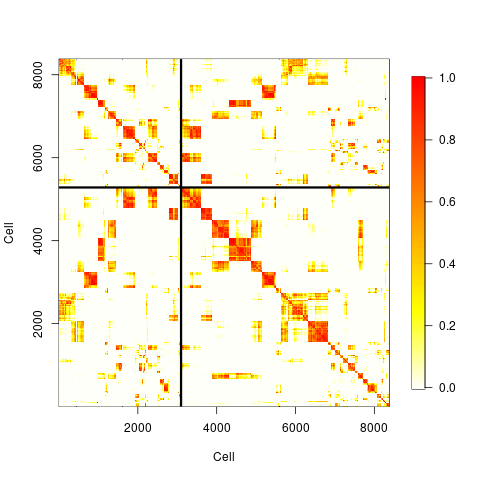}}
    \caption{Left: posterior similarity matrix without distinguishing between cells from different datasets. Right: posterior similarity matrix, grouped by HET and HOM.}
    \label{fig:psm_allchain2}
\end{figure}

A heatmap of the posterior similarity matrix in Figure \ref{fig:psm_allchain2} provides a visualization of the clustering structure and its uncertainty, both within and across the control and mutant datasets. The estimated clustering which minimizes the variation of information contains 22 clusters, 19 of which are shared between HET and HOM; while this clustering estimate is clearly observed in Figure \ref{fig:psm_allchain2}, there is also some apparent uncertainty on  whether to further split some clusters. 

\begin{figure}[h]
\centering
\subfigure{\includegraphics[width=0.9\textwidth]{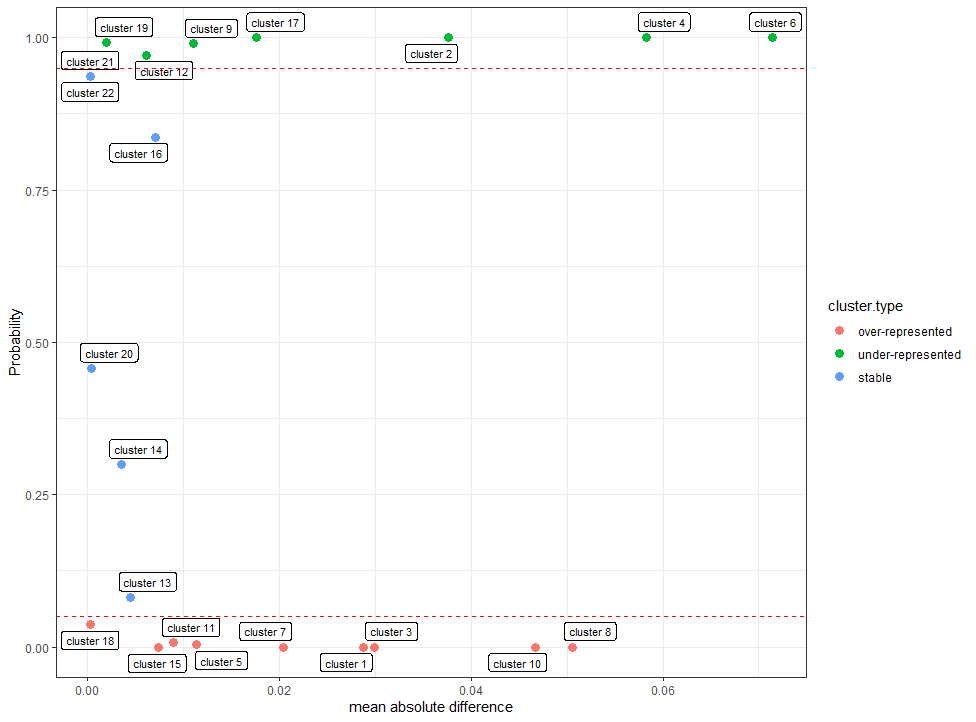}}
\caption{Detecting differences in cell-subtype proportions between the control and mutant groups. The x-axis shows the posterior mean absolute difference in the cell-subtype proportions. The y-axis shows the posterior probability of having a larger proportion when PAX6 is present. Horizontal red dashed lines are the thresholds for cell-subtype classification as stable, under or over represented when PAX6 is deleted.}
\label{fig:difference_in_pjd_plot1}
\end{figure}

To further study the clustering estimate and its differences between the control and mutant groups, we determine which cell subtypes have a high posterior probability of differing in the proportions (Section \ref{sec:differences_in_pj}). Specifically, we define as cell subtype as \text{stable} if no difference is detected, \textit{over-represented} in the mutant group if the proportion is larger in the mutant group with high posterior probability, and \textit{under-represented} in the mutant group if the  proportion is smaller in the mutant group with high posterior probability. 
Following this approach, cell subtypes $1,3,5,7,8,10,11,15,18$ are \textit{over-represented} in the mutant group, cell subtypes $13,14,16,20,22$ are \text{stable} and cell subtypes $2,4,6,9,12,17,19,21$ are \textit{under-represented} in the mutant group. Figure \ref{fig:difference_in_pjd_plot1} provides a visualization of the mean absolute difference in the cell-subtype proportions against the posterior probability of having a larger proportion when PAX6 is present. 
We note that although several cell subtypes are either over or under represented when PAX6 is deleted, the mean absolute difference in the proportions tends to be small. This suggests that  PAX6 may play a smaller role at this early stage in the development (day E13.5).

Additional figures examining the posterior relationship between the mean expressions and dispersions for each cell subtype are provided in Web Appendix \ref{sec:real_results}. The following subsections provide a further analysis of the patterns that characterize each cell subtype.

\subsection{Posterior Estimated Latent Counts}

\begin{figure}[!t]
\centering
\makebox{\includegraphics[width=\textwidth]{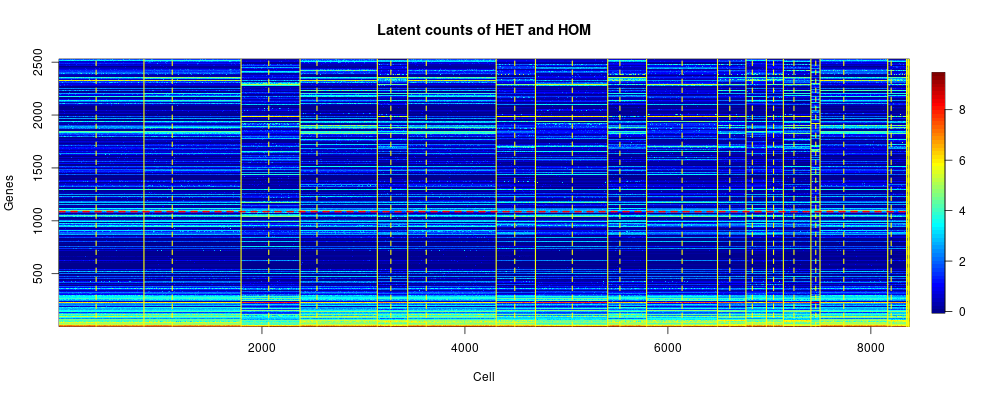}}
\caption{Heat-map of posterior estimated latent gene-counts for HET and HOM. Genes are reordered by global DE tail probabilities and genes above the red horizontal line are global DE. Cells for reordered by the cell subtype, with cells from different subtypes  separated by solid lines and cells from different datasets separated by dashed lines.}
\label{fig:real_data_summary_2}
\end{figure}

We compute the posterior estimated latent counts for all cells and compare between cell subtypes. Figure \ref{fig:real_data_summary_2} provides a heat-map of the estimated latent counts; cells are ordered by the cell subtypes, with solid vertical lines separating cells from different subtypes and dashed vertical lines separating HET and HOM within cell subtype. Genes are reordered by global DE tail probabilities, with global DE genes above the horizontal line. Corresponding figures for the observed counts are shown in Web Appendix \ref{sec:global_marker_genes}. For each gene, posterior estimated latent counts and observed counts for cells within each cell subtype are similar, and clear differences are observed across cells from different subtypes.

\begin{figure}[!h]
    \centering
    \subfigure{\includegraphics[width=0.45\textwidth]{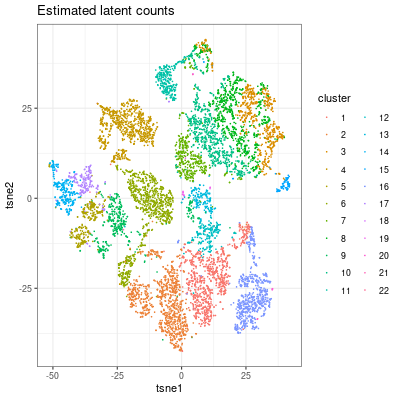}}
    \subfigure{\includegraphics[width=0.45\textwidth]{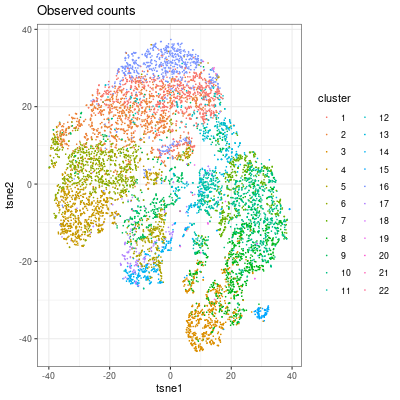}}
    \caption{t-SNE plot for the posterior estimated latent counts (left) and observed counts (right). Cells from different subtypes are shown in different colours.}
    \label{fig:latent_observed_tsne_allgenes}
\end{figure}

In addition, we use t-SNE (a commonly used dimensional reduction method for visualising gene expressions) to visualize similarities between cells within each subtype and differences across sybtypes (Figure \ref{fig:latent_observed_tsne_allgenes}). Applying t-SNE to the posterior estimated latent counts for genes which are global DE and DD shows a clear separation between cell subtypes (Web Appendix \ref{sec:global_marker_genes}).

\subsection{Global Marker Genes}

\begin{figure}[!t]
\centering
\makebox{\includegraphics[width=1\textwidth]{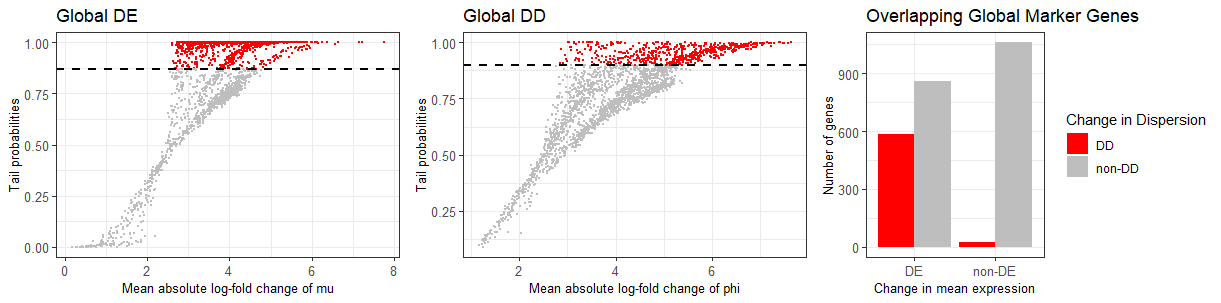}}
\caption{Relationship between mean absolute LFCs and tail probabilities, and a summary of the number of genes that are global markers.}
\label{fig:global_marker_summary}
\end{figure}

\begin{figure}[!t]
    \centering
    \subfigure{\includegraphics[width=0.9\textwidth]{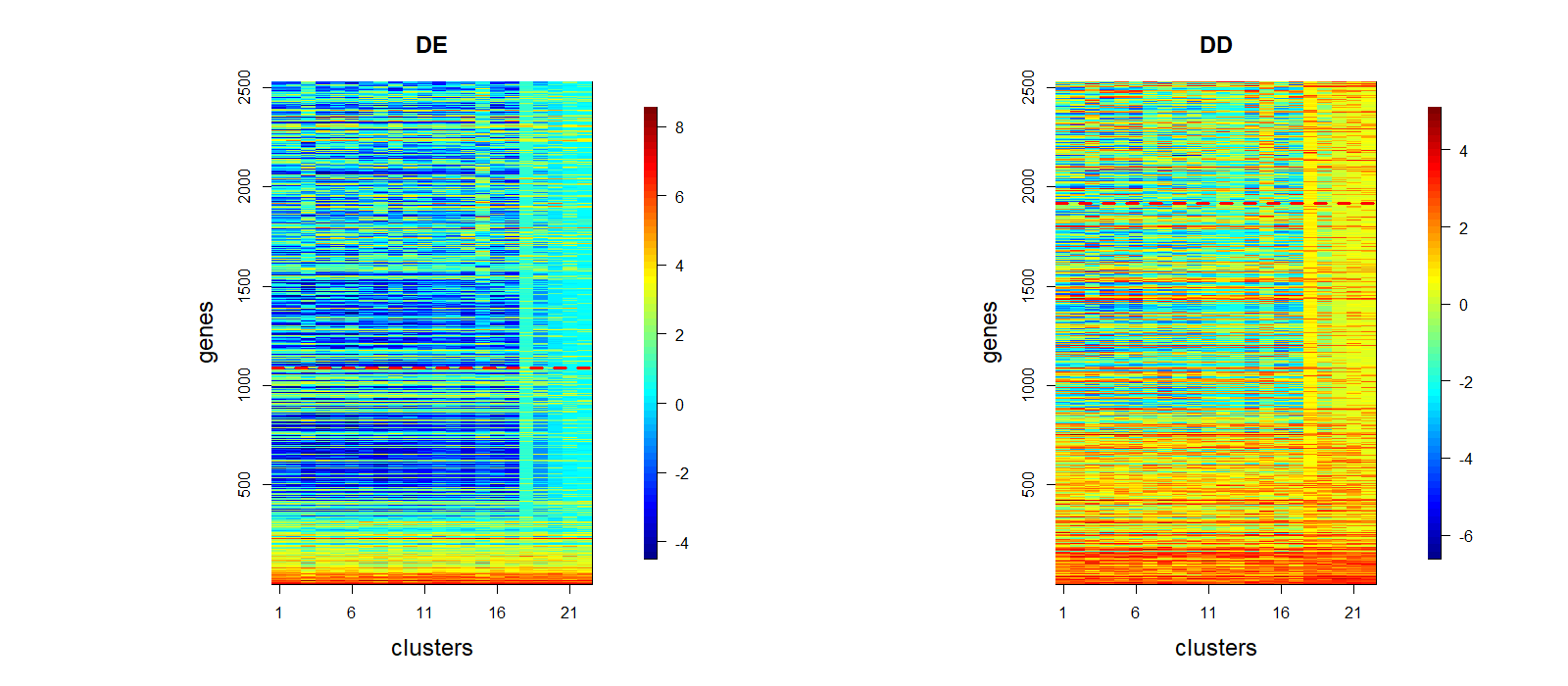}}
    \caption{Heat-maps of estimated unique parameters (mean expression on the left and dispersion on the right) on the log scale, with columns representing cell subtypes and rows representing genes. Genes are reordered by tail probabilities; tail probabilities decrease as we move down the rows, with the horizontal dashed lines separating global marker genes from non-marker genes.}
    \label{fig:heatmap_global1}
\end{figure}

The detected global marker genes for differential expression and dispersion  (with $\tau_0 = 2.5$ and $\omega_0 = 2.5$) are summarized in Figures \ref{fig:global_marker_summary} and \ref{fig:heatmap_global1} and \ref{sec:global_marker_genes}. In this case, 57\% of genes are global markers for DE and 24\% of genes are global markers for DD. To visualize the detected global marker genes across cell subtyes, we include heat-maps of the posterior mean of the subtype-specific parameters, with rows representing genes (reordered by gene-wise tail probabilities) and columns representing cell subtyes. 
We observe that the rare cell subtypes, namely the over-represented cluster 18, the under-represented cluster 21, and the stable clusters 20 and 22, have slightly higher expression levels for genes that are lowly expressed in other cell subtypes and, in general, higher dispersion parameters across most of the the global DD genes and some of the global non-DD genes.

\subsection{Local Marker Genes}
\begin{figure}[!t]
    \centering
    \subfigure{\includegraphics[width=1\textwidth]{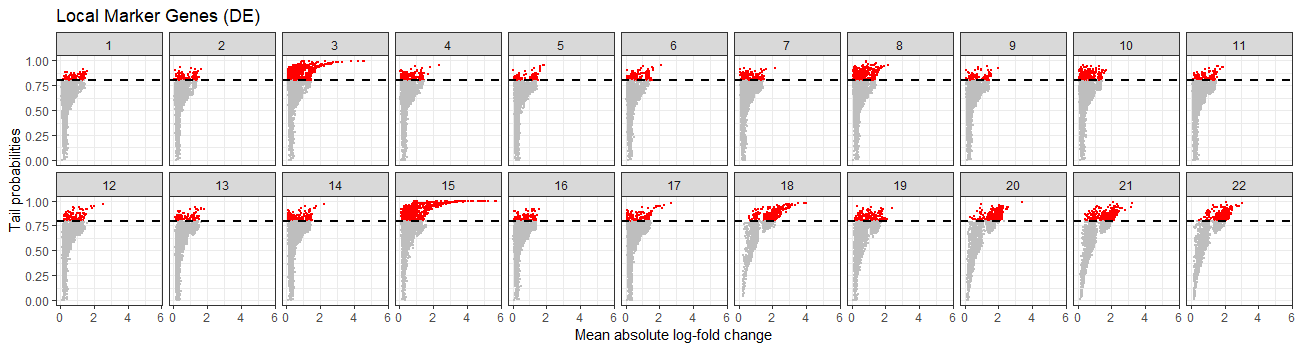}}
    \subfigure{\includegraphics[width=1\textwidth]{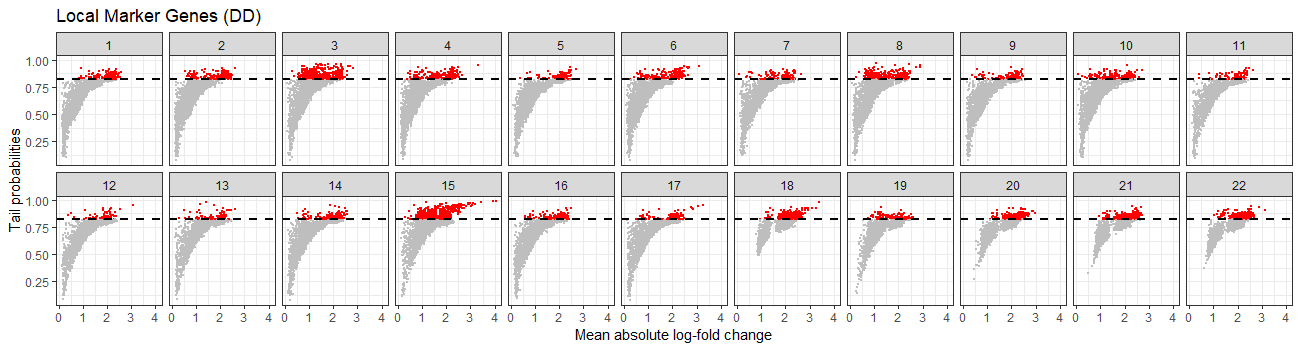}}
    \subfigure{\includegraphics[width=0.45\textwidth]{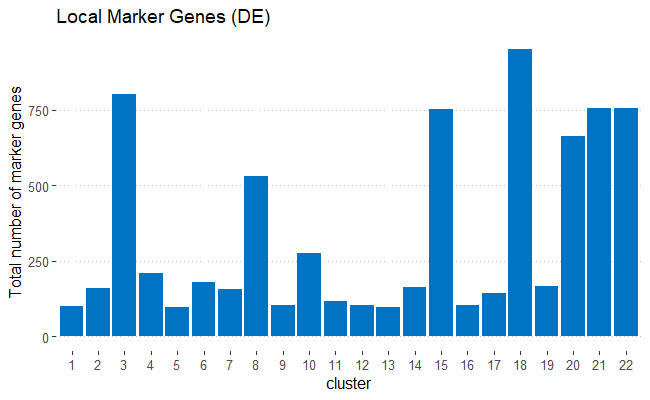}}
    \subfigure{\includegraphics[width=0.45\textwidth]{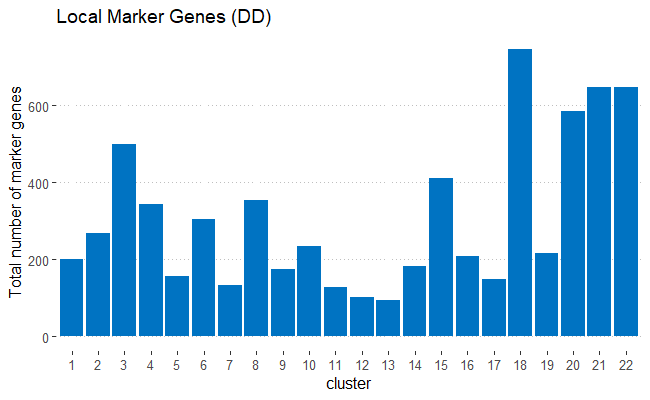}}
    \caption{Detection of local marker genes. Top rows: plots of tail probabilities against mean absolute LFCs for each cell subtype, with mean expression in the top row and dispersion in the middle row. Bottom row: a summary of the number of local DE (left) and DD genes (right).}
    \label{fig:local_marker_summary_1}
\end{figure}

For each cell subtype, we detect local marker genes  to identify distinct gene expression patterns in the current cell subtype in comparison with all other subtypes. Threshold values for the LFC are set to $0.8$. Figure \ref{fig:local_marker_summary_1} plots the absolute LFC against tail probabilities together with summary plots.
We observe that the over-represented cell subtypes $3,8,15$ and the rare cell subtypes $18,20,21,22$ have high numbers of local marker genes in terms of both DE and DD. For DE, $1752$ genes are classified as local DE genes for more than one cell subtype and gene \textit{Neurod4} is classified as a local DE gene for the most cell subtypes ($16$ cell subtypes). For DD, $1234$ genes are classified as local DD genes for more than one cell subtype and genes \textit{Kcnma1} and \textit{Scgn} are classified as a local DD gene for the most cell subtypes ($18$ cell subtypes). Heat-maps of the estimated mean expressions and dispersions for local marker genes are shown in Web Appendix \ref{sec:local_marker_genes}. We observe that the local marker DE genes for the over-represented cell subtype $15$ tend to be more highly expressed in this subtype. For the rare cell subtypes $18 - 22$, the local DE genes tends to be lowly expressed in all other cell subtypes, with slightly higher expression in cell subtype $15$, and most of the local DD genes have higher dispersions (less over-dispersion). In addition, for the remaining cell subtypes $1-14$, $16$ and $17$, most local DD genes have smaller dispersions, thus, these genes are over-dispersed in the corresponding cell subtypes.

Further findings and discussions on specific \textit{important} genes provided by the research group of Prof. Price  can be found in Web Appendix \ref{Web Appendix D}.

\subsection{Posterior Predictive Checks} \label{sec:ppc_summary}

\begin{figure}[!t]
\centering
\subfigure{\includegraphics[width=\textwidth]{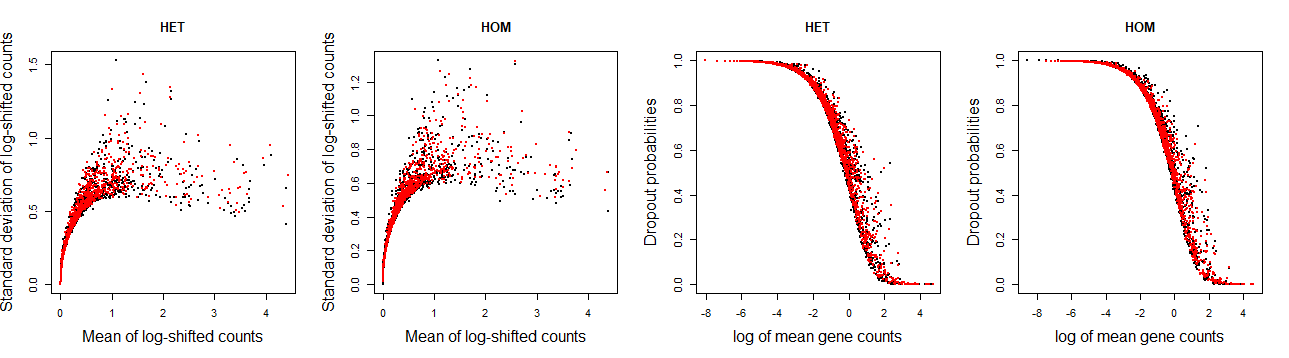}}
\caption{Posterior predictive checks for the experimental data. Comparing the relationships between key statistics (namely, the mean, standard deviation, and dropout probability) for the observed dataset (in red) and the replicated dataset (in black).}
\label{fig:posterior_predictice_check}
\end{figure}

\begin{figure}[!t]
\centering
\subfigure{\includegraphics[width=0.9\textwidth]{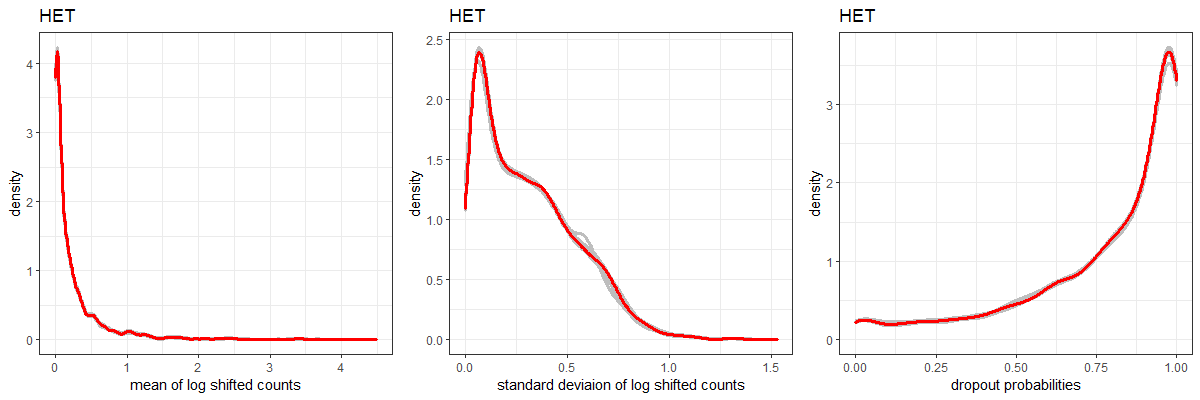}}
\subfigure{\includegraphics[width=0.9\textwidth]{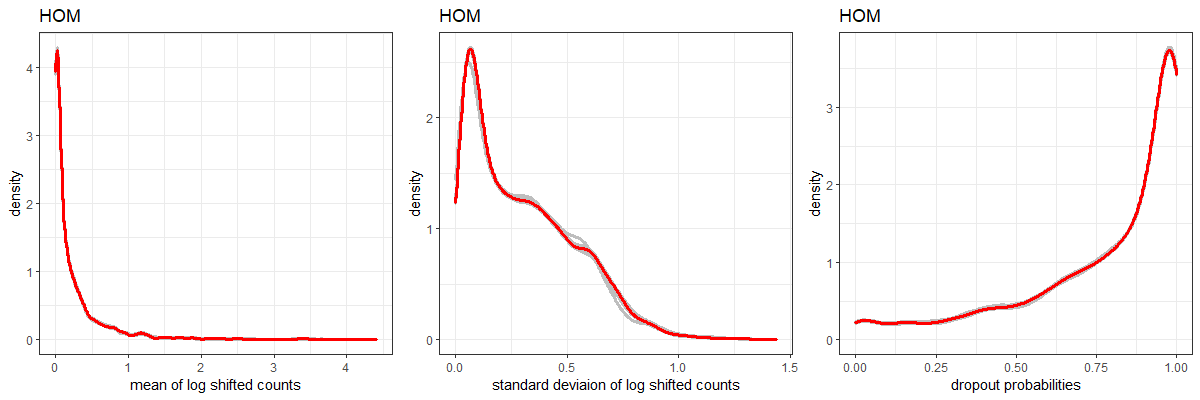}}
\caption{Comparison of the KDE of statistics from the observed and replicated datasets, with observed dataset in red and replicated datasets in grey.}
\label{fig:posterior_predictice_check_multiple}
\end{figure}

To assess the fit of the proposed NormHDP model to the experimental data, we carry out  mixed posterior predictive checks, as described in Section \ref{sec:ppp}. For a single replicated dataset, we compare key statistics, namely, the mean and standard deviation of the log shifted counts and the proportion of dropouts for each gene. %, for the observed and replicated dataset. Namely, the statistics are the mean, standard deviation, and proportion of dropouts for each gene. 
The statistics of the replicated dataset match well the observed data, highlighting the sensible fit of the model to the data (Figure \ref{fig:posterior_predictice_check}). For multiple replicates, we compare the KDEs of these statistics between the observed dataset and replicated datasets; the KDEs are similar, further supporting the model fit (Figure \ref{fig:posterior_predictice_check_multiple}).

\section{Discussion} \label{sec:discussion}

In this article, we constructed a novel hierarchical Bayesian model that integrates multiple RNA-seq datasets for shared clustering and identification of cell types. Moreover, it addresses the  challenges of  normalization and imputation in scRNA-seq by building on the \textit{bayNorm} model. In this high-dimensional setting, the proposed model allows borrowing of information and detection of differences in cell-type proportions across datasets as well as measures of uncertainty in the estimated clustering and detection of marker genes to characterize the patterns within each cell type. In simulated datasets, our proposed NormHDP model is robust and able to recover the true parameters and clustering, as well detect correctly the marker genes. 

Our work was motivated by experimental scRNA-seq data collected to shed light on the role of PAX6 in prenatal development. By applying NormHDP, we can identify cell types that form in this early stage of  development and gain an understanding of the differences in cell type presence/absence when PAX6 is deleted.
In the experimental data, the model estimates a total of 22 cell types, with four rare cell types. Among the identified cell types, nine are over-represented and seven are under-represented when PAX6 is deleted. However, although differences between the control and mutant groups are detected with a high posterior probability, the posterior mean absolute value of the difference tends to be small, suggesting that PAX6 plays a smaller role at this early stage in the development (E13.5). Following this, our colleagues at the Centre for Discovery Brain Sciences have collected additional data at day E14.5. In ongoing work, initial results suggest a stronger role of PAX6 at this slightly later stage of development, and a comprehensive analysis will be conducted in future work to integrate all datasets. 
More generally, experiments now routinely collect multiple scRNA-seq datasets, and the proposed model is relevant in such applications for shared clustering and borrowing of information and understanding differences across datasets.

For posterior inference, we have developed a Gibbs sampling algorithm, which produces asymptotically exact posterior samples. 
However, in high-dimensional settings, such as scRNA-seq, where the number of cells and genes are typically in the thousands, such algorithms suffer from mixing convergence issues. To combat this and improve computational speed, we have employed a parallel consensus clustering approach.
However, a larger number of iterations may be required, and in ongoing work, we are also developing a variational Bayes approximation \citep{hughes2015reliable} for faster, approximate inference.
Other potential model extensions of interest include cluster-specific mean-dispersion relationships for increased flexibility and incorporation of covariate information (such as cell-specific latent time) \citep{bergen2020generalizing}, as well as priors for clustering beyond the HDP \citep{argiento2020}.

\section*{Acknowledgements}

We thank Dr. Tan, Prof. Price, Prof. Mason, Dr. Kozic and their team for providing  the datasets, as well as for the insightful descriptions, motivations, suggestions and comments. We also thank Dr. Catalina Vallejos at the MRC Human Genetics Unit, University of Edinburgh for valuable suggestions and comments. \vspace*{-8pt}

\section*{Supporting Information}
Appendices referenced in Sections \ref{s:model}, \ref{sec:post}, \ref{sec:sim}, \ref{sec:real}, and \ref{sec:discussion} are available online.  All code for model implementation and analysis is publicly available through the Github repository (\url{https://github.com/jinluliu550/normHDP}), along with the simulated data (Section \ref{sec:sim}).
\vspace*{-8pt}

\bibliographystyle{plainnat}
\bibliography{main}

\end{document}

\maketitle

In Web Appendix \ref{Web Appendix A}, we provide further model insights on the capture efficiencies, describe the MCMC algorithm for posterior inference, explain the derivation of \textit{bayNorm} estimates of the capture efficiencies, and provide pseudo-code for replicating data from the mixed posterior predictive.
In Web Appendix \ref{Web Appendix B}, we include the details of simulated examples, including data generation, results and additional experiment to investigate misspecification of the mean capture efficiency. 
In Web Appendix \ref{web_appendix_PAX6}, we provide further details on the PAX6 data analysis, including the the filtering process; MCMC summaries; \textit{global} marker genes; \textit{local} marker genes; t-SNE plots and posterior predictive checks. 
In Web Appendix \ref{Web Appendix D}, we show the results corresponding to the set of \textit{important} genes.

\pagebreak

\pagebreak
\tableofcontents

\appendix
%\appendixpage
\addappheadtotoc
\setcounter{table}{0}
\setcounter{figure}{0}
\renewcommand{\thetable}{\Alph{section}.\arabic{table}}
\renewcommand{\thefigure}{\Alph{section}.\arabic{figure}}
\renewcommand{\figurename}{Web Figure}
\renewcommand{\tablename}{Web Table}

\pagebreak

\section{Posterior Inference and Model Insights} \label{Web Appendix A}

\subsection{Gene Dependence through Capture Efficiencies}\label{sec:gene_dep}

\begin{figure}[!t]
    \centering
    \subfigure[Independent: $\E(\beta_c)=0.06$]{\includegraphics[width=0.3\textwidth]{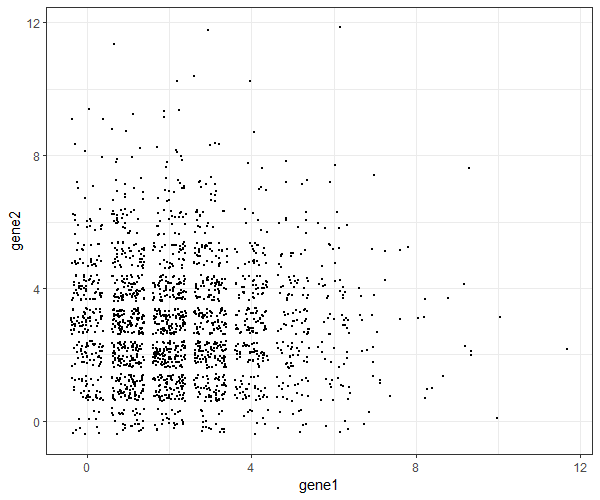}}
    \subfigure[$\E(\beta_c)=0.06$, $\mathbb{V}(\beta_c)=0.0004$]{\includegraphics[width=0.3\textwidth]{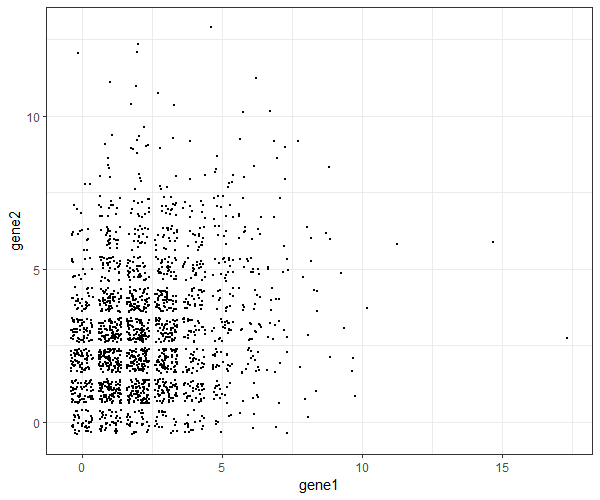}}
    \subfigure[$\E(\beta_c)=0.06$, $\mathbb{V}(\beta_c)=0.05$]{\includegraphics[width=0.3\textwidth]{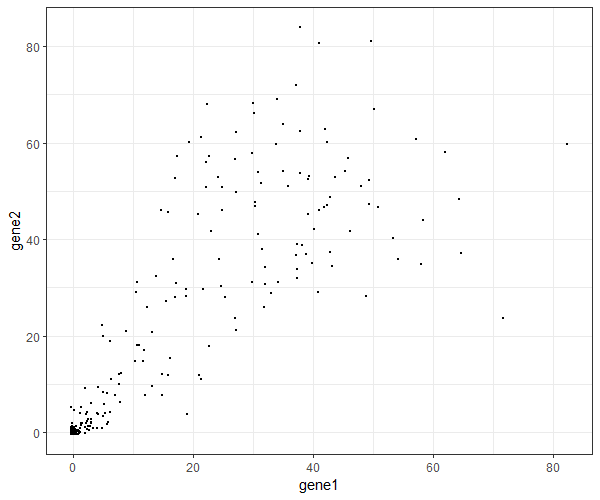}}
    \subfigure[Independent: $\E(\beta_c)=0.12$]{\includegraphics[width=0.3\textwidth]{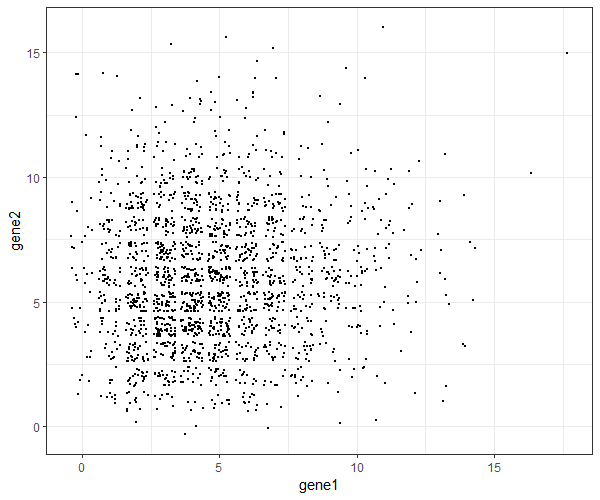}}
    \subfigure[$\E(\beta_c)=0.12$, $\mathbb{V}(\beta_c)=0.0004$]{\includegraphics[width=0.3\textwidth]{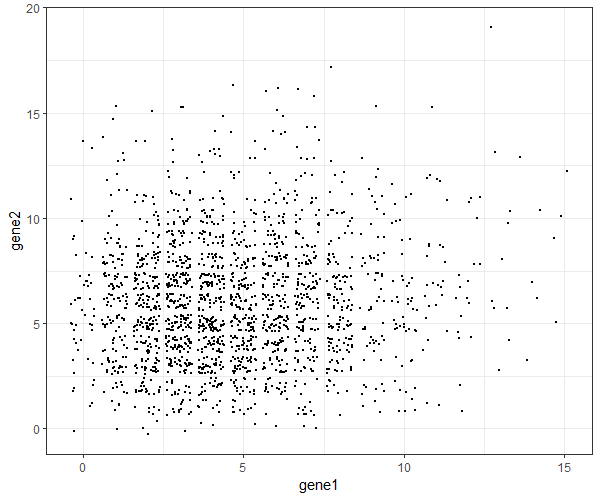}}
    \subfigure[$\E(\beta_c)=0.12$, $\mathbb{V}(\beta_c)=0.05$]{\includegraphics[width=0.3\textwidth]{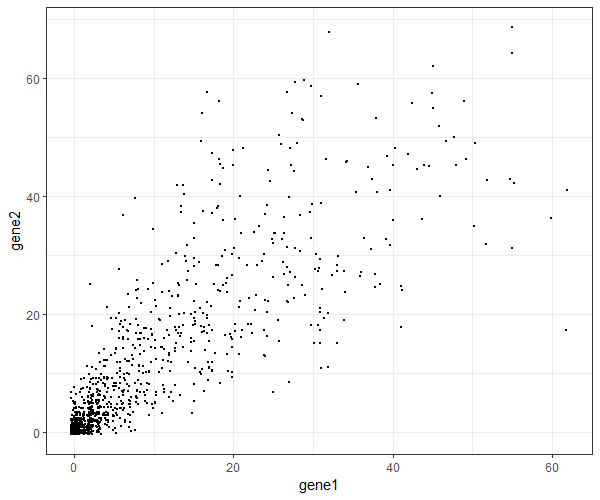}}
    \subfigure[Independent: $\E(\beta_c)=0.2$]{\includegraphics[width=0.3\textwidth]{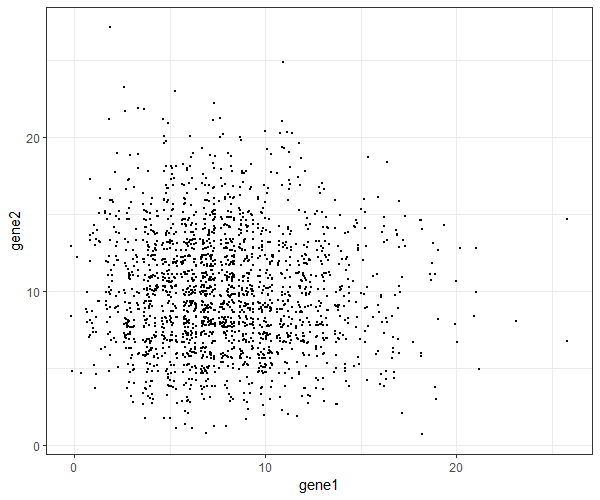}}
    \subfigure[$\E(\beta_c)=0.2$, $\mathbb{V}(\beta_c)=0.0004$]{\includegraphics[width=0.3\textwidth]{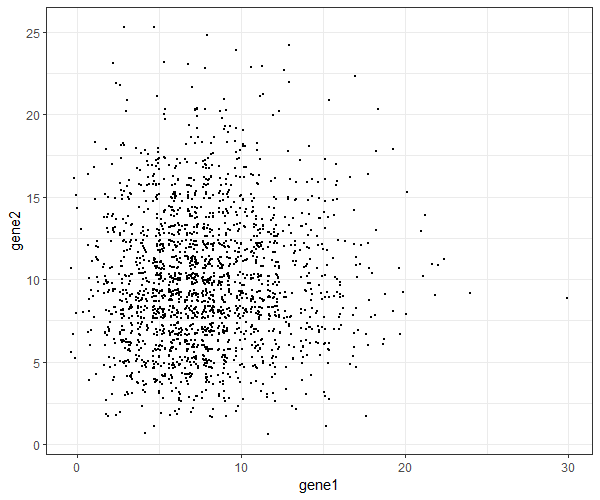}}
    \subfigure[$\E(\beta_c)=0.2$, $\mathbb{V}(\beta_c)=0.05$]{\includegraphics[width=0.3\textwidth]{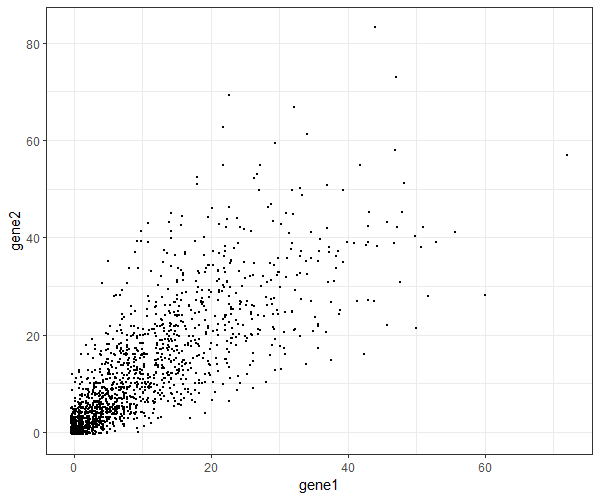}}
    \caption{Simulated counts for $C=2,000$ cells and two genes to examine dependence induced through the capture efficiencies. Rows correspond to increasing prior mean of the capture efficiencies, and column correspond to increasing prior variance.}
    \label{fig:dependence}
\end{figure}

The cell-specific capture efficiencies play an important role in normalization and imputation to obtain the latent counts $y_{c,g,d}^0$. Following other fields, where under-reporting of counts also occurs, we alleviate identifiability issues by employing an informative, empirical prior based on the \textit{bayNorm} estimates and global mean capture efficiency. As suggested by one reviewer, the capture efficiencies also play an interesting role in inducing dependence across the genes. Specifically,  for each cell,
dependence across genes is obtained after integrating over the capture efficiency $\beta_{c,d}$:
$$ p(y_{c,\cdot,d} \mid z_{c,d},\bm{\mu}_{j},\bm{\phi}_{j} ) = \int \prod_{g=1}^G p(y_{c,g,d} \mid z_{c,d},\mu_{j,g},\phi_{j,g},\beta_{c,d} ) \pi(\beta_{c,d}) d \beta_{c,d},$$
where $\beta_{c,d} \sim \text{Beta}(a_d^\beta, b_d^\beta)$. 
While the integral above is not available in closed form, we can simulate to understand better the induced dependence. For two genes with mean expression and dispersion set to $(40,10)$ and $(50,20)$ respectively, we simulate $y_{c,g}$ for $C=2,000$ cells and under different scenarios for the hyperparameters $a_d^\beta$ and $b_d^\beta$. The results are shown in Web Figure \ref{fig:dependence}, with rows corresponding to increasing  prior mean of the capture efficiencies, equal to $0.06, 0.12, 0.2$, and columns corresponding to increasing prior variance, equal to $0, 0.0004,0.05$. For larger variance, higher dependence is evident, but we note that for the middle column (with variance similar to the empirical variance from the bayNorm estimates), no clear dependence structure is observed.

\clearpage

\subsection{MCMC} 
In this section, we describe the MCMC algorithm for posterior inference with the NormHDP model. In the following, we assume the relationship between the mean and dispersion parameters is linear on the log scale. Let $\bZ=(z_{c,d})_{c=1,d=1}^{C_d,D}$;  $\bY=(y_{c,d,g})_{c=1,d=1, g=1}^{C_d,D,G}$;  $\bp_d=(p^J_{1,d},\ldots, p^J_{J,d})$; $\bp=(p^J_{1},\ldots, p^J_{J})$;  $\bbeta=(\beta_{c,d})_{c=1,d=1}^{C_d,D}$.  The posterior of interest is:
\begin{align*}
\pi(\bZ,\bp_d,&\bp,\bmu^*_{1:J},\bphi^*_{1:J},\bbeta,\alpha,\alpha_0,\bbv,a_\phi^2 |\bY)  \propto \prod_{j=1}^J \prod_{(c,d): z_{c,d}=j} \prod_{g=1}^G \NegB(y_{c,d,g} \mid \mu_{j,g}^* \beta_{c,d}, \phi_{j,g}^* ) \\
	&\quad  *  \prod_{j=1}^J \prod_{d=1}^D p_{j,d}^{N_{j,d}} *\prod_{d=1}^D\Dir(\bp_{d} \mid \alpha \bp)* \Dir\left(\bp \mid \frac{\alpha_0}{J},\ldots, \frac{\alpha_0}{J}\right) \\
	&\quad * \prod_{j=1}^J   \prod_{g=1}^G  \text{LN}(\mu^*_{j,g}  \mid m_u, a_u^2)  \text{LN} ( \phi^*_{j,g} \mid b_0 +b_1 \log(\mu_{j,g}^*), a_\phi^2) \\
	&\quad * \prod_{d=1}^D \prod_{c=1}^{C_d} \Be(\beta_{c,d} \mid a^{\beta}_{d}, b^{\beta}_{d}) *  \Gam(\alpha \mid 1,1) * \Gam(\alpha_0 \mid 1,1) \\
	&\quad * \Norm(\bbv \mid \mathbf{m}_b, a_\phi^2 \mathbf{V}_b) *  \IG(a_\phi^2 \mid \nu_1, \nu_2),
\end{align*}
where  
$N_{j,d}=\sum_{c=1}^{C_d} \1(z_{c,d}=j)$ is the number of cells in component $j$ in dataset $d$. A Markov chain Monte Carlo (MCMC) algorithm is developed for full posterior inference. The algorithm is a Gibbs sampler which produces asymptotically exact samples from the posterior by iteratively sampling the parameters in blocks corresponding to the:
\begin{itemize}
	\item allocation variables $\bZ| \bp_d,\bmu^*_{1:J},\bphi^*_{1:J},\bY, \bbeta$,
	\item dataset-specific component probabilities $\bp_d | \bZ,\bp,\alpha$,
	\item component probabilities $\bp| \bp_{1:D},\alpha_0$,
	\item unique parameters $\bmu^*_{j}, \bphi^*_{j} \mid \bZ,\bbv,a_\phi^2,\bY, \bbeta$,
	\item concentration parameters $\alpha | \bp_{1:D},\bp$ and  $\alpha_0 |\bp$,
	\item mean-dispersion hyperparameters $ \bbv,a_\phi^2 | \bmu^*_{1:J}, \bphi^*_{1:J}$,
	\item capture efficiencies $\bbeta | \bY,\bZ, \bmu^*_{1:J}, \bphi^*_{1:J}$.
\end{itemize}
The complexity of the algorithm is dominated by the update of the allocation variables, which is of order $\mathcal{O}(\text{sum}(C)JG)$. In the following subsections, we describe the update for each block of parameters.

\subsubsection{Mean-dispersion Hyperparameters} \label{sec:mean_dispersion_hyper}

The full conditional distribution for the mean-dispersion hyperparameters in the linear case is: 
\begin{align}
	&\pi(\bbv, \alpha_\phi^2 | \boldsymbol{\mu}_{1:J}^*, \boldsymbol{\phi}_{1:J}^*) \propto \Norm (\bbv | \mathbf{m}_b, \alpha_\phi^2 \mathbf{V}_b) * \IG (\alpha_\phi^2 | v_1, v_2) * \prod_{j=1}^J \prod_{g=1}^G \logNorm (\phi_{j,g}^* | b_0 + b_1 \log(\mu_{j,g}^*), \alpha_\phi^2) \nonumber \\
	&\quad \quad \propto \left( \frac{1}{\alpha_\phi^2} \right)^{v_1+2+\frac{JG}{2}} \exp \Bigg( -\frac{1}{\alpha_\phi^2} \Bigg[\frac{1}{2} \sum_{j=1}^J \sum_{g=1}^G (\ln(\phi_{j,g}^*)-b_0-b_1\log(\mu_{j,g}^*))^2 + v_2 \nonumber \\
 &\quad \quad \quad \quad \quad + \frac{1}{2}[(b_0-m_{b0})^2 + (b_1-m_{b1})^2] \Bigg] \Bigg).
	\label{eq:b_marginal}
\end{align}
Following eq. \eqref{eq:b_marginal}, the full conditional for $\bbv$ conditioning on $\alpha_\phi^2$ is:
\begin{align*}
	\pi(\bbv &| \boldsymbol{\mu}_{1:J}^*, \boldsymbol{\phi}_{1:J}^*, \alpha_\phi^2) \propto \exp \Bigg( -\frac{1}{2\alpha_\phi^2} \Bigg[ \sum_{j=1}^J \sum_{g=1}^G (\ln(\phi_{j,g}^*)-b_0-b_1\log(\mu_{j,g}^*))^2\\
 &\quad \quad \quad \quad + [(b_0-m_{b0})^2 + (b_1-m_{b1})^2] \Bigg] \Bigg).
\end{align*}
Hence we have the full conditional:
\begin{align}
	\bbv | \boldsymbol{\mu}_{1:J}^*, \boldsymbol{\phi}_{1:J}^*, \alpha_\phi^2 \sim \Norm (\Tilde{\mathbf{m}}_b, \alpha_\phi^2 \Tilde{\mathbf{V}}_b),
	\label{eq:b_standard}
\end{align}
where
\begin{align*}
	\Tilde{\mathbf{m}}_b &= \left( \sum_{j=1}^J \Tilde{\bmu}_j^T \Tilde{\bmu}_j + I \right)^{-1} \left( \sum_{j=1}^J  \Tilde{\bmu}_j^T \ln(\bphi_j^*) + \mathbf{m}_b \right), \\
	\Tilde{\mathbf{V}}_b &= \left( \sum_{j=1}^J \Tilde{\bmu}_j^T \Tilde{\bmu}_j + I \right)^{-1},
\end{align*}
and
\begin{equation*}
	\ln(\boldsymbol{\phi}_j^*) = 
	\begin{bmatrix}
		\ln (\phi_{j,1}^*) \\
		\vdots \\
		\ln (\phi_{j,G}^*)
	\end{bmatrix}, \quad
	\Tilde{\boldsymbol{\mu}}_j =
	\begin{bmatrix}
		1 & \log (\mu_{j,1}^*) \\
		\vdots & \vdots \\
		1 & \log (\mu_{j,G}^*)
	\end{bmatrix}. %, \quad
	%\textbf{b} =
	%\begin{bmatrix}
	%    b_0 \\
	%    b_1
	%\end{bmatrix}.
\end{equation*}

\noindent And following eq.  \eqref{eq:b_marginal}, the full conditional for $\alpha_\phi^2$ is:
\begin{align*}
	\pi(\alpha_\phi^2 &| \boldsymbol{\mu}_{1:J}^*, \boldsymbol{\phi}_{1:J}^*) = \int \pi( \bbv, \alpha_\phi^2 | \boldsymbol{\mu}_{1:J}^*, \boldsymbol{\phi}_{1:J}^*) \, d\bbv \\
	&\propto \int \left(\frac{1}{\alpha_\phi^2}\right)^{v_1+1} \exp \left( -\frac{v_2}{\alpha_\phi^2}\right) \left( \frac{1}{\alpha_\phi^2}\right)^{JG/2} \left( \frac{1}{\alpha_\phi^2}\right) \\
	&*\exp \left( -\frac{1}{2\alpha_\phi^2} \left[ (\bbv-\Tilde{\mathbf{m}}_b)^T \Tilde{\mathbf{V}}_b^{-1} (\bbv-\Tilde{\mathbf{m}}_b) - \Tilde{\mathbf{m}}_b^T \Tilde{\mathbf{V}}_b^{-1} \Tilde{\mathbf{m}}_b + \sum_{j=1}^J \ln(\bphi_j^*)^T  \ln(\bphi_j^*) +
 \sum \mathbf{m}_b^2 \right] \right) \, d\bbv.
\end{align*}
%We simplify the above Equation by considering the cumulative density of an Inverse-Gamma distribution to have:
Thus, we have: 
\begin{align}
	\alpha_\phi^2 | \boldsymbol{\mu}_{1:J}^*, \boldsymbol{\phi}_{1:J}^* \sim \IG \left(\Tilde{v_1}, \Tilde{v_2}\right),
	\label{eq:alpha_phi_standard}
\end{align}
where
\begin{align*}
	\Tilde{v_1} &= v_1 + JG/2, \\
	\Tilde{v_2} &= v_2 + \frac{1}{2} \left( \sum_{j=1}^J \ln(\bphi_j^*)^T  \ln(\bphi_j^*) - \Tilde{\mathbf{m}}_b^T \Tilde{\mathbf{V}}_b^{-1} \Tilde{\mathbf{m}}_b + \sum \mathbf{m}_b^2 \right).
\end{align*}
For Gibbs sampling, at each iteration, we first simulate $\alpha_\phi^2$ from the Inverse-Gamma distribution in eq.  \eqref{eq:alpha_phi_standard} and then conditioned on this value, simulate $\bbv$ from the Normal distribution in eq. \eqref{eq:b_standard}. Details are given in Algorithm \ref{alg:sim_hyper}.

\begin{algorithm}
	\caption{Simulation of Mean-dispersion Hyperparameters}\label{alg:sim_hyper}
	\begin{algorithmic}
		\Require $\bmu_{1:J}^*, \bphi_{1:J}^*$
		\Ensure $\alpha_\phi^2, \textbf{b}$
		\State $A \gets \text{Identity matrix of dimension 2}$;
		\State $B \gets \text{Matrix of $\mathbf{m}_b$}$;
		\State $C \gets 0$;
		\For{$j = 1, \dots, J$}
		\State $A = A + \Tilde{\bmu}_j^T \Tilde{\bmu}_j$; \quad $B = B + \Tilde{\bmu}_j^T \ln (\bphi_j^*)$; \quad $C = C + \ln (\bphi_j^*)^T \ln (\bphi_j^*)$;
		\EndFor
		\State $\Tilde{\mathbf{V}}_b \gets A^{-1}$; \quad $\Tilde{\mathbf{m}}_b \gets \Tilde{\mathbf{V}}_b A$; \quad $\Tilde{v}_1 \gets v_1 + JG/2$; \quad $\Tilde{v}_2 \gets v_2 + 1/2 * (C - \Tilde{\mathbf{m}}_b^T \Tilde{\mathbf{V}}_b^{-1}\Tilde{\mathbf{m}}_b + \sum \mathbf{m}_b^2)$;
		\State Simulate $\alpha_\phi^2$ and $\bbv$ from the NIG in eq. \ref{eq:alpha_phi_standard} and \ref{eq:b_standard} using the above parameters.
	\end{algorithmic}
\end{algorithm}

\subsubsection{Allocation Variables} \label{sec:allocation}

The full conditional for the allocation variables is:
\begin{align*}
	\pi(\bZ | \bP_{1:D}, \bmu_{1:J}^*, \bphi_{1:J}^*, \bY, \bbeta) &\propto \prod_{j=1}^J \prod_{(c,d):z_{c,d} =j} \prod_{g=1}^G \NegB (y_{c,g,d} | \mu_{j,g}^* \beta_{c,d}, \phi_{j,g}^*)  \prod_{j=1}^J \prod_{d=1}^D p_{j,d}^{N_{j,d}},
\end{align*}
where $N_{j,d}$ is the total number of cells in component $j$ in dataset $d$. Thus, since the allocations are conditionally independent, we have the full conditional of $z$ for cell $c$ in dataset $d$ as:
\begin{align}
	z_{c,d} \sim \Cat \left(\pi_{c,d,1}, \dots, \pi_{c,d,J} \right),
	\label{eq:z_fullcond}
\end{align}
where $\pi_{c,d,j} =  \pi(z_{c,d} =j \mid \bP_{1:D}, \bmu_{1:J}^*, \bphi_{1:J}^*, \bY, \bbeta )$. To avoid numerical errors, we employ the log-sum trick to compute the probabilities:
\begin{align*}
	\pi_{c,d,j}  =
	\frac{\exp(\log(\Tilde{\pi}_{c,d,j})+ \log(K)}{\sum_{j=1}^J \exp(\log(\Tilde{\pi}_{c,d,j})+\log(K))}.
\end{align*}
where 
\begin{align*}
	\Tilde{\pi}_{c,d,j} &= \prod_{g=1}^G \NegB (y_{c,g,d} \mid \mu_{j,g}^* \beta_{c,d}, \phi_{j,g}^*) p_{j,d},\\
	\log(K) &= -\max_j \log(\Tilde{\pi}_{c,d,j}).
\end{align*}
Details are given in Algorithm \ref{alg:sim_allocation}.
\begin{algorithm}
	\caption{Simulation of Allocation Variables}\label{alg:sim_allocation}
	\begin{algorithmic}
		\Require $\textbf{P}_{1:D}, \bmu_{1,J}^*, \bphi_{1,J}^*, \by, \bbeta$
		\Ensure $\bZ$
		\State $Z \gets \text{a list of length} \; D$;
		\For{$d = 1, \dots, D$ and $c = 1, \dots, C_d$}
		\State Compute $\Tilde{\pi}_{c,d,j}$ for each $j$; and $K$,
		\State Use $\Tilde{\pi}_{c,d,j}$ and $K$ to compute $\pi_{c,d,j}$ and simulate $z_{c,d}$ from the Categorical distribution in eq. \eqref{eq:z_fullcond}.
		\EndFor
	\end{algorithmic}
\end{algorithm}

\subsubsection{Dataset-specific Component Probabilities} \label{sec:p_jd}

The full conditional distribution for the dataset-specific component probabilities is:
\begin{align*}
	\pi(\bp_d | \bZ, \bp, \alpha) & \propto \prod_{j=1}^J p_{j,d}^{N_{j,d}} \Dir( \bp_d | \alpha \bp) 
	\propto \prod_{j=1}^J p_{j,d}^{N_{j,d} + \alpha p_j - 1}.
\end{align*}
Hence, $\textbf{p}_d | \textbf{Z}, \textbf{p}, \alpha$ follows a Dirichlet distribution with parameters equals to $N_{j,d} + \alpha p_j$, for $j=1, \ldots, J$. 
Details are given in Algorithm \ref{alg:sim_pd}. 

\begin{algorithm}[H]
	\caption{Simulation of Dataset-specific Component Probabilities}\label{alg:sim_pd}
	\begin{algorithmic}
		\Require $\bZ, \bP, \alpha$
		\Ensure $\bp_{1:D}$
		\State $\bp_{1:D} \gets \text{Matrix with J rows and D columns}$;
		\For{$d = 1, \dots, D$}
		\State $\Tilde{\bm{\alpha}} \gets \text{table}(\bZ_d) + \alpha \bp$
		\State Simulate $\bp_d$ from the Dirichlet distribution using the updated parameters $\Tilde{\bm{\alpha}}$.
		\EndFor
	\end{algorithmic}
\end{algorithm}

\subsubsection{Component Probabilities} \label{sec:p_j}

The full conditional distribution for the component probabilities is:
\begin{align}
	&\pi(\bp | \bp_{1:D}, \alpha_0, \alpha) \propto \Dir(\bp | \frac{\alpha_0}{J}, \dots, \frac{\alpha_0}{J}) *\prod_{d=1}^D \Dir(\bp_d | \alpha \bp) \nonumber \\
	&\quad \quad \propto \left[\prod_{j=1}^J p_j^{\frac{\alpha_0}{J} - 1} \right] \prod_{d=1}^D \frac{1}{B(\alpha\bp)} \prod_{j=1}^J p_{j,d}^{\alpha p_j}
	*\textbf{1} \left(p_j > 0, \forall j, \quad \sum_{j=1}^J p_j = 1 \right).
	\label{eq:sim:component_probabilities}
\end{align}
As the full conditional distribution has no closed-form, we will use adaptive Metropolis-Hastings \citep{griffin2013advances}  to obtain posterior samples of $\bp$.  The log of the full conditional in eq. \eqref{eq:sim:component_probabilities} can be written as:
\begin{align*}
	\log \pi (\bp|\dots) = \sum_{j=1}^J \left[ \left(\frac{\alpha_0}{J} - 1\right) \log (p_j) \right] + \sum_{d=1}^D \sum_{j=1}^J \left[\alpha p_j \log (p_{j,d}) - \log \Gamma(\alpha p_j) \right] + \text{const.}
\end{align*}

\paragraph{Adaptive Metropolis-Hastings for $\bp$} \label{sec:amh1}
In the following, we describe the steps of the adaptive Metropolis-Hastings algorithm. 
\begin{enumerate}
	\item Note that since $p_j > 0$ for all $j = 1, \dots, J$ and $\sum_{j=1}^J p_j = 1$, we apply the following transformation:
	\begin{equation*}
		\bp \in \triangle^{J-1} \mapsto \bx \in \R^{J-1}
	\end{equation*}
	where
	\begin{equation*}
		x_j = t(p_j) :=  \log \left(\frac{p_j}{p_J}\right), \quad j = 1, \dots, J-1.
	\end{equation*}
	Note that the reverse of the transformation is:
	\begin{equation*}
		p_j = \frac{e^{x_j}}{1+\sum_{j=1}^{J-1} e^{x_j}}, \quad j = 1, \dots, J-1.
	\end{equation*}
	\item For $n \leq 100$, $\bx_{new}$ is simulated from $\Norm (\bx_{old}, \textbf{I}_d)$. If $n > 100$, $\bx_{new}$ is simulated from $ \Norm (\bx_{old}, 2.4^2/d * (\bSigma_{n-1} + \epsilon  \textbf{I}_d))$, where $\bSigma_{n-1}$ is the current estimate of covariance structure of $\bx$ based on the first $n-1$ samples; $d$ is the length of the parameters of interest, i.e. $d=J-1$; and an epsilon is small constant, i.e. $\epsilon = 0.01$.
	\item To avoid re-computing $\bSigma_n$ at each iteration, we compute $\bSigma_n$ based on two statistics: $\Tilde{\bS}_{n}$ and $\textbf{m}_{n}$, which can be sequentially updated. These statistics are defined as:
	\begin{align*}
		\Tilde{\bS}_n &= 
		\begin{bmatrix}
			\sum_{i=1}^{n} x_{1,i}^2 & \sum_{i=1}^{n} x_{1,i}x_{2,i} & \cdots & \sum_{i=1}^{n} x_{1,i}x_{J-1,i} \\
			\vdots  & \vdots  & \ddots & \vdots \\
			\sum_{i=1}^{n} x_{J-1,i} x_{1,i} & \sum_{i=1}^{n} x_{J-1,i}x_{2,i} & \cdots & \sum_{i=1}^{n} x_{J-1,i}^2
		\end{bmatrix}, \\
		\textbf{m}_n &= [\sum_{i=1}^{n} x_{1,i}, \dots,\sum_{i=1}^{n} x_{J-1,i}]^T.
	\end{align*}
	%where $m_j(n-1)$ is the mean of $j^{th}$ parameter based on all $n-1$ samples. 
	We can express $\bSigma_n$ as:
	\begin{align*}
		\bSigma_n = \frac{1}{n-1}\Tilde{\bS}_n - \frac{n}{n-1}\textbf{m}_n \textbf{m}_n^T,
	\end{align*}
	where we sequentially update the required statistics:
	\begin{align*}
		\Tilde{\bS}_n &= \Tilde{\bS}_{n-1} + \textbf{x}_n \textbf{x}_n^T; \\
		\textbf{m}_n &= \left(1-\frac{1}{n}\right)\textbf{m}_n + \frac{1}{n} \textbf{x}_n.
	\end{align*}
	\item To evaluate the proposal distribution, we are required to compute the Jacobian of the transformation. More specifically, the proposal is:
	\begin{align*}
		q_n(\bp_{new} | \bp_{old}) = q_n \left(\bt(\bp_{new}) | \bt(\bp_{old})\right) |J_{\bt(\bp_{new})}|,
	\end{align*}
	where
	\begin{align*}
		J_{\textbf{t}(\textbf{p})} &= 
		\begin{pmatrix}
			\frac{dt_1}{dp_1} & \frac{dt_2}{dp_1} & \cdots & \frac{dt_{J-1}}{dp_1} \\
			\vdots  & \vdots  & \ddots & \vdots  \\
			\frac{dt_1}{dp_{J-1}} & \frac{dt_2}{dp_{J-1}} & \cdots & \frac{dt_{J-1}}{dp_{J-1}}
		\end{pmatrix} \\
		&=
		\begin{pmatrix}
			\frac{1}{p_J} & \dots & \frac{1}{p_J} \\
			\vdots & \ddots & \vdots \\
			\frac{1}{p_J} & \dots & \frac{1}{p_J}
		\end{pmatrix} + 
		\begin{pmatrix}
			\frac{1}{p_1} & 0 &\dots & 0 \\
			0 & \frac{1}{p_2} &\vdots & 0 \\
			\vdots & \vdots & \ddots & 0 \\
			0 & 0 & \dots & \frac{1}{p_{J-1}}
		\end{pmatrix} = B + A.
	\end{align*}
	Since $\det(A + B) = \det(A) + \det(B) + Tr(A^{-1} B)\det(A)$, and in our case, we have $\det(B) = 0$ and $\det(A) = \prod_{j=1}^{J-1} \frac{1}{p_j}$, hence $\det(A+B) = \prod_{j=1}^{J-1} \frac{1}{p_j} + [1-p_J] \prod_{j=1}^{J} \frac{1}{p_j} = \prod_{j=1}^{J} \frac{1}{p_j}$. And taking the log of the determinant of the Jacobian, we have
 \begin{align*}
     \log \det J_{\textbf{t}(\bP)} = \log \left[\prod_{j=1}^J \frac{1}{p_j} \right] = -\sum_{j=1}^J \log (p_j).
 \end{align*}
	\item Next, we compute the acceptance probability:
	\begin{align*}
		\alpha(\textbf{p}_{new}, \textbf{p}_{old}) &= \min \{1, \exp(\Tilde{\alpha}(\bp_{new}, \bp_{old}))\},
	\end{align*}
	where
	\begin{align*}
		\Tilde{\alpha}(\bp_{new}, \bp_{old}) &= \log \Bigg[ \frac{\pi(\bp_{new})}{\pi(\bp_{old})}
		\frac{q_n(\bp_{old} | \bp_{new})}{q_n(\bp_{new} | \bp_{old})} \Bigg] = \log \Bigg[\frac{\pi(\bp_{new})}{\pi(\bp_{old})}  \frac{|J_{\textbf{t}(\bp_{old})}|}{|J_{\textbf{t}(\bp_{new})}|} \Bigg]\\
		&= \log \Bigg[\frac{\pi(\bp_{new})}{\pi(\bp_{old})} \Bigg] -\sum_{j=1}^J \log (p_{old,j}) + \sum_{j=1}^J \log (p_{new,j}) ,
	\end{align*}
	and $\pi(\cdot)$ is the density of the target distribution.
\end{enumerate}

\subsubsection{Concentration Parameters} \label{sec:alpha}

The full conditional distribution for $\alpha$ is:
\begin{align}
	\pi(\alpha | \bp_{1:D}, \bp) &\propto \Gam(\alpha | 1,1) *\prod_{d=1}^D \Dir(\bp_d | \alpha \bp)
	\propto \exp(-\alpha) \prod_{d=1}^D \frac{1}{B(\alpha \bp)} \prod_{j=1}^J p_{j,d}^{\alpha p_j} \nonumber \\
	& \propto \exp(-\alpha) \prod_{d=1}^D \frac{\Gamma(\alpha)}{\prod_{j=1}^J \Gamma(\alpha p_j)} \prod_{j=1}^J p_{j,d}^{\alpha p_j} * \textbf{1} (\alpha > 0).
	\label{eq:alpha_conditional}
\end{align}
The full conditional distribution for $\alpha_0$ is:
\begin{align}
	\pi(\alpha_0 | \bp) &\propto \Gam(\alpha_0 | 1,1) * \Dir \left(\bp | \frac{\alpha_0}{J}, \dots, \frac{\alpha_0}{J}\right) 
	\propto \exp(-\alpha_0) \frac{1}{B(\frac{\alpha_0}{J})} \prod_{j=1}^J p_j^{\frac{\alpha_0}{J}} \nonumber\\
	& \propto \exp(-\alpha_0) \frac{\Gamma(\alpha_0)}{[\Gamma(\frac{\alpha_0}{J})]^J} \prod_{j=1}^J p_j^{\frac{\alpha_0}{J}} * \textbf{1} (\alpha_0 > 0).
	\label{eq:alpha_0_conditional}
\end{align}
We obtain no closed-form distributions for both concentration parameters $\alpha$ and $\alpha_0$, hence we apply adaptive Metropolis-Hastings to obtain posterior samples. The log of the full condition in eq. \eqref{eq:alpha_conditional} is:
\begin{align*}
	\log \pi(\alpha | \bp_{1:D}, \bp) = -\alpha + D\log(\Gamma(\alpha)) + \sum_{d=1}^D \sum_{j=1}^J \left[ \alpha p_j \log(p_{j,d}) - \log(\Gamma(\alpha p_j)) \right] + \text{const.},
\end{align*}
and in eq. \eqref{eq:alpha_0_conditional} is:
\begin{align*}
	\log \pi(\alpha_0 | \bp) = -\alpha_0 + \log(\Gamma(\alpha_0)) - J\log(\Gamma(\frac{\alpha_0}{J})) + \sum_{j=1}^J \frac{\alpha_0}{J} \log(p_j)+ \text{const}.
\end{align*}

\paragraph{Adaptive Metropolis-Hastings for $\alpha$}
\begin{enumerate}
	\item Since $\alpha$ is always greater than zero, we apply the following transformation to map $\alpha$ to the real axis:
	\begin{align*}
		\alpha \in \R_+ \mapsto x \in \R,
	\end{align*}
	where $x = t(\alpha) = \log (\alpha)$.
	\item We apply an adaptive random walk in the transformed space, where the variance of proposal is adapted in similar fashion to Web Appendix \ref{sec:amh1}, with the sample variance computed by sequentially updating the required statistics.  
	\item To compute the Jacobian of the transformation, we differentiate $t(\alpha)$ with respect to $\alpha$:
	\begin{align*}
		\frac{d t(\alpha)}{d\alpha} = \frac{d\log(\alpha)}{d\alpha} = \frac{1}{\alpha}.
	\end{align*}
	\item Hence the acceptance probability simplifies to:
	\begin{align*}
		\alpha(\alpha_{new}, \alpha_{old}) = \min \{1, \Tilde{\alpha}(\alpha_{new}, \alpha_{old}) \},
	\end{align*}
	where
	\begin{align*}
		\Tilde{\alpha}(\alpha_{new}, \alpha_{old}) = \log \Bigg[ \frac{\pi(\alpha_{new})}{\pi(\alpha_{old})}
		\frac{q_n(\alpha_{old} | \alpha_{new})}{q_n(\alpha_{new} | \alpha_{old})} \Bigg] = \log
		\Bigg[ \frac{\pi(\alpha_{new})}{\pi(\alpha_{old})}\frac{\alpha_{new}}{\alpha_{old}} \Bigg].
	\end{align*}
\end{enumerate}

\subsubsection{Unique Parameters} \label{sec:mu_and_phi}

The full conditional distribution for $\bmu^*$ and $\bphi^*$ is:
\begin{align*}
	&\pi(\bmu_j^*, \bphi_j^* | \bZ, \bbv, \alpha_{\phi}^2, \bY, \bbeta) \propto \prod_{g=1}^G \logNorm (\mu_{j,g}^* | m_u, \alpha_{\mu}^2) \; \logNorm (\phi_{j,g}^* | b_0 + b_1 \log(\mu_{j,g}^*), \alpha_{\phi}^2) \\
	&\quad \quad * \prod_{(c,d):z_{c,d} =j} \prod_{g=1}^G \NegB (y_{c,g,d} | \mu_{j,g}^* \beta_{c,d}, \phi_{j,g}^*).
\end{align*}
Since the genes are conditionally independent, we can write the full condition for $\mu_{j,g}^*$ and $\phi_{j,g}^*$ as:
\begin{align}
	&\pi(\mu_{j,g}^*, \phi_{j,g}^* | \bZ, \bbv, \alpha_{\phi}^2, \bY, \beta) \propto \nonumber \\
	&\quad \quad  \prod_{(c,d):z_{c,d} =j} \binom{y_{c,g,d} + \phi_{j,g}^* - 1}{\phi_{j,g}^*-1} \left( \frac{\phi_{j,g}^*}{\mu_{j,g}^* \beta_{c,d} + \phi_{j,g}^*}\right)^{\phi_{j,g}^*} \left( \frac{\mu_{j,g}^*}{\mu_{j,g}^* \beta_{c,d} + \phi_{j,g}^*}\right)^{y_{c,g,d}} \nonumber \\
 &\quad \quad \quad \left( \frac{1}{\mu_{j,g}^* \phi_{j,g}^*} \right) \exp \left( -\frac{1}{2 \alpha_{\mu}^2} (\ln{\mu_{j,g}^*}- m_u)^2 -\frac{1}{2 \alpha_{\phi}^2} (\ln{\phi_{j,g}^*} - (b_0 + b_1\log\mu_{j,g}^*))^2\right).
	\label{eq:unique_parameters_conditional}
\end{align}
The full conditional distribution has no closed-form, hence we apply adaptive Metropolis-Hastings to obtain posterior samples. We first compute the log-likelihood of eq. \eqref{eq:unique_parameters_conditional}:
\begin{align*}
	&\log \pi\left(\mu_{j,g}^*, \phi_{j,g}^* | \dots\right) = -\log (\mu_{j,g}^*\phi_{j,g}^*) -\frac{1}{2 \alpha_{\mu}^2} (\ln{\mu_{j,g}^*}-m_u)^2 \\
 &\quad -\frac{1}{2 \alpha_{\phi}^2} (\ln{\phi_{j,g}^*} - (b_0 + b_1\log\mu_{j,g}^*))^2 \\
	&\quad + \sum_{(c,d):z_{c,d}=j} \Bigg[ \log \binom{y_{c,g,d} + \phi_{j,g}^* - 1}{\phi_{j,g}^*-1} + \phi_{j,g}^* \log \left( \frac{\phi_{j,g}^*}{\mu_{j,g}^* \beta_{c,d} + \phi_{j,g}^*}\right) \\
 &\quad \quad \quad \quad \quad + y_{c,g,d} \log \left( \frac{\mu_{j,g}^*}{\mu_{j,g}^* \beta_{c,d} + \phi_{j,g}^*}\right) \Bigg] + \text{const}.
\end{align*}

\paragraph{Adaptive Metropolis Hastings for $\mu$ and $\phi$}
\begin{enumerate}
	\item Since $(\mu^*,\phi^*)$ are always positive, we transform to the real axis:
	\begin{align*}
		(\mu^*,\phi^*) \in \R_+^2 \mapsto \bx \in \R^2,
	\end{align*}
	where $t_1(\mu^*,\phi^*) = x_1 = \log(\mu^*)$ and $t_2(\mu^*,\phi^*) = x_2 = \log(\phi^*)$.
	\item  We apply an adaptive random walk in the transformed space, where the covariance matrix of proposal is adapted in similar fashion to Web Appendix \ref{sec:amh1}, with the sample covariance matrix computed by sequentially updating the required statistics.
	\item The Jacobian matrix of the transformation is:
	\begin{align*}
		J_{\bt(\mu^*, \phi^*)} = 
		\begin{pmatrix}
			\frac{dt_1}{d\mu^*} & \frac{dt_1}{d\phi^*} \\
			\frac{dt_2}{d\mu^*} & \frac{dt_2}{d\phi^*}
		\end{pmatrix} = 
		\begin{pmatrix}
			\frac{1}{\mu^*} & 0 \\
			0 & \frac{1}{\phi^*}
		\end{pmatrix};
	\end{align*}
	and the log determinant of the Jacobian is:
	\begin{align*}
		\log \det J_{\bt(\mu^*, \phi^*)} = \log \left[\frac{1}{\mu^*\phi^*}\right] = -\log(\mu^*)-\log(\phi^*).
	\end{align*}
	
	\item The corresponding acceptance probability is given by:
	\begin{align*}
		\alpha ((\mu^*,\phi^*)_{new}, &(\mu^*,\phi^*)_{old}) = \min \{1, \exp[\Tilde{\alpha} ((\mu^*,\phi^*)_{new}, (\mu^*,\phi^*)_{old})]\},
	\end{align*}
	where
	\begin{align*}
		\Tilde{\alpha} ((\mu^*,\phi^*)_{new}, (\mu^*,\phi^*)_{old}) &= \log \Bigg[\frac{\pi((\mu^*,\phi^*)_{new})}{\pi((\mu^*,\phi^*)_{old})} \Bigg]  -\log(\mu^*_{old})\\
  &\quad - \log(\phi^*_{old})+\log(\mu^*_{new})-\log(\phi^*_{new}).
	\end{align*}
	The above process is repeated for all unique parameters corresponding to occupied components. For non-occupied components, we can sample directly from the prior.
\end{enumerate}

Note the random walk used here is specific to each cluster and gene $(j,g)$. For empty components, i.e. in the case when $\sum_{d=1}^D N_{j,d} = 0$, we sample the mean expression and dispersion from the prior. The adaptive covariance matrix is updated in either case.

\subsubsection{Capture Efficiencies} \label{sec:beta}

The full conditional distribution for $\bbeta$ is:
\begin{align*}
	\pi(\bbeta | \bY, \bZ, \mu_{1:J}^*, \phi_{1:J}^*) &\propto \prod_{j=1}^J \prod_{(c,d):z_{c,d} =j}  \prod_{g=1}^G \NegB (y_{c,g,d} | \mu_{j,g}^* \beta_{c,d}, \phi_{j,g}^*) \\
 &\quad * \prod_{d=1}^D \prod_{c=1}^{C_d} \Be (\beta_{c,d} | a_d^\beta, b_d^\beta).
\end{align*}
Since all $\beta$ are conditionally independent, they can be sampled in parallel, and we can write the full conditional distribution for cell $c$ and gene $g$ as:
\begin{align*}
	&\pi(\beta_{c,d} | \bY, \bZ, \mu_{1:J}^*, \phi_{1:J}^*) \propto \Be (\beta_{c,d} | a_d^\beta, b_d^\beta) \prod_{g=1}^G  \NegB (y_{c,g,d} | \mu_{j,g}^* \beta_{c,d}, \phi_{j,g}^*) \nonumber \\
	&\quad \propto \left[ \prod_{g=1}^G \left( \frac{1}{ \phi_{j,g}^* + \mu_{j,g}^* \beta_{c,d}} \right)^{\phi_{j,g}^* + y_{c,g,d}} (\beta_{c,d})^{y_{c,g,d}} \right] \\
 &\quad \quad \quad *(\beta_{c,d})^{a_d^\beta -1} (1-\beta_{c,d})^{b_d^\beta - 1} \; \textbf{1}(\beta_{c,d} \in [0,1]).
\end{align*}
As no closed-form is obtained, we apply adaptive Metropolis-Hastings to obtain posterior samples. First, the full-conditional on the log-scale is:
\begin{align*}
	\log \pi (\beta_{c,d}|\dots) &= (a_d^\beta - 1)\log (\beta_{c,d}) + (b_d^\beta - 1) \log(1-\beta_{c,d}) \\
	&\quad \quad - \sum_{g=1}^G (\phi_{j,g}^* + y_{c,g,d}) \log ( \phi_{j,g}^* + \mu_{j,g}^* \beta_{c,d}) - y_{c,g,d}\log(\beta_{c,d}) + \text{const.}
\end{align*}

\paragraph{Adaptive Metropolis-Hastings for $\beta$}
\begin{enumerate}
	\item We apply a logit transformation to transform $\beta$ into $x$, where $x$ belongs to the real axis:
	\begin{align*}
		\beta \in [0,1] \mapsto x \in \R,
	\end{align*}
	where $t(\beta) = \log (\frac{\beta}{1-\beta}) = x$.
	\item We apply an adaptive random walk in the transformed space, where the variance of proposal is adapted in similar fashion to Web Appendix \ref{sec:amh1}, with the sample variance computed by sequentially updating the required statistics.
	\item To Jacobian of the transformation is:
	\begin{align*}
		\frac{dt(\beta)}{d\beta} = \frac{d}{d\beta} (\log(\beta)-\log(1-\beta)) = \frac{1}{\beta(1-\beta)},
	\end{align*}
	and the log of the determinant of the Jacobian is:
	\begin{align*}
		\log \left( \frac{1}{\beta(1-\beta)} \right) = -\log(\beta)-\log(1-\beta).
	\end{align*}
	
	\item The acceptance probability is given by:
	\begin{align*}
		\alpha(\beta_{new}, \beta_{old}) = \min \{1, \exp [\Tilde{\alpha}(\beta_{new}, \beta_{old})] \},
	\end{align*}
	where
	\begin{align*}
		\Tilde{\alpha}(\beta_{new}, \beta_{old}) &= \log \Bigg[\frac{\pi(\beta_{new})}{\pi(\beta_{old})} \Bigg] + \log(\beta_{new}) + \log(1-\beta_{new}) \\
  &\quad \quad - \log(\beta_{old}) - \log(1-\beta_{old}).
	\end{align*}
\end{enumerate}

\subsection{MCMC with Fixed Clustering}\label{sec:mcmc_fixedclust}

After obtaining the optimal clustering which minimizes the posterior expected VI, we subsequently rerun the MCMC algorithm described above, but omitting the step described in Web Appendix \ref{sec:allocation} and fixing the allocation variables to the optimal estimate. This allows us to analyze the patterns and uncertainty within each cluster. We note that updating the allocation variables is the most expensive step in the MCMC algorithm; thus, this subsequent run is much faster.

\subsection{bayNorm Estimates of the Capture Efficiencies} \label{sec:baynorm_beta}

In the bayNorm approach, capture efficiencies are estimated using the following approach:
\begin{align*}
	\hat{\beta}_{c,d}^{\text{bay}} = \frac{\sum_{g=1}^G Y_{c,g,d}}{\frac{1}{C_d} \sum_{c=1, g=1}^{C_d, G} Y_{c,g,d}} \times \lambda,
\end{align*}
where $\lambda$ is the mean of estimated capture efficiencies. Under the default setting of bayNorm, $\lambda = 0.06$. The estimates are used to construct empirical priors for the capture efficiencies in our NormHDP model.

\subsection{Mixed Posterior Predictive Checks} \label{sec:ppc_pseudo_code}

The pseudo-code below demonstrates the steps taken to replicate data from the mixed posterior predictive.

\begin{algorithm}[H]
	\caption{Replicating data for mixed posterior predictive checks}\label{alg:ppc}
	\begin{algorithmic}
		\Require Number of replicated datasets $M$, Cell allocations $\bZ$, Posterior draws $(\bmu^{*\, (t)}_{1:J},  \bbeta^{(t)}, \bbv^{(t)},a_\phi^{2\,(t)})$ for $t=1,\ldots, T$.
  \For{$m = 1, \dots M$}
		\State Index $t_m \gets$ randomly sampled from $1:T$.
  \State Set parameters based on the $t$th posterior draw: $$\bmu^{*}_{1:J} = \bmu^{*\, (t_m)}_{1:J}, \bbeta = \bbeta^{(t_m)}, \bbv = \bbv^{(t_m)}, \alpha_\phi^2 = a_\phi^{2\,(t_m)}.$$
  \For{$j=1, \ldots, J$ and $g = 1, \ldots, G$}
  \State Simulate $\bphi^*_{j,g} \sim \logNorm(b_0 +b_1\log(\mu^*_{j,g}),  \alpha_\phi^2)$. 
  \EndFor
  \For{$c=1, \ldots, C_d$, $g=1, \ldots, G$ and $d=1,\ldots, D$}
            \State Simulate replicated data: $y_{c,g,d}^{\text{rep},m} \mid z_{c,d} = j \indsim \NegB(\mu^*_{j,g}\beta_{c,d}, \phi^*_{j,g})$. 
            \EndFor
  \EndFor
	\end{algorithmic}
\end{algorithm}

\subsection{Compare LFC in the residual overdispersion} \label{subsec:residual_overdispersion}

We can also classify global marker genes based on the residuals of dispersions \citep{basics3}, where the residual $\epsilon_{j,g}$ is defined as the difference between estimated dispersion $\phi_{j,g}$ and the fitted dispersion based on the regression trend, i.e.: in the linear case, we have $$\epsilon_{j,g} = \phi_{j,g} - (b_0 + b_1 \log \mu_{j,g}^*)$$ and the conditionally differentially dispersed genes are the ones with a large proportion of $|\epsilon_{j,g} - \epsilon_{j',g}|$ being greater than a threshold value between some cluster pairs $(j,j')$.

\pagebreak

\section{Simulations} \label{Web Appendix B}

\subsection{Simulation 1: Data Generation} \label{sec:sim1}

In the first simulation, the two datasets are simulated using the following model:
\begin{align*}
	y_{c,g,d}|z_{c,d} = j, \mu_{j,g}^*, \beta_{c,d}, \phi_{j,g}^* &\sim \NegB(\mu_{j,g}^*\beta_{c,d}, \phi_{j,g}^*), \\
	z_{c,d}|(p_{1,d}^J, \cdots, p_{J,d}^J) &\sim \Cat(p_{1,d}^J, \cdots, p_{J,d}^J), \\
	\mu_{j,g}^* &\sim \logNorm (0, \alpha_\mu^2), \\
	\phi_{j,g}^* | \mu_{j,g}^* &\sim \logNorm (b_0 + b_1 \log(\mu_{j,g}^*), \alpha_\phi^2), \\
	\beta_{c,d} &\sim \Be (a_d^\beta, b_d^\beta),
\end{align*}
where we set $b_1 = 0; b_2 = 3; \alpha_\mu^2 = 1; \alpha_\phi^2 = 1$ and $a_d^\beta = 1; b_d^\beta = 0.5$ for both datasets. The first dataset contains $C_1= 50$ cells and the second contains $C_2= 100$ cells, with $G= 50$ genes. We assume there are $J=3$ clusters, with true cell proportions $(p_{1,1},p_{2,1}, p_{3,1}) =  (0.6,0.4,0)$ for dataset 1 and $(p_{1,2},p_{2,2}, p_{3,2}) = (0.4,0,0.6)$ for dataset 2.

\subsection{Simulation 2:  Data Generation} \label{sec:sim2}

In the second simulation, we assume a non-linear relationship between the mean expressions and dispersions, to assess robustness to misspecification. The two datasets are generated as in Simulation 1 (Web Figure \ref{sec:sim1}), however, we assume the following relationship:
\begin{align*}
	\phi_{j,g}^* | \mu_{j,g}^* \sim \logNorm (4-2/\mu_{j,g}, \alpha_\phi^2),
\end{align*}
and values of $\alpha_\mu^2$; $\alpha_\phi^2$; $a_d^\beta$ and $b_d^\beta$ are set to be the same as in Simulation 1.

\subsection{Simulation 3: Data Generation} \label{sec:sim3}

\begin{figure}[!t]
	\centering
	\subfigure{\includegraphics[width=0.45\textwidth]{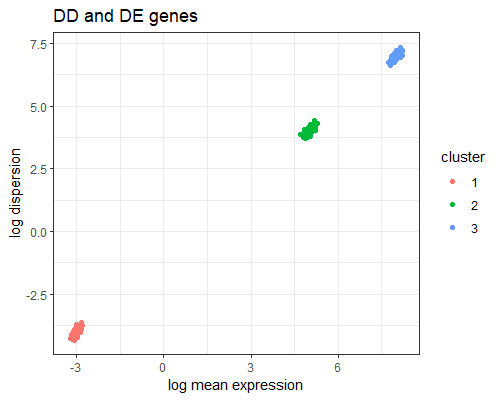}}
 \subfigure{\includegraphics[width=0.45\textwidth]{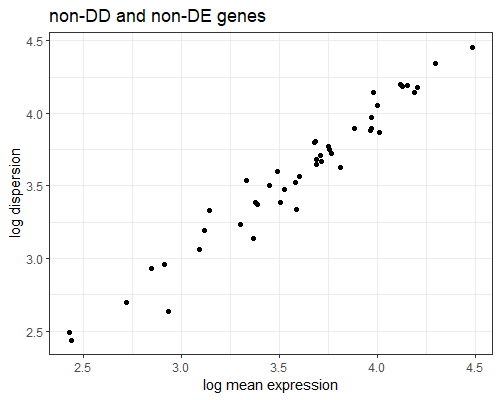}}
	\caption{True mean-dispersion relationships for DE and DD genes (left) and non-DE (also non-DD) genes (right). On the left, different colors are used to represent mean expressions and dispersions from different clusters.}
	\label{Fig:DD_DE_test_corr}
\end{figure}

In Simulation 3, we test the ability of our algorithm to detect global differentially expressed and dispersed genes. In this case, we have $G = 150$ genes; $C_1 = 300$ cells in dataset 1 and $C_2 = 400$ cells in dataset 2. We assume that there are $J = 3$ clusters, with dataset-specific allocation proportions $(0.8,0.2,0)$ and $(0.8,0,0.2)$ for dataset 1 and 2, respectively. For simplicity, we assume the first 70 percent of the genes are DD and DE, specifically, these are the genes with indices 1 to 105.

Based on the results from the real data, non-DE genes tend to be highly expressed. Thus, we set the true mean for non-DE genes $\mu_{j,g}^*$ to $\mu_g^*$, with $\mu_g^* \sim \logNorm(3.5,0.5)$. Instead, for DE genes, we set the true mean $\mu_{j,g}^* \sim \logNorm(m_j, 0.1)$, where $m_{1:J} = (-3,5,8)$. We assume all DE genes are also DD and vice versa. For non-DD (and non-DE) genes, we set the true dispersion $\phi_{j,g}^* = \phi_g^*$ with $\phi_g^* \sim \logNorm(b_0 + b_1\log(\mu_g^*), 0.1)$, where $b_0 = -1$ and $b_1 = 1$. For DD (and DE) genes, we set the true dispersion $\phi_{j,g}^* \sim \logNorm(b_0 + b_1\log(\mu_{j,g}^*), 0.1)$. The true relationships between mean expressions and dispersions are shown in  Web Figure \ref{Fig:DD_DE_test_corr}.

\clearpage

\subsection{Simulations 1 and 2: Results} \label{sec:sim1_and_2_results}

For both simulations, we investigate prior sensitivity by comparing general priors based on standard hyperparameter values with empirical priors based on hyperparameters specified using initial \textit{bayNorm} estimates (as described in the main article). In addition, to enhance flexibility, we consider both linear and quadratic relationships in the prior model for the mean-variance relationship. This results in four settings for the proposed NormHDP model: linear model with general priors (GL), quadratic model with general priors (GQ), linear model with empirical priors (EL) and quadratic model with empirical priors (EQ).

\begin{figure}[!t]
	\centering
	\subfigure[Trace plot of $\alpha$.]{\includegraphics[width=0.8\textwidth]{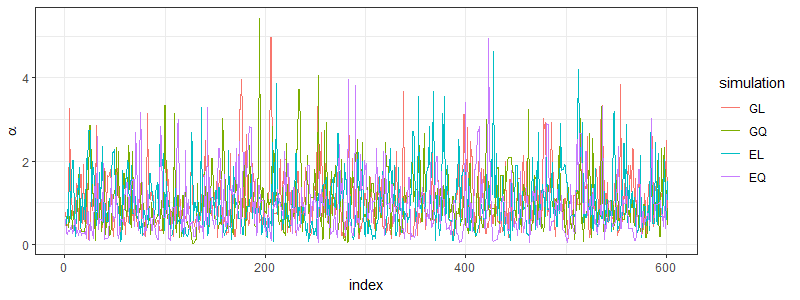}}
	\subfigure[Trace plot of $\alpha_0$.]{\includegraphics[width=0.8\textwidth]{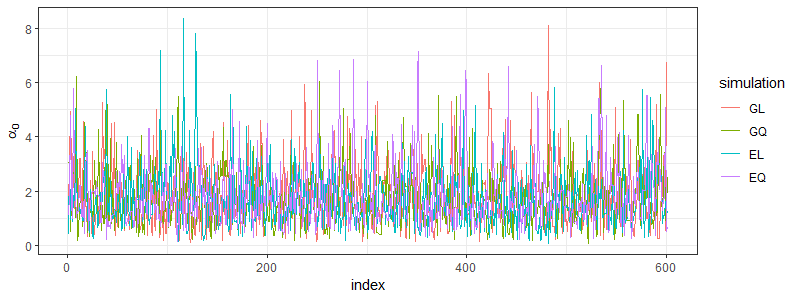}}
	\subfigure[Trace plot of $\alpha_\phi^2$.]{\includegraphics[width=0.8\textwidth]{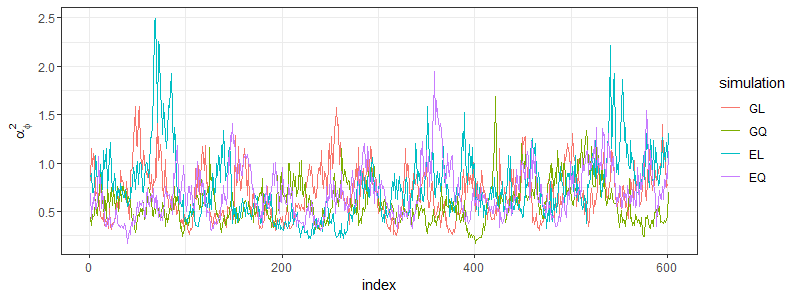}}
	\caption{Traceplots of the concentration parameters and regression parameter $\alpha_\phi^2$ for Simulation 2. Colors represent the chains under the different prior and model choices.}
	\label{fig:traceplot_sim2}
\end{figure}

Traceplots of concentration parameters and $\alpha_\phi^2$ for Simulation 1 and Simulation 2 provided in  Web Figure \ref{fig:traceplot_sim1} and  \ref{fig:traceplot_sim2} suggest convergence:
\begin{figure}[!t]
	\centering
	\subfigure[Trace plot of $\alpha$.]{\includegraphics[width=0.8\textwidth]{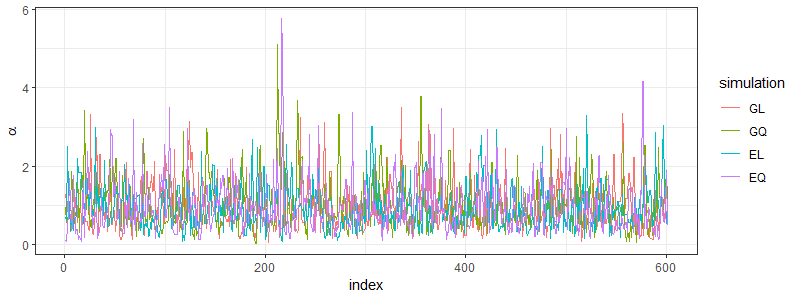}}
	\subfigure[Trace plot of $\alpha_0$.]{\includegraphics[width=0.8\textwidth]{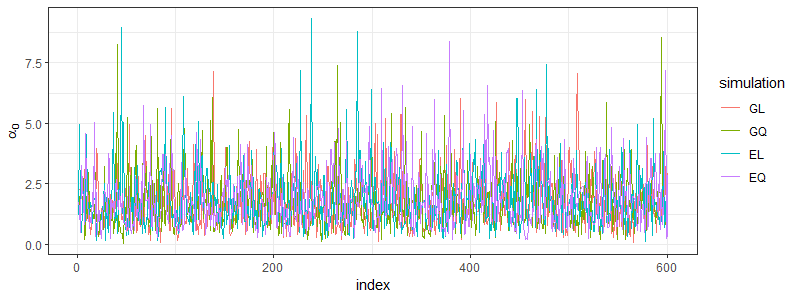}}
	\subfigure[Trace plot of $\alpha_\phi^2$.]{\includegraphics[width=0.8\textwidth]{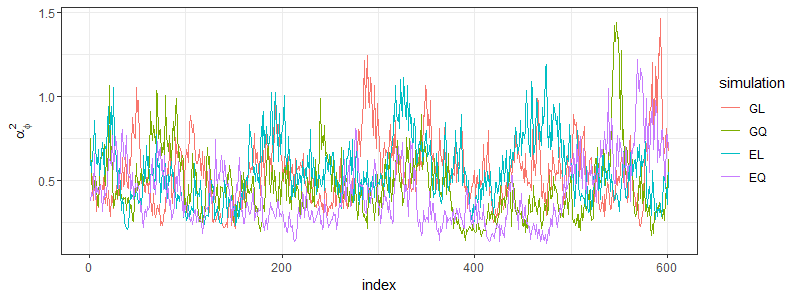}}
	\caption{Traceplots of the concentration parameters and regression parameter $\alpha_\phi^2$ for Simulation 1. Colors represent the chains under the different prior and model choices.}
	\label{fig:traceplot_sim1}
\end{figure}

\begin{figure}[!t]
	\centering
	\subfigure{\includegraphics[width=0.3\textwidth]{figures/simulation1_true_psm.png}}
 \subfigure{\includegraphics[width=0.3\textwidth]{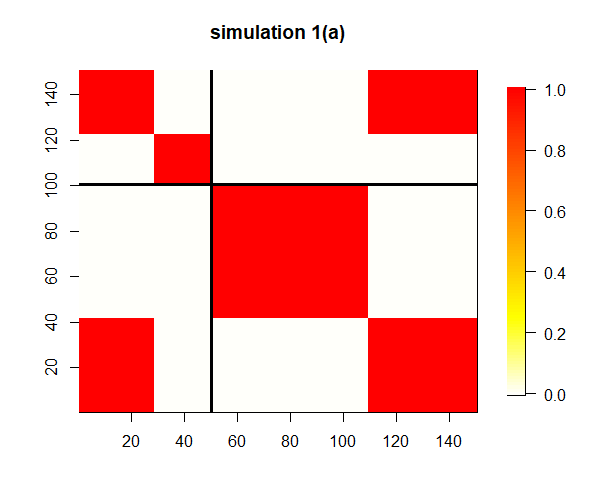}}
 \subfigure{\includegraphics[width=0.3\textwidth]{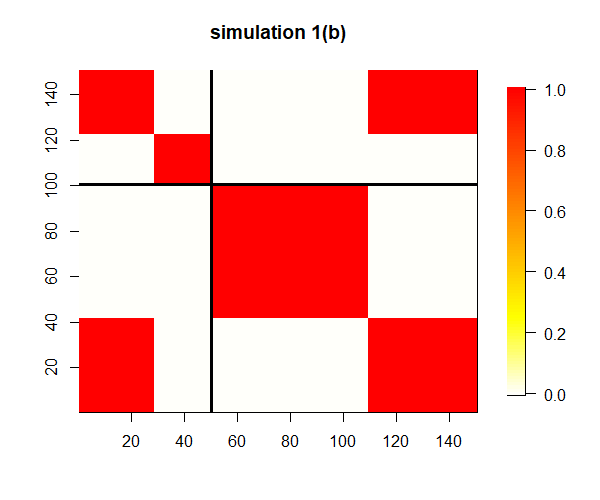}}
 \subfigure{\includegraphics[width=0.3\textwidth]{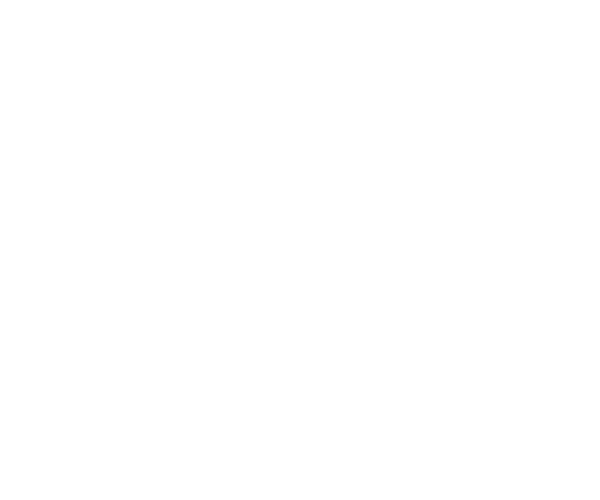}}
 \subfigure{\includegraphics[width=0.3\textwidth]{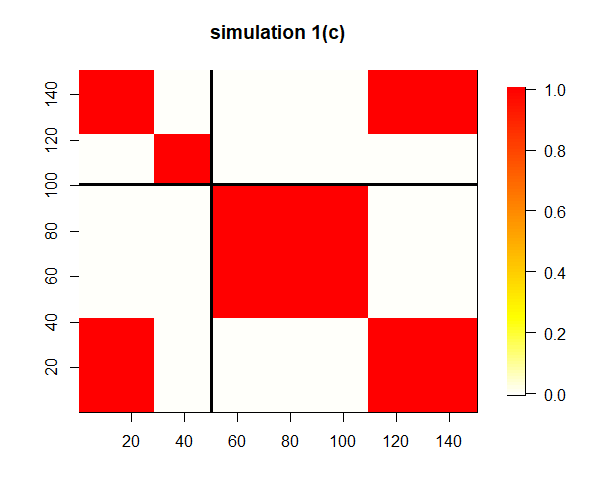}}
 \subfigure{\includegraphics[width=0.3\textwidth]{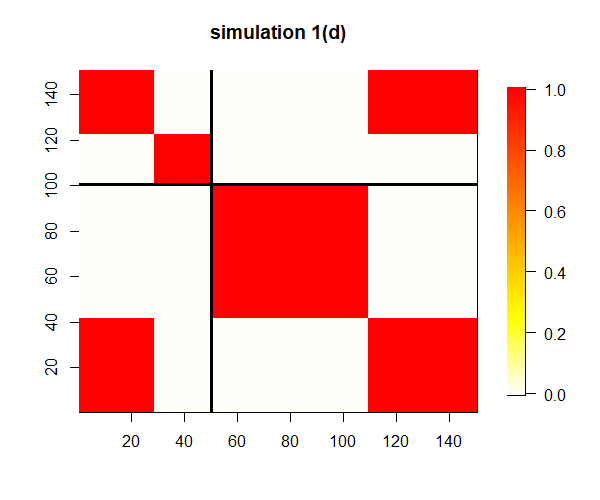}}
	\caption{Comparison of the true and posterior similarity matrix for Simulation 1, with the true similarity matrix (left) and the posterior similar matrix with (a) the linear model and general prior, (b) the linear model and empirical  prior, (c) the quadratic model and general prior, and (d) the quadratic model and empirical prior.}
	\label{fig:sim1_psm_copy}
\end{figure}

\begin{figure}[!t]
	\centering
	\subfigure{\includegraphics[width=0.3\textwidth]{figures/simulation2_true_psm.png}}
 \subfigure{\includegraphics[width=0.3\textwidth]{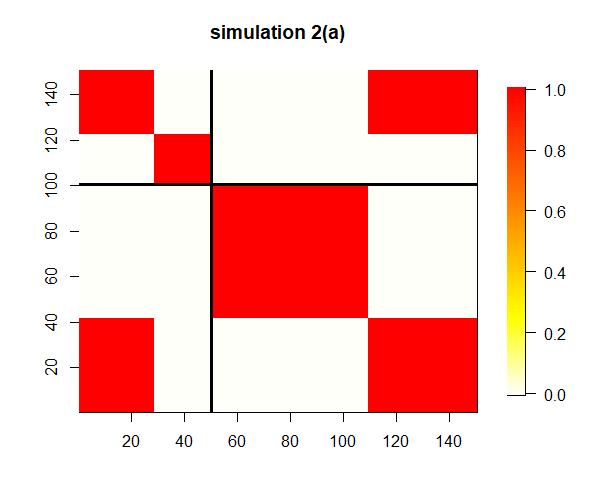}}
 \subfigure{\includegraphics[width=0.3\textwidth]{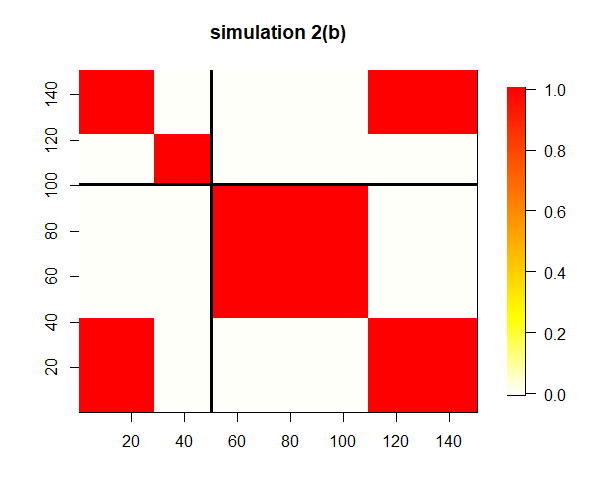}}
 \subfigure{\includegraphics[width=0.3\textwidth]{figures/psm_simulation1_empty.png}}
 \subfigure{\includegraphics[width=0.3\textwidth]{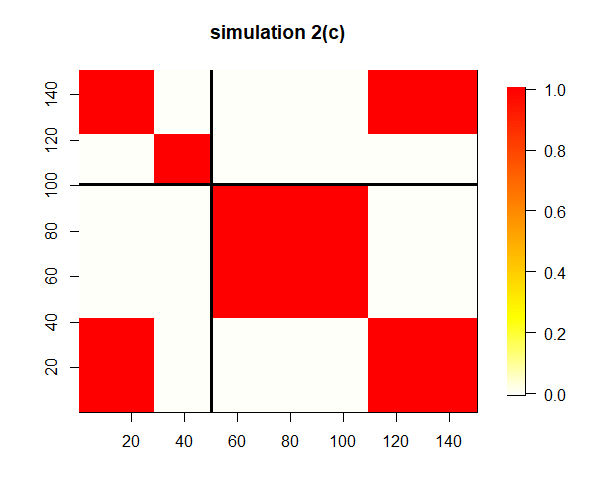}}
 \subfigure{\includegraphics[width=0.3\textwidth]{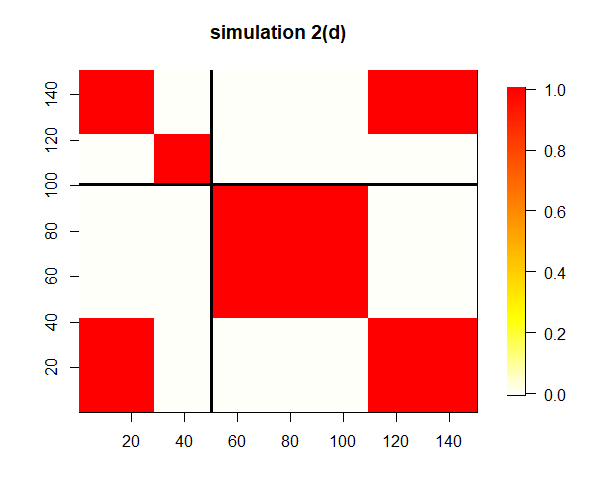}}
	\caption{Comparison of the true and posterior similarity matrix for Simulation 2, with the true similarity matrix (left) and the posterior similar matrix with (a) the linear model and general prior, (b) the linear model and empirical  prior, (c) the quadratic model and general prior, and (d) the quadratic model and empirical prior.}
	\label{fig:sim2_psm_copy}
\end{figure}

In Web Figure \ref{fig:sim1_psm_copy} and \ref{fig:sim2_psm_copy}, we compare the true similarity matrix with the posterior similarity matrix obtained from the different prior (general vs. empirical) and model (linear vs. quadratic) choices, for Simulation 1 and 2 respectively. In all cases, the results highlight that the true clustering structure is well recovered.

\begin{figure}[!t]
	\centering
	\subfigure{\includegraphics[width=0.45\textwidth]{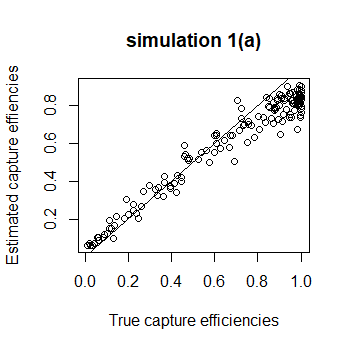}}
        \subfigure{\includegraphics[width=0.45\textwidth]{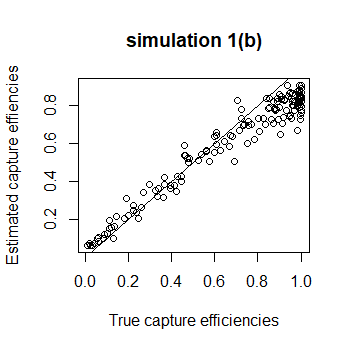}}
        \subfigure{\includegraphics[width=0.45\textwidth]{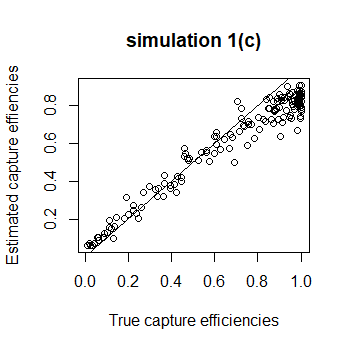}}
         \subfigure{\includegraphics[width=0.45\textwidth]{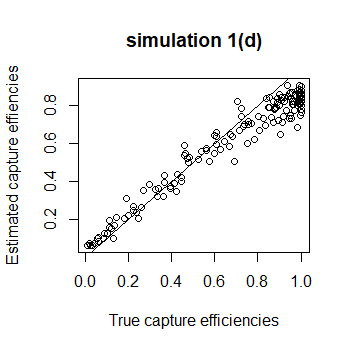}}
	\caption{Comparison between the true capture efficiencies and the posterior mean of capture efficiencies for Simulation 1 with (a) the linear model and general prior, (b) the linear model and empirical  prior, (c) the quadratic model and general prior, and (d) the quadratic model and empirical prior..} \label{fig:sim1_capeff}
\end{figure}

\begin{figure}[!t]
	\centering
	\subfigure{\includegraphics[width=0.45\textwidth]{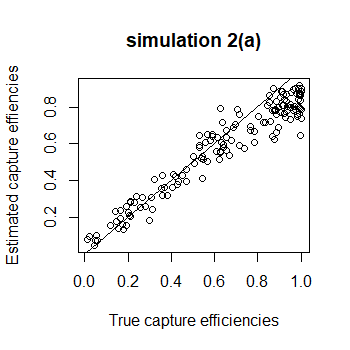}}
        \subfigure{\includegraphics[width=0.45\textwidth]{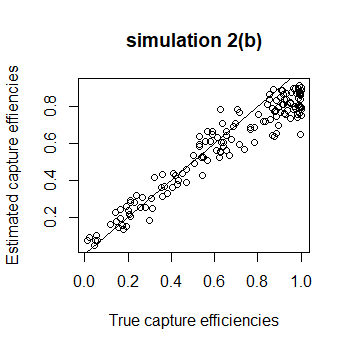}}
        \subfigure{\includegraphics[width=0.45\textwidth]{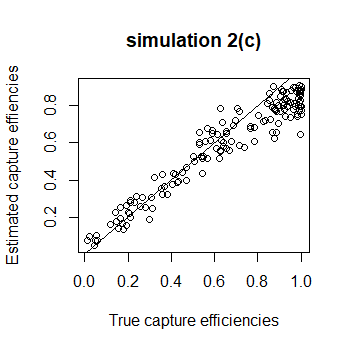}}
         \subfigure{\includegraphics[width=0.45\textwidth]{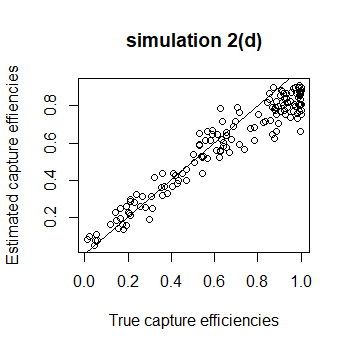}}
	\caption{Comparison between the true capture efficiencies and the posterior mean of capture efficiencies for Simulation 2 with (a) the linear model and general prior, (b) the linear model and empirical  prior, (c) the quadratic model and general prior, and (d) the quadratic model and empirical prior..} \label{fig:sim2_capeff}
\end{figure}

Web Figure \ref{fig:sim1_capeff} and \ref{fig:sim2_capeff} compares  the true capture efficiencies with estimated capture efficiencies for Simulation 1 and Simulation 2, respectively. The figures highlight that the informative prior choice helps to mitigate indentifiability issues. 

\clearpage

\subsection{Simulation 3: Results} \label{sec:sim3_results}

\begin{figure}[!t]
	\centering
	\includegraphics[width=\textwidth]{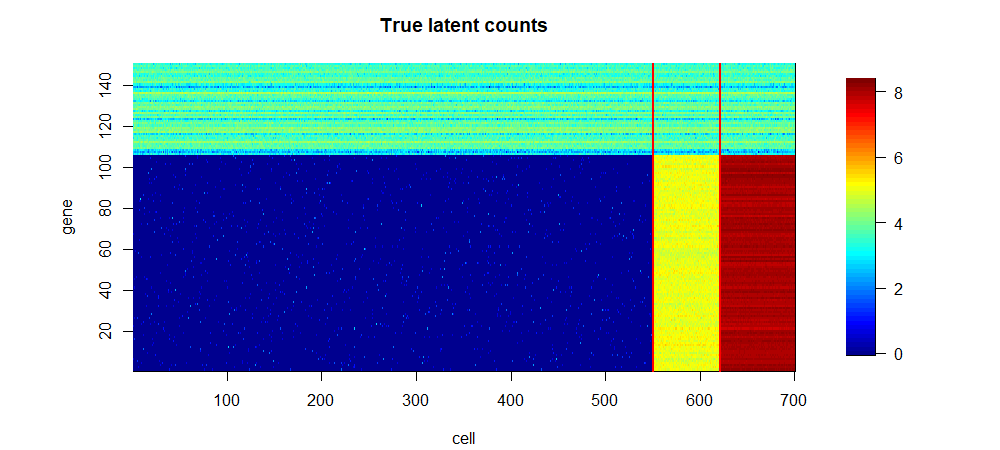}
	\caption{Heat map of true latent gene-counts, with rows representing genes and columns representing cells. Cells are reordered by the true clustering, with cells from different clusters separated by vertical lines.}
	\label{fig:latent_count_by_clust}
\end{figure}

\begin{figure}[!t]
	\centering
	\includegraphics[width=\textwidth]{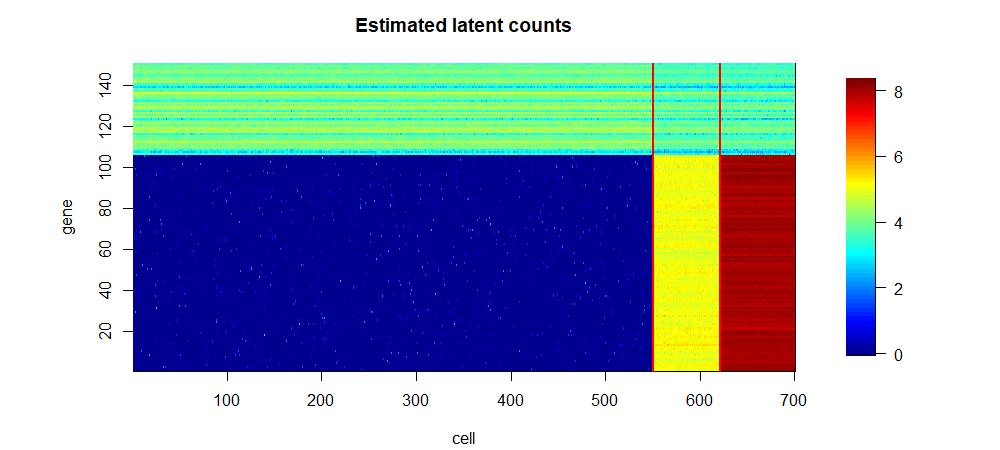}
	\caption{Heat map of posterior estimated latent gene-counts, with rows representing genes and columns representing cells. Cells are reordered by the posterior estimated clustering, with cells from different clusters separated by vertical lines.}
	\label{fig:latent_count_est}
\end{figure}

The true latent counts and the posterior estimated latent counts are illustrated in Web Figures \ref{fig:latent_count_by_clust} and \ref{fig:latent_count_est}, respectively, for Simulation 3. A comparison of the figures demonstrates that the latent counts are well recovered. 

\clearpage

\subsection{Additional Simulations with Mis-specified Mean Capture Efficiencies} \label{sec:sim_add}

\begin{figure}[!t]
	\centering
	\includegraphics[width=10cm]{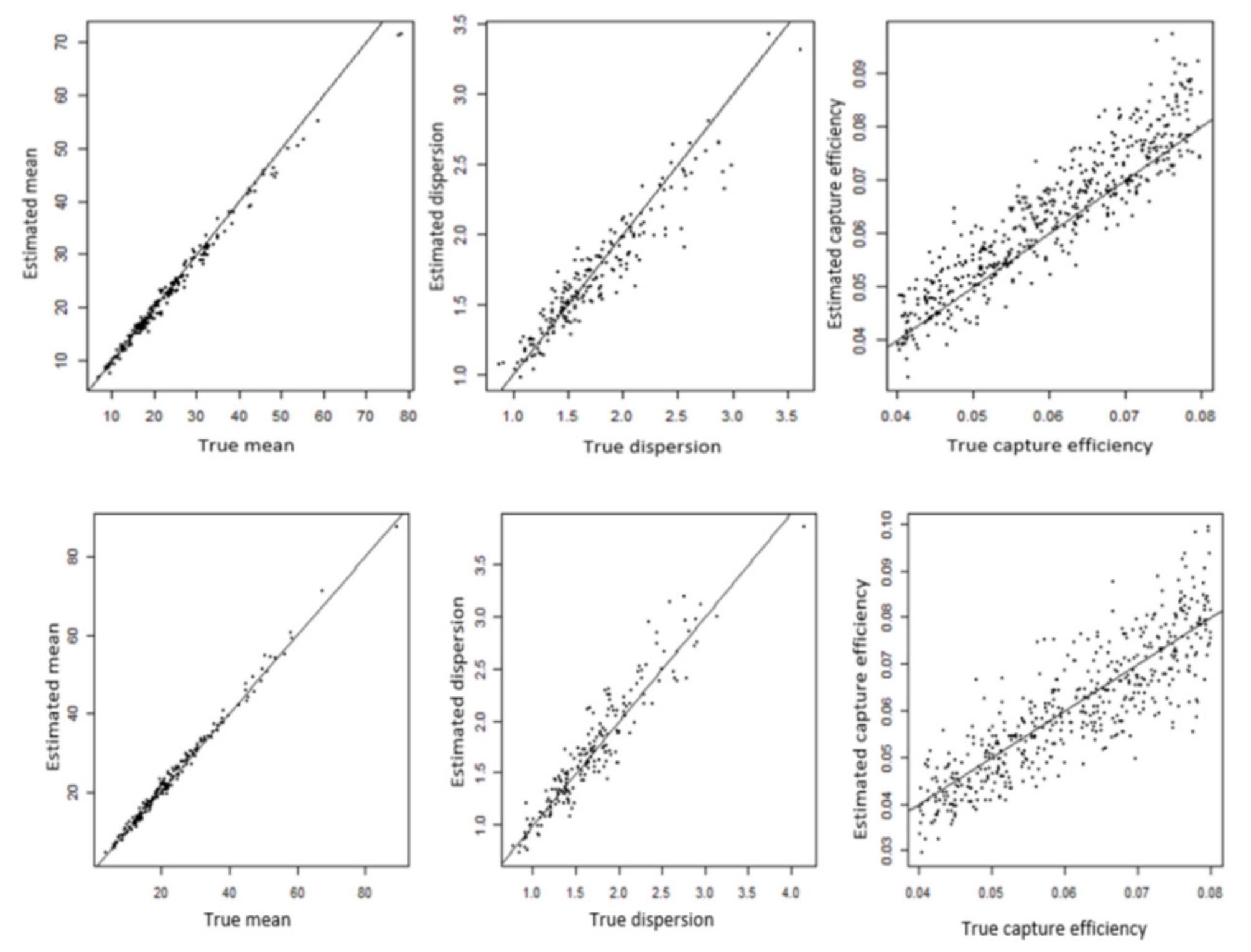}
	\caption{Comparison between our Bayesian approach (top row) and \textit{bayNorm} (bottom row) in recovering the  mean-expression, dispersion and capture efficiency (left to right) for dataset 1.}
	\label{fig:identifiable_case1}
\end{figure}

\begin{figure}[!t]
	\centering
	\includegraphics[width=10cm]{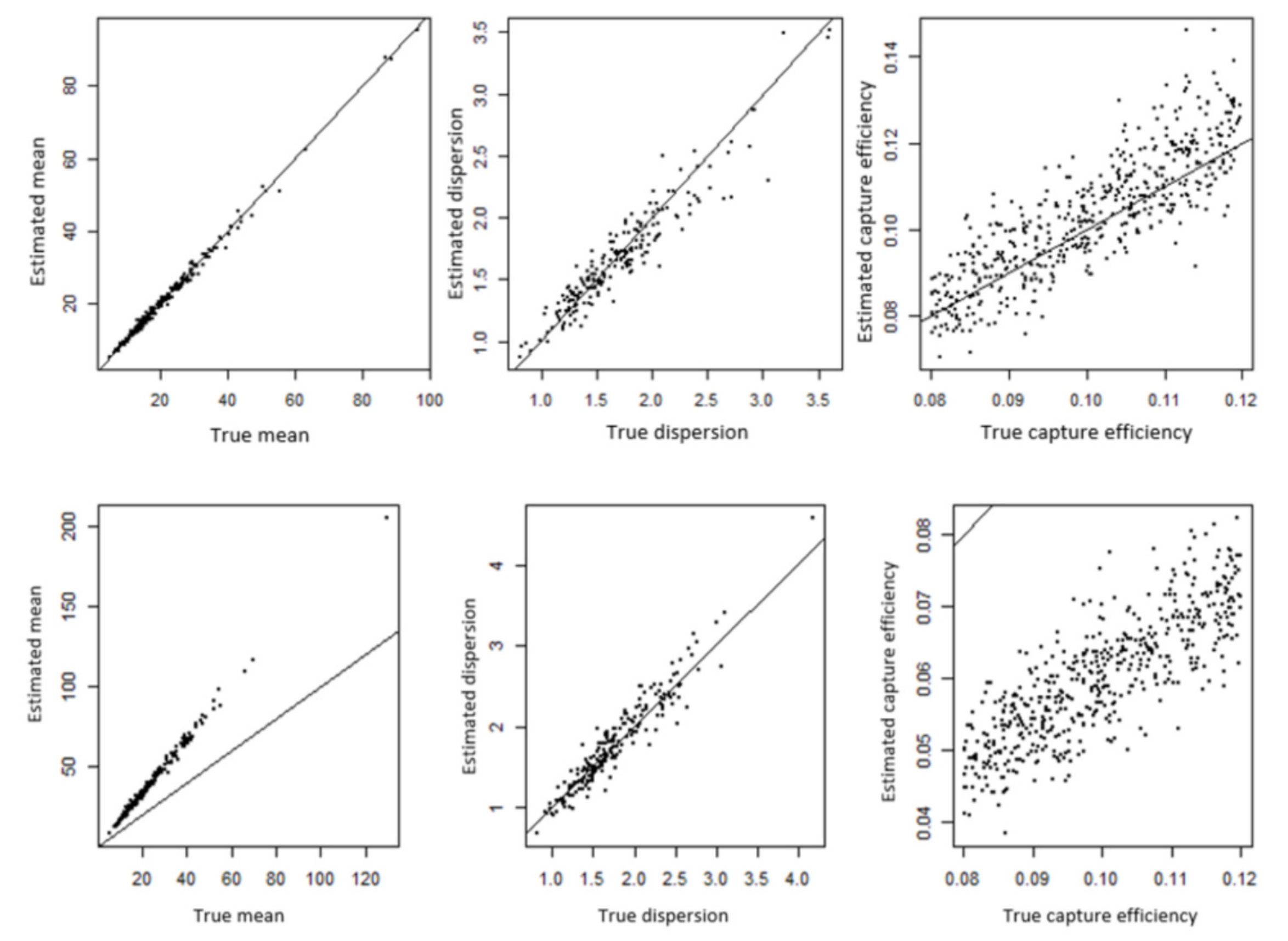}
	\caption{Comparison between our Bayesian approach (top row) and \textit{bayNorm} (bottom row) in recovering the  mean-expression, dispersion and capture efficiency (left to right) for dataset 2.}
	\label{fig:identifiable_case2}
\end{figure}

In the previous simulations, we used an informative prior for the capture efficiencies to mitigate identifiability issues. In addition, we carried out further simulations to investigate the effect of misspecification of the mean capture efficiency in our informative prior. Specifically, we set the true mean capture efficiency for dataset 1 and 2 as $0.06$ and $0.10$, respectively, and simulate the true  unique parameters and capture efficiencies under the following model:
\begin{align*}
	\beta_{c,1} &\sim \Unif (0.04,0.08), \\
	\beta_{c,2} &\sim \Unif (0.08,0.12), \\
	\phi_g &\sim \logNorm (3, 0.5), \\
	\mu_g &\sim \logNorm (-1+0.5\log(\phi_g), 0.1).
\end{align*}
We compare the \textit{bayNorm} estimates of the capture efficiencies and unique parameters with our empirical Bayesian approach. 
Results for dataset 1 and 2 are shown in Web Figure \ref{fig:identifiable_case1} and \ref{fig:identifiable_case2}, respectively. In the first dataset, the mean capture efficiency is correctly specified and both approaches are able to recover the true mean expression, dispersion, and capture efficiencies (Web Figure \ref{fig:identifiable_case1}). However, in the second dataset, the mean capture efficiency is misspecified; while bayNorm underestimates the capture efficiencies and overestimates the mean expressions, our empirical Bayesian approach is more robust to such minor misspecifications.  

\clearpage

\section{PAX6 Data Analysis}\label{web_appendix_PAX6}

\subsection{Filtering Process} \label{sec:filter}

%We use the following method to select cells and genes in the filtering process.
Similar to the approach taken in \citep{seurat}, we filter the raw dataset based on empirical statistics to remove extreme observations, and we select the top $2,000$ genes with the largest cell-to-cell variability:
\begin{enumerate}
	\item Genes that are expressed by less than 5 cells are excluded (threshold chosen to remove half of the genes).
	\item Cells with less than 2000 expressed genes are excluded (threshold chosen to remove cells with gene expressions less than 1.5 standard deviations below the empirical mean).
	\item Cells with greater than 3.7 percent mitochondrial counts are excluded (threshold chosen to remove cells with mitochondrial counts greater than 1.5 standard deviations above than the empirical mean).
	\item Cells with gene counts greater than 17500 are excluded (threshold chosen to remove cells with gene counts greater than 1.5 standard deviations above than the empirical mean).
	\item 2000 genes with the greatest variability are selected.
\end{enumerate}

\begin{figure}[!t]
	\centering
	\includegraphics[width=14cm]{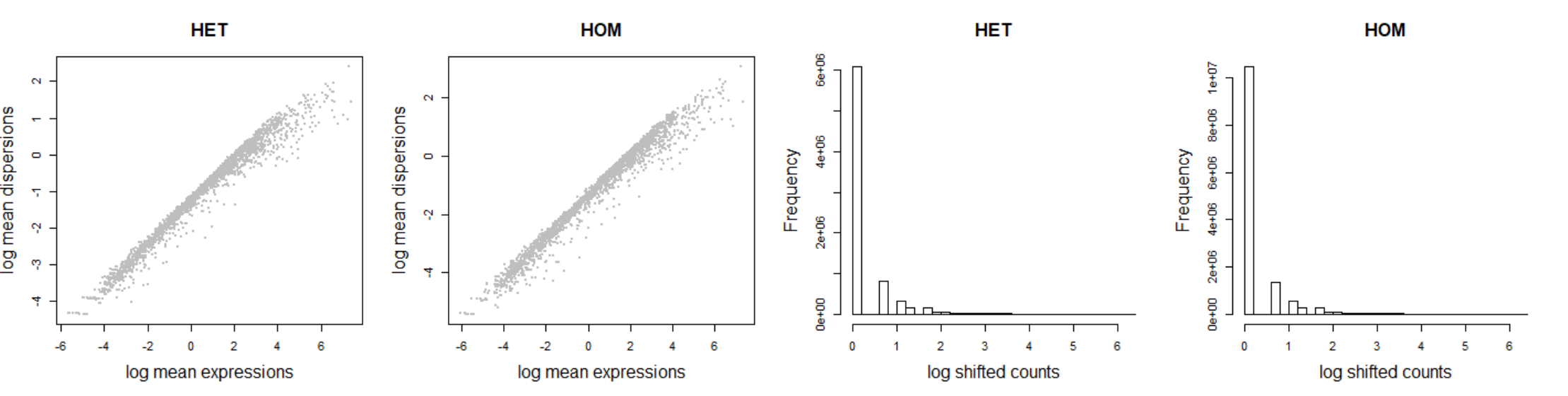}
	\caption{Left two plots: empirical relationship between the log mean expressions and log dispersions. Right two plots: histogram of shifted latent gene-counts.}
	\label{fig:E135_empirical}
\end{figure}

Finally, we use the union of genes remaining after applying all above filters to each dataset to form the final processed data. Note that all the `important' genes are also included for each dataset (Web Table \ref{tab:important_genes}). The empirical relationship between  \textit{bayNorm} estimates of the mean expressions and dispersions and the distribution of log shifted counts based on the final processed data are shown in Web Figure \ref{fig:E135_empirical}.

\subsection{Results} \label{sec:real_results}

\begin{figure}[!t]
	\centering
	\includegraphics[width=0.9\textwidth]{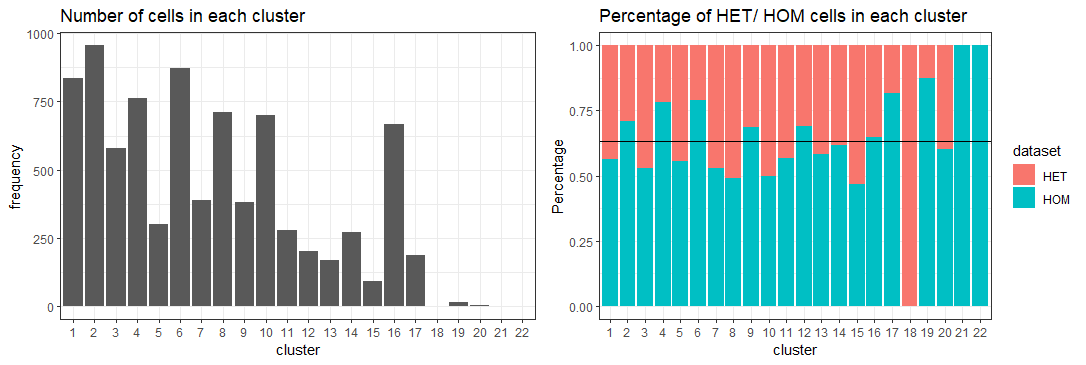}
	\caption{Left: the number of cells within each cluster. Right: proportion of HET and HOM cells within each cluster, the horizontal line indicates the overall proportion of HOM cells across both datasets.}
	\label{fig:z_frequency}
\end{figure}

\begin{figure}[!t]
	\centering
	\subfigure[Trace plot of $\alpha_0$.]{\includegraphics[width=0.45\textwidth]{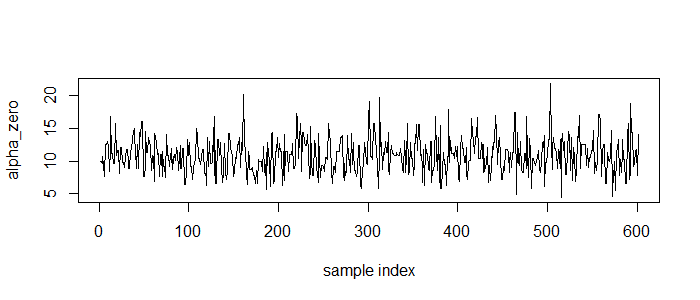}}
	\subfigure[Trace plot of $\alpha$.]{\includegraphics[width=0.45\textwidth]{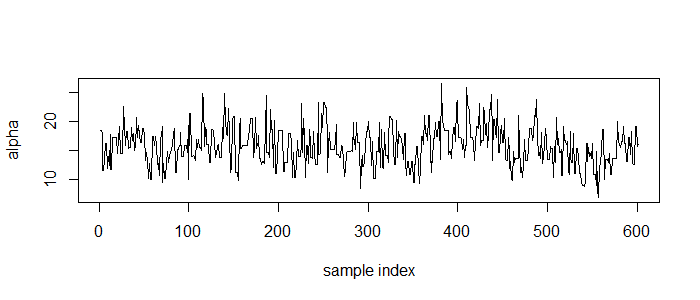}}
	\caption{Trace plot of concentration parameters.}
	\label{fig:traceplot_concentration_parameters}
\end{figure}

\begin{figure}[!t]
	\centering
	\subfigure[Trace plot of $\alpha_\phi^2$.]{\includegraphics[width=0.45\textwidth]{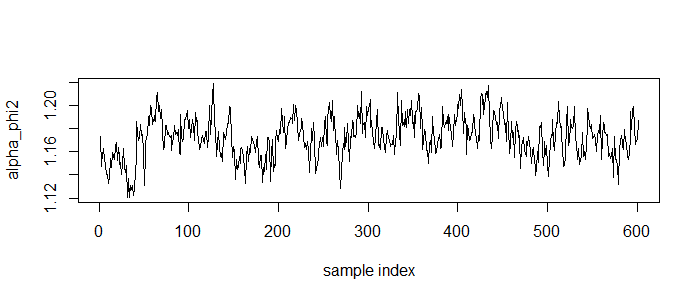}}
	\subfigure[Trace plot of $b_0$.]{\includegraphics[width=0.45\textwidth]{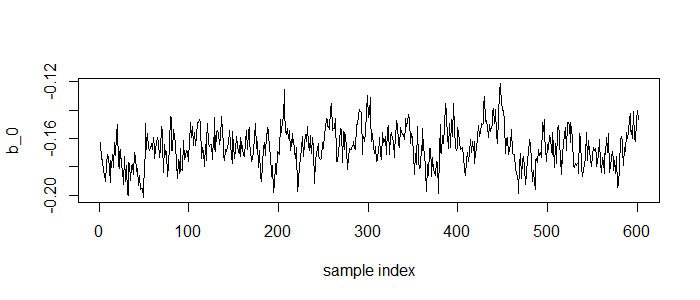}}
	\subfigure[Trace plot of $b_1$.]{\includegraphics[width=0.45\textwidth]{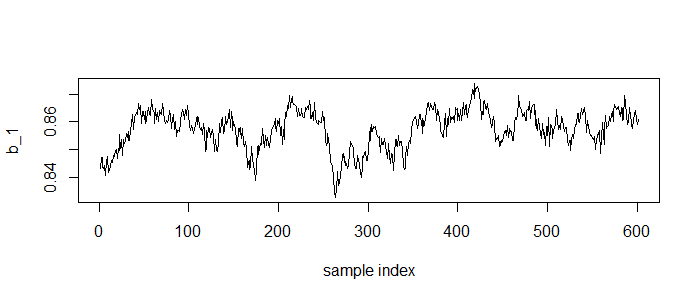}}
	\subfigure[Trace plot of $b_2$.]{\includegraphics[width=0.45\textwidth]{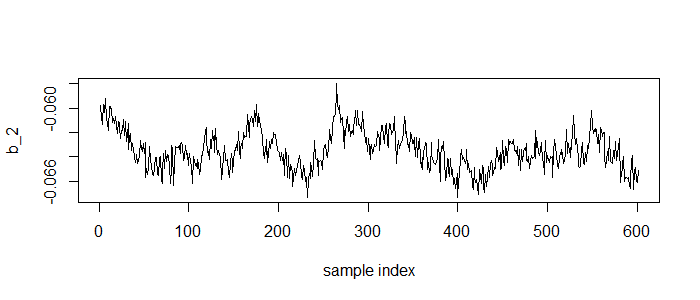}}
	\caption{Trace plot of regression parameters.}
	\label{fig:traceplot_regression_parameters}
\end{figure}

\begin{figure}[!t]
	\centering
	\subfigure[Density of $\alpha$]{\includegraphics[width=0.3\textwidth]{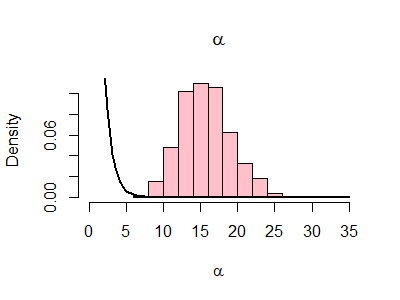}}
        \subfigure[Density of $\alpha_0$]{\includegraphics[width=0.3\textwidth]{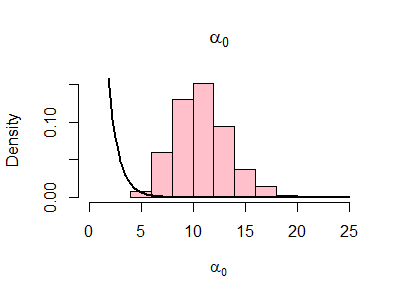}}
        \subfigure[Density of $\alpha_\phi^2$]{\includegraphics[width=0.3\textwidth]{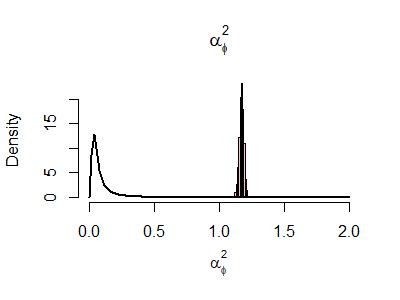}}
	\caption{Comparison between prior and posterior of $\alpha, \alpha_0$ and $\alpha_\phi^2$ (left to right). Prior densities are shown with black lines and posterior densities are shown with histograms.}
	\label{fig:alpha_prior_posteropr_comp}
\end{figure}

\begin{figure}[!t]
	\centering
	\subfigure[Density of $b_0$]{\includegraphics[width=0.3\textwidth]{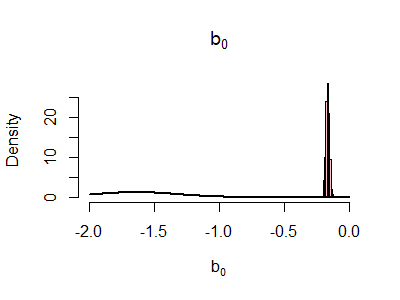}}
        \subfigure[Density of $b_1$]{\includegraphics[width=0.3\textwidth]{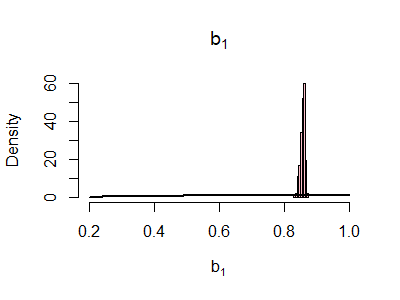}}
        \subfigure[Density of $b_2$]{\includegraphics[width=0.3\textwidth]{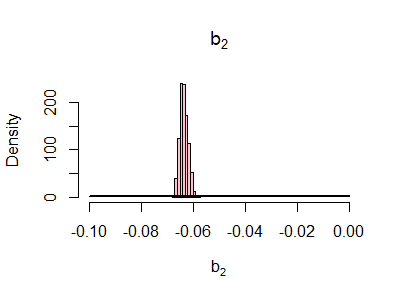}}
	\caption{Comparison between priors and posteriors of $b_0, b_1$ and $b_2$ (left to right). Prior densities are shown with black lines and posterior densities are shown with histograms.}
	\label{fig:regression_coefficient_comp}
\end{figure}

\begin{figure}[!t]
	\centering
	\includegraphics[width=13cm]{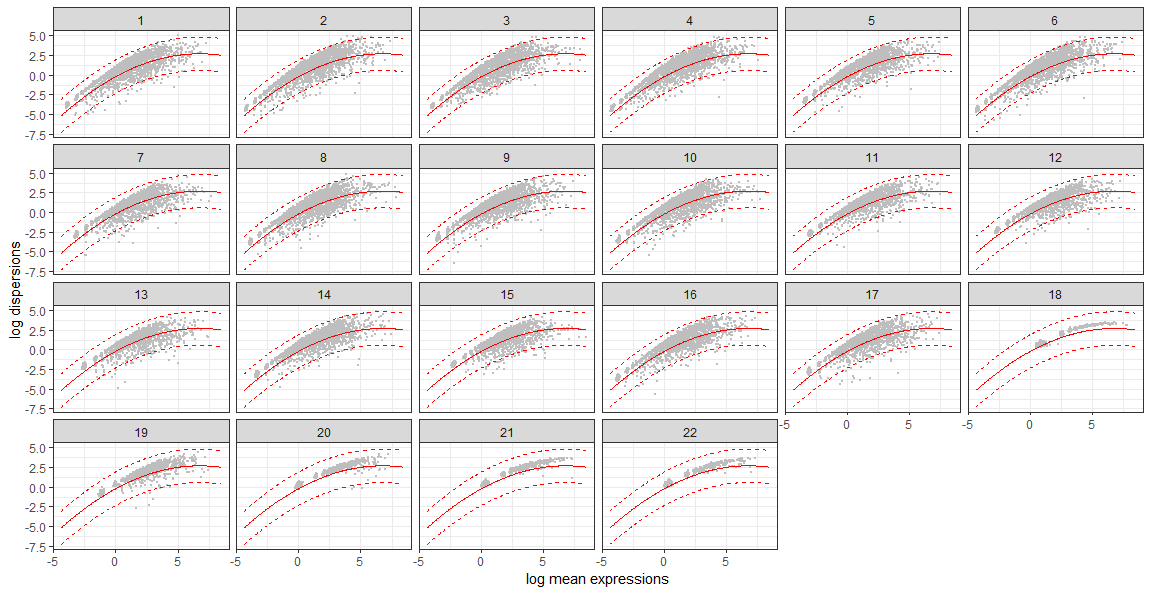}
	\caption{Posterior estimated relationships between the mean expressions and dispersions for each cluster. The posterior means of the mean expressions and dispersions are plotted. The red dashed lines are the lower and upper bound of the 95 percent credible band obtained by considering posterior estimates of regression parameters $\bbv$ and $\alpha_\phi^2$. The red solid line is the posterior estimated relationship.}
	\label{Fig:est_post_corr}
\end{figure}

\begin{figure}[!t]
	\centering
	\includegraphics[width=13cm]{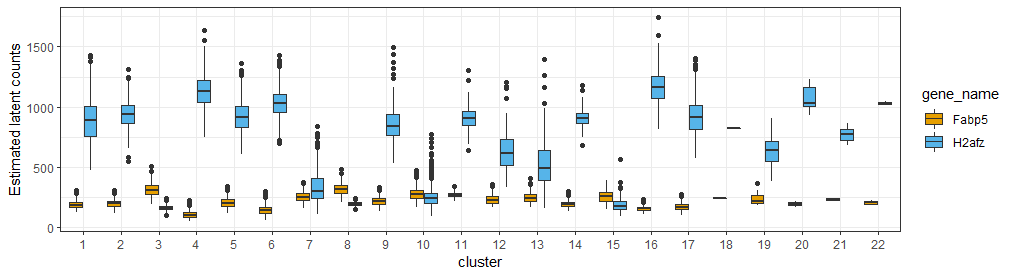}
	\caption{Boxplots of posterior estimated latent counts across cells for genes Fabp5 and H2afz.}
	\label{fig:latent_counts}
\end{figure}

\begin{figure}[!t]
\centering
	\subfigure{\includegraphics[width=0.7\textwidth]{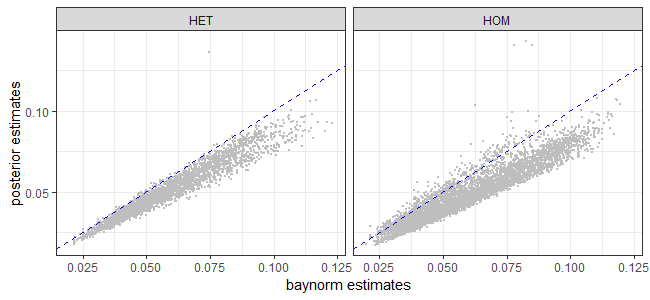}}

        \subfigure{\includegraphics[width=0.9\textwidth]{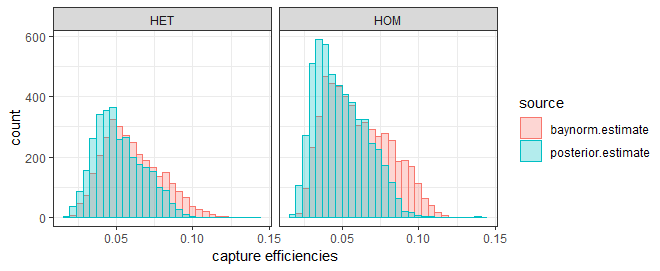}}
	\caption{Comparison of the bayNorm and posterior mean of the capture efficiencies. For the top two plots: bayNorm estimates are plotted against posterior estimates, and the dashed lines represents when the two values are equivalent. For the bottom two plots: histograms are drawn to compare distribution of the posterior mean capture efficiencies, with the bayNorm and posterior estimates in blue and red, respectively.}
	\label{Fig:prior_posterior_beta}
\end{figure}

\begin{figure}[!t]
	\centering
	\includegraphics[width=\textwidth]{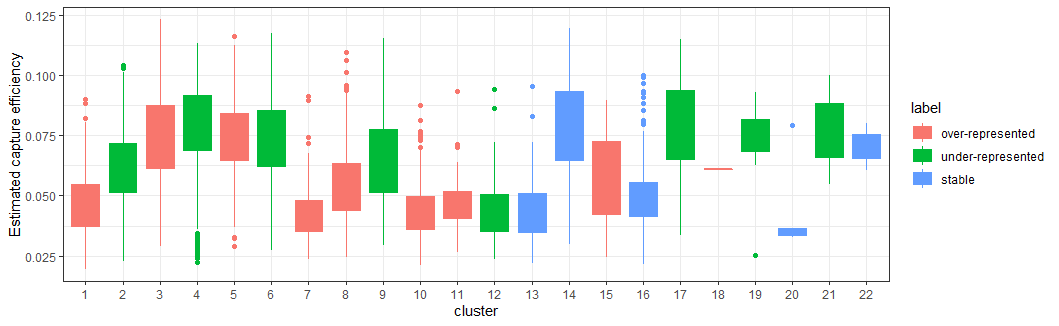}
	\caption{box plots of the estimated capture efficiencies for cells in each cluster.}
	\label{fig:estimated_capture_efficiencies}
\end{figure}

\begin{figure}[!t]
	\centering
	\includegraphics[width=\textwidth]{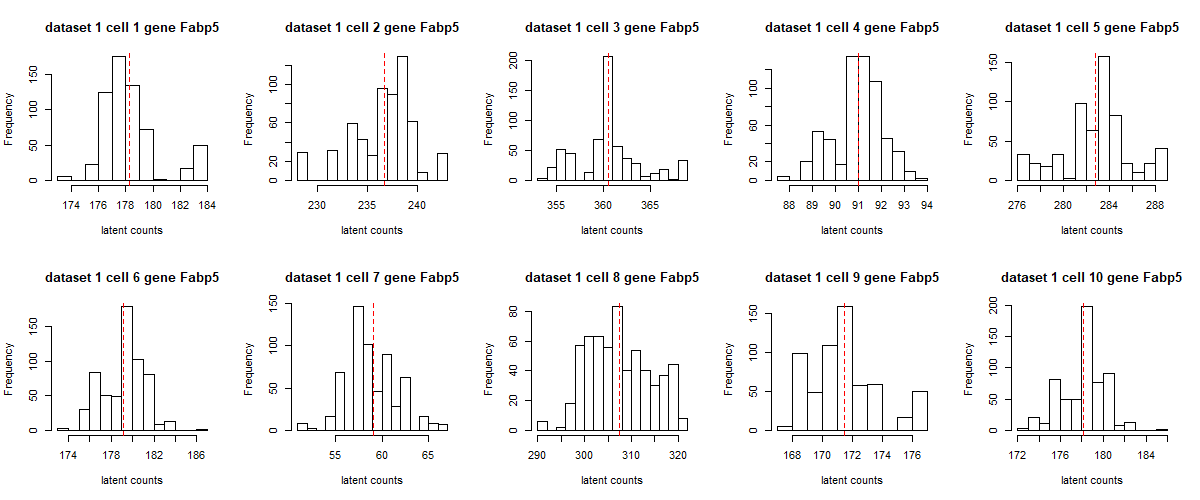}
	\caption{Histograms of the posterior estimated latent counts for cell $1$ to $10$ for gene 'Fabp5' in dataset 1. For each MCMC iteration, we compute the mean estimated latent count. The red vertical line indicates the overall mean latent count, averaged across all MCMC iterations, for a given gene and cell.}
	\label{fig:Fabp5_latent}
\end{figure}

\begin{figure}[!t]
	\centering
	\includegraphics[width=\textwidth]{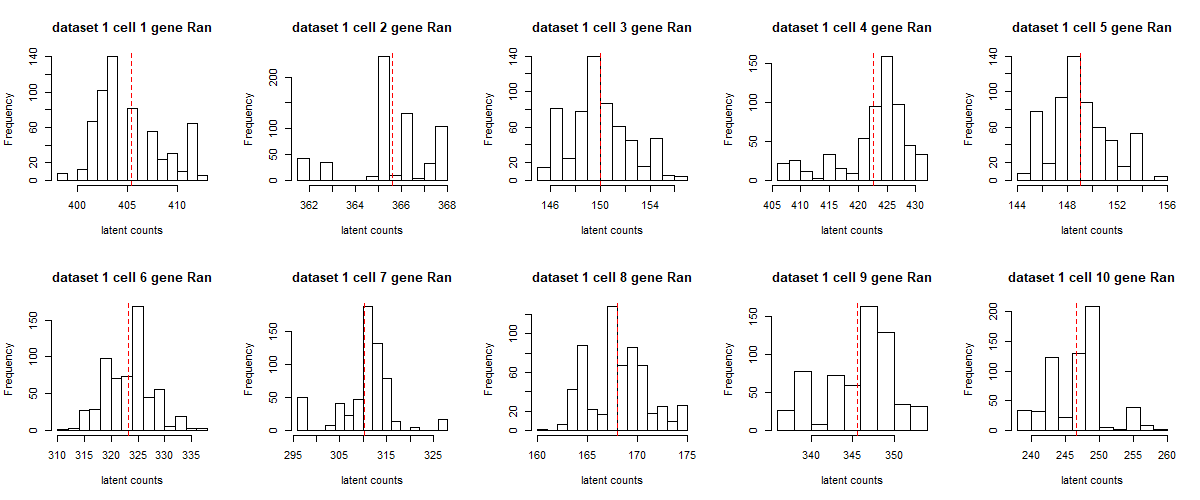}
	\caption{Histograms of the posterior estimated latent counts for cell $1$ to $10$ for gene 'Ran' in dataset 1. For each MCMC iteration, we compute the mean estimated latent count. The red vertical line indicates the overall mean latent count, averaged across all MCMC iterations, for a given gene and cell.}
	\label{fig:Ran_latent}
\end{figure}

For the posterior estimate of clustering, we show the number of genes in each cluster and the proportion of HET/HOM genes in each cluster in Web Figure \ref{fig:z_frequency}. The traceplots of concentration parameters and hyperparameters are shown in Web Figures \ref{fig:traceplot_concentration_parameters} and \ref{fig:traceplot_regression_parameters}, respectively, suggest convergence. The posterior and prior distributions of these parameters are compared in Web Figures \ref{fig:alpha_prior_posteropr_comp} and \ref{fig:regression_coefficient_comp}, highlighting the influence of the data. The relationship between the posterior estimated mean expressions and dispersions on the log-scale for each cluster are shown in Web Figure \ref{Fig:est_post_corr}; the relationships are similar across clusters. In addition, we compare the posterior estimated latent counts of two selected genes, namely `Fabp5' and `H2afz' across different clusters in Web Figure \ref{fig:latent_counts}. For gene `Fabp5', there are no evident differences between the estimated latent counts across clusters, whereas for gene `H2afz', differences across clusters are more apparent. Within each cluster, we observe greater variability of posterior estimated latent counts across cells for `H2afz'. For the capture efficiencies, we compare the bayNorm estimates and NormHDP posterior mean estimates of the capture efficiencies in Web Figure \ref{Fig:prior_posterior_beta} and present boxplots of the posterior mean  capture efficiencies for cells in each cluster in Web Figure \ref{fig:estimated_capture_efficiencies}. We observe that bayNorm tends to produce slightly larger estimates of the capture efficiencies compared with our model (Web Figure \ref{Fig:prior_posterior_beta}). In addition, when comparing across clusters (Web Figure \ref{fig:estimated_capture_efficiencies}), there is no evident difference, with the exception of the small group of cells in cluster $20$, which have lower capture efficiencies.  Another important aspect of our model is the quantification of  uncertainty in the estimated latent counts;  we show for two gene `Fabp5' and `Ran' the range of uncertainty in the posterior estimated latent counts in Web Figures \ref{fig:Fabp5_latent} and \ref{fig:Ran_latent}.

\clearpage

\subsection{Global Marker Genes} \label{sec:global_marker_genes}

\begin{figure}[!t]
    \centering
    \subfigure{\includegraphics[width=0.9\textwidth]{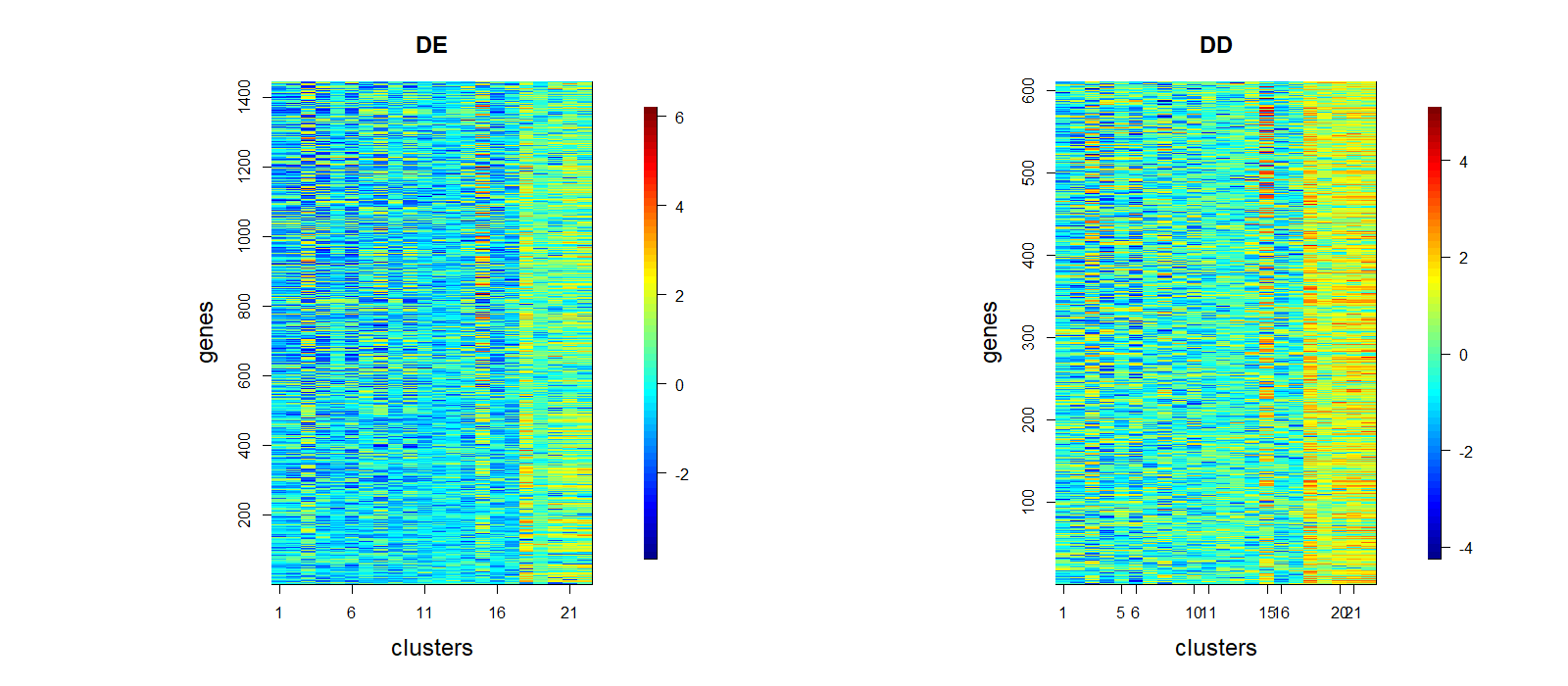}}
    \caption{Heat-maps of estimated relative unique parameters on the log-scale, with columns representing clusters and rows representing genes. Only global marker genes are included in the heat-maps. The relative value is defined as the estimated value minus the average across all clusters for that gene.}
    \label{fig:heatmap_global2}
\end{figure}

\begin{figure}[!t]
	\centering
	\includegraphics[width=\textwidth]{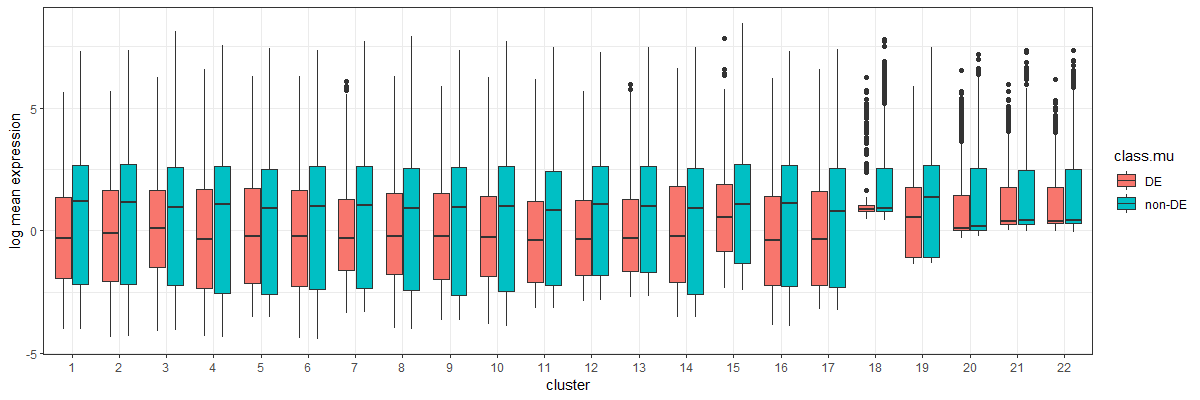}
	\caption{Comparison of the posterior estimated log mean expressions between global DE and non-DE genes for each cluster.}
	\label{Fig:global_marker_mu_dist}
\end{figure}

\begin{figure}[!t]
	\centering
	\includegraphics[width=\textwidth]{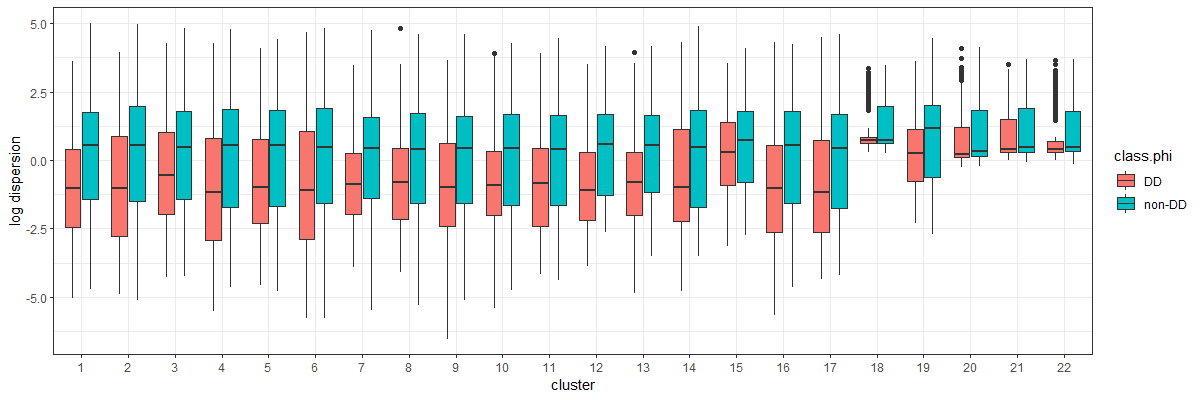}
	\caption{Comparison of the posterior estimated log dispersions between global DD and non-DD genes for each cluster.}
	\label{Fig:global_marker_phi_dist}
\end{figure}

\begin{figure}[!t]
	\centering
	\subfigure{\includegraphics[width=0.9\textwidth]{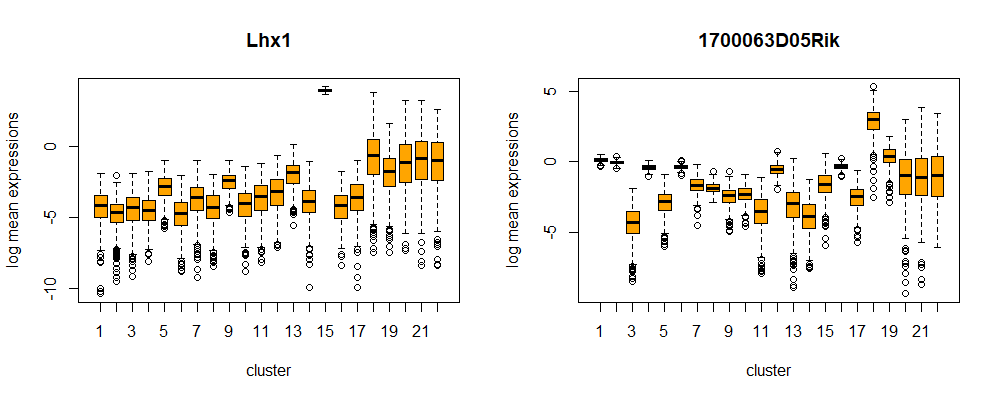}}
        \subfigure{\includegraphics[width=0.9\textwidth]{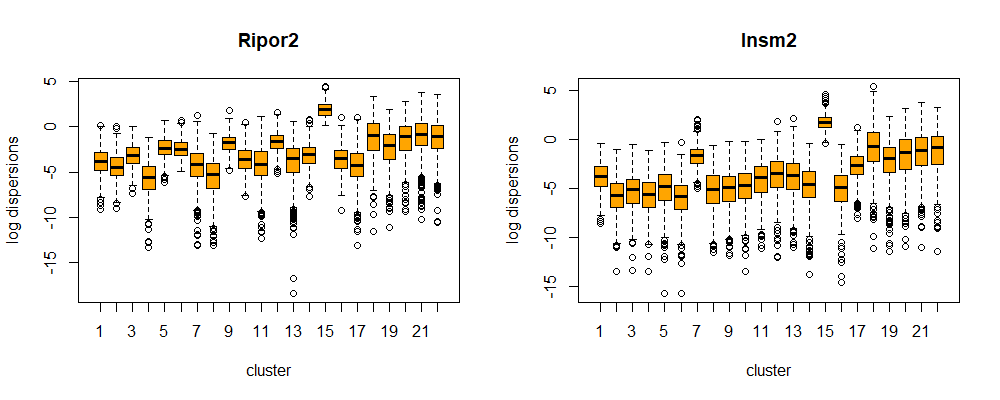}}
	\caption{Posterior of the unique parameters of global marker genes. Lhx1 and 1700063D05Rik are examples of global DE genes. Ripor2 and Insm2 are examples of global DD genes.}
	\label{Fig:global_marker_genes_example}
\end{figure}

\begin{figure}[!t]
	\centering
	\includegraphics[width=13cm]{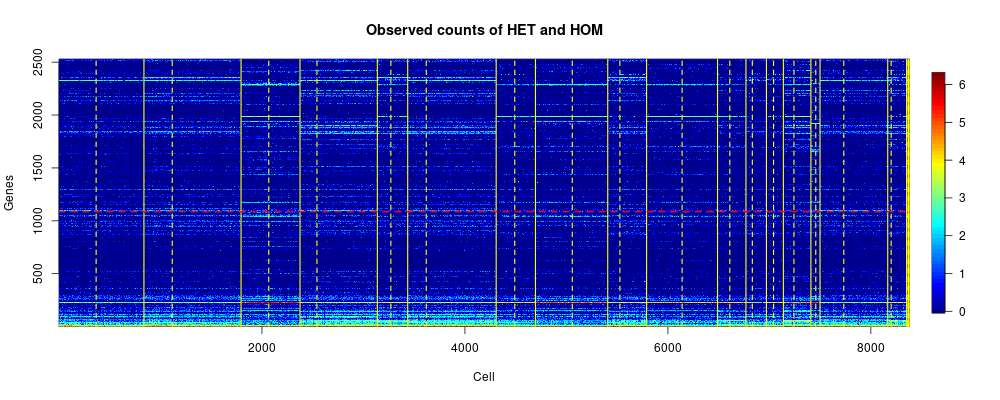}
	\caption{Heat-map of the  observed counts for HET and HOM data. Cells from different clusters are separated by yellow vertical lines. Cells from different datasets are separated by yellow dashed lines. DE and non-DE genes are separated by the horizontal line such that genes above the horizontal line are global marker genes.}
	\label{Fig:observed_counts_heatmap_DE}
\end{figure}

\begin{figure}[!t]
	\centering
	\includegraphics[width=13cm]{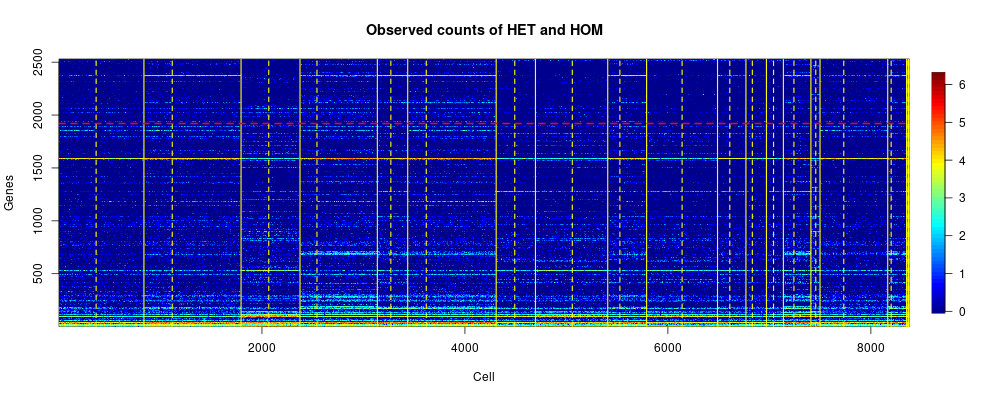}
	\caption{Heat-map of observed counts for HET and HOM data. Cells from different clusters are separated by yellow vertical lines. Cells from different datasets are separated by yellow dashed lines. DD and non-DD genes are separated by the horizontal line such that genes above the horizontal line are global marker genes.}
	\label{Fig:observed_counts_heatmap_DD}
\end{figure}

\begin{figure}[!t]
	\centering
	\subfigure{\includegraphics[width=0.45\textwidth]{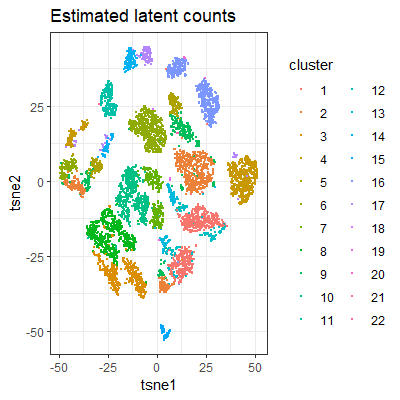}}
        \subfigure{\includegraphics[width=0.45\textwidth]{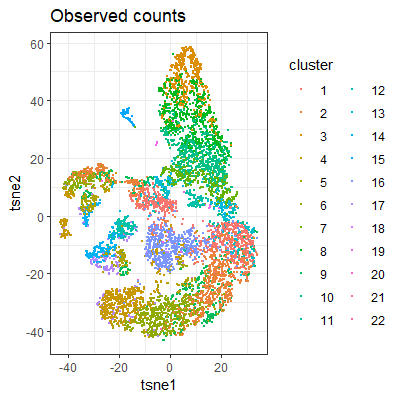}}
	\caption{t-SNE plot based on the estimated latent counts (left) and the observed counts (right) with global marker genes only. Cells are colored by cluster membership.}
	\label{fig:latent_observed_tsne_DEDD}
\end{figure}

Based on the heat-maps, we show that the posterior estimated mean expressions and dispersions tend to be lower for global DE and DD genes in comparison to the non-DE and non-DD genes for each cluster, apart from the small clusters $18$ and $20-22$ (Web Figure \ref{fig:heatmap_global2}, \ref{Fig:global_marker_mu_dist} and \ref{Fig:global_marker_phi_dist}). In addition, we compare the posterior of the mean expression and dispersion within cluster for some global marker genes that are identified as DE or DD in Web Figure \ref{Fig:global_marker_genes_example}. Further, heatmaps of the observed gene-counts with genes reordered by tail probabilities are presented in Web Figure \ref{Fig:observed_counts_heatmap_DE} and \ref{Fig:observed_counts_heatmap_DD} for DE and DD, respectively; genes above the horizontal red line are identified as global markers. t-SNE plots with only the global marker genes are shown in Web Figure \ref{fig:latent_observed_tsne_DEDD} for both the observed and posterior estimated latent counts; separation between clusters is more evident in the t-SNE plot based on the latent counts.

\begin{table}[!t]
	\centering
	\caption{Top 20 Global DE genes}
	\scalebox{0.9}{\begin{tabular}{llllllllll}
			\hline
			Lhx1 & 1700063D05Rik & Nr2f2 & Lhx1os & Trp73 & Dlx2 & Ppp2r2c & Gpr88 & Gm27199 & Foxo1 \\ 
			Gm30648 & Elfn2 & Insm2 & Psd & Kcnj5 & Ramp1 & Lhx5 & Mab21l1 & Rspo2 & AI593442 \\ \hline
	\end{tabular}}
	\label{tab:global.de.top20}
\end{table}

\begin{table}[!t]
	\centering
	\caption{Top 20 Global DD genes}
	\scalebox{0.9}{\begin{tabular}{llllllllll}
			\hline
			Ripor2 & Spdl1 & Trp73 & Kif14 & Wwtr1 & Cxcl12 & Ank3 & Etv1 & Cenpe & Nrp1 \\
			Plk4 & Mfng & Zic1 & Cdkn1a & Nt5dc2 & Nusap1 & Rtn1 & Elavl3 & Neurod2 & Eomes \\ \hline
	\end{tabular}}
	\label{tab:global.dd.top20}
\end{table}

\clearpage

\subsection{Local Marker Genes} \label{sec:local_marker_genes}

\begin{figure}[!t]
	\centering
	\includegraphics[width=\textwidth]{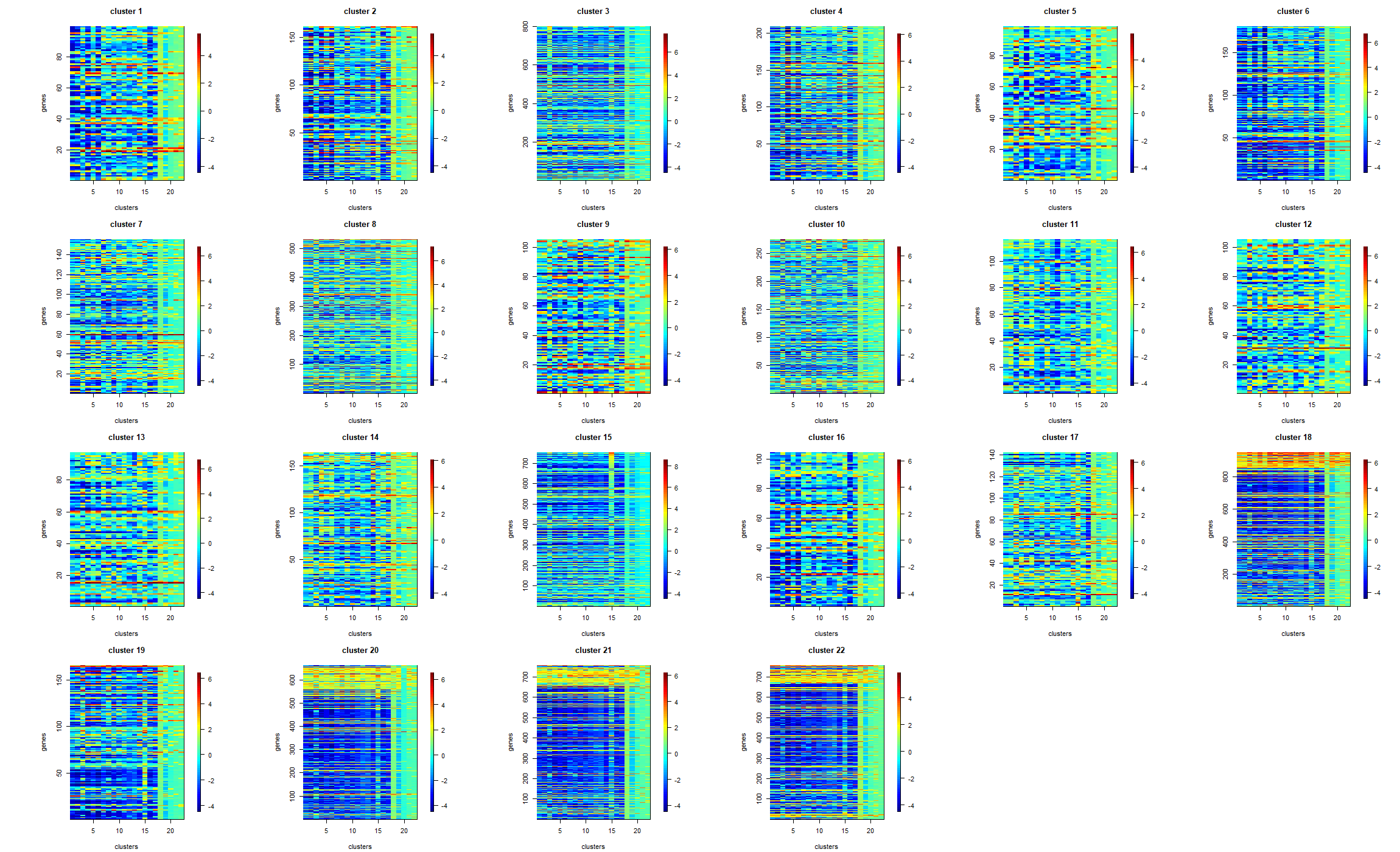}
	\caption{Heat-maps to compare estimated unique parameters of local marker genes for each cluster. Columns in each heat map represent clusters and rows represent genes. The first two rows are the estimated mean expressions for the local DE genes and the last two rows are the estimated dispersions for the local DD genes. For all heat-maps, rows are reordered by local tail probabilities, hence genes on the top rows have a higher probability of being locally DE.} %Genes with different local features are shown with different colours}
\label{fig:local_marker_summary_2}
\end{figure}

\begin{figure}[!t]
	\centering
	\includegraphics[width=\textwidth]{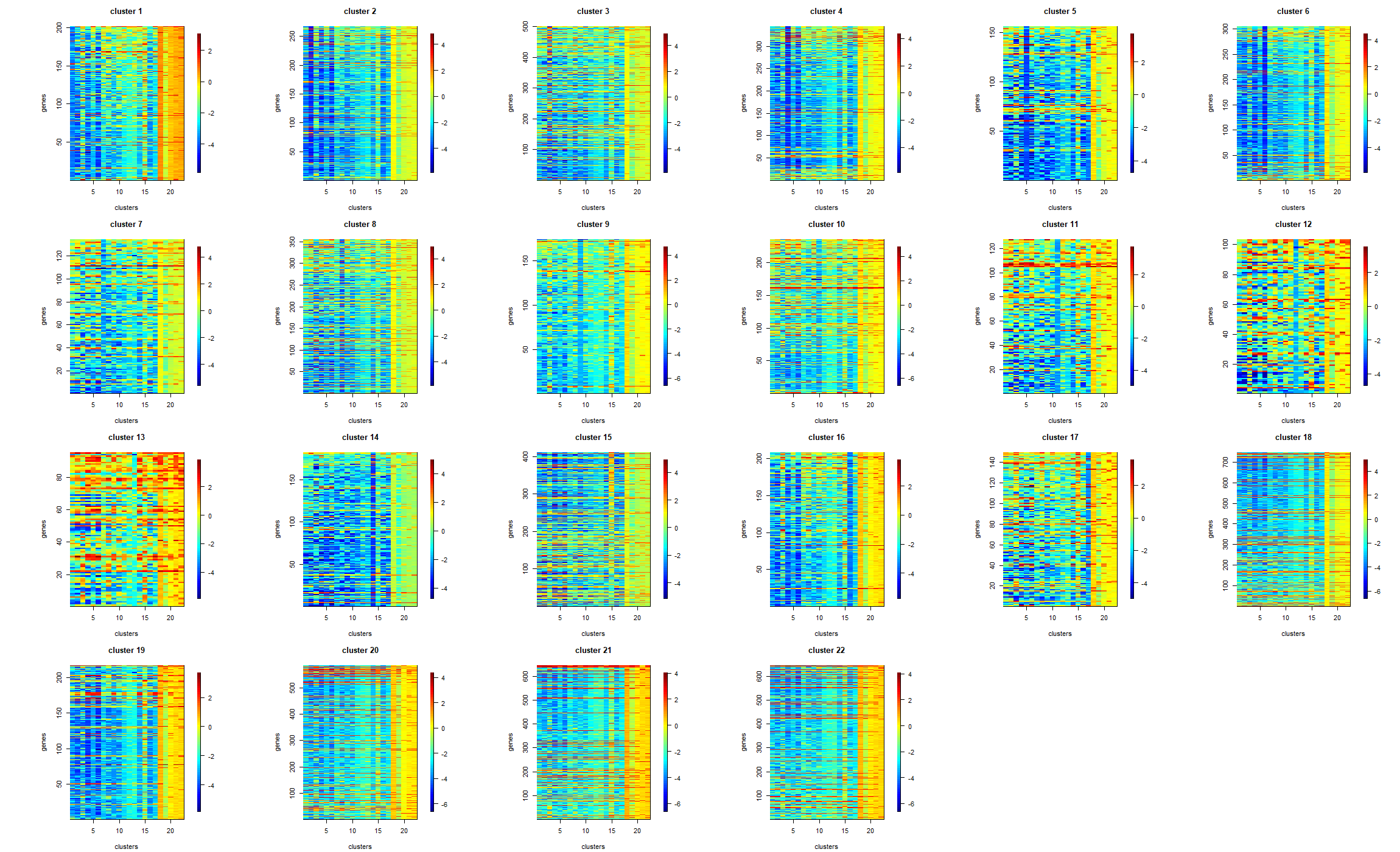}
	\caption{Heat-maps to compare estimated unique parameters of local marker genes for each cluster. Columns in each heat map represent clusters and rows represent genes. The first two rows are the estimated mean expressions for the local DE genes and the last two rows are the estimated dispersions for the local DD genes. For all heat-maps, rows are reordered by local tail probabilities, hence genes on the top rows have a higher probability of being locally DD.}
	\label{fig:local_marker_summary_3}
\end{figure}

\begin{figure}[!t]
	\centering
	\includegraphics[width=\textwidth]{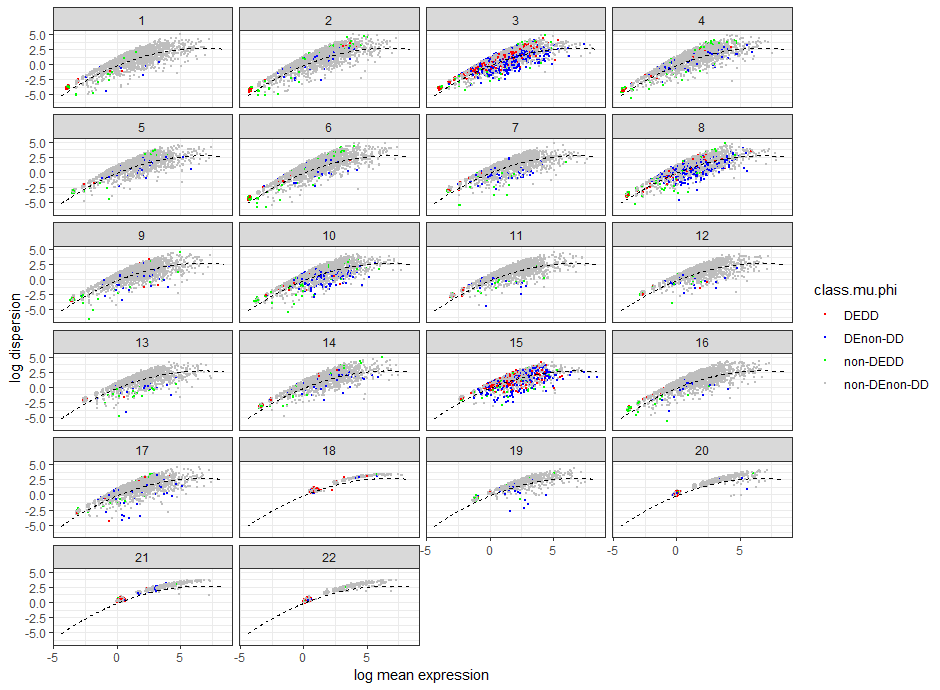}
	\caption{Posterior estimated relationships between the mean expressions and dispersions for each cluster. The posterior means of the mean expressions and dispersions are plotted. To compare the local marker genes for each cluster, we highlight local DE and DD genes in red, local DE and non-DD genes in green, and local DD and non-DE genes in blue.} %Genes with different local features are shown with different colours}
\label{fig:local_marker_genes_position}
\end{figure}

\begin{figure}[!t]
\centering
\includegraphics[width=\textwidth]{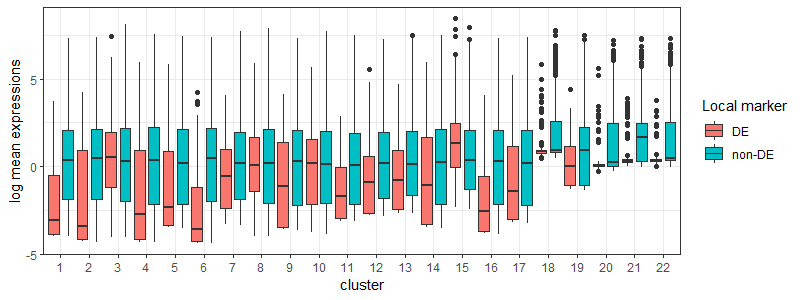}
\caption{Distribution of posterior estimates of mean expressions for local DE and non-DE genes for each cluster.}
\label{fig:local_marker_mu_dist}
\end{figure}

\begin{figure}[!t]
\centering
\includegraphics[width=\textwidth]{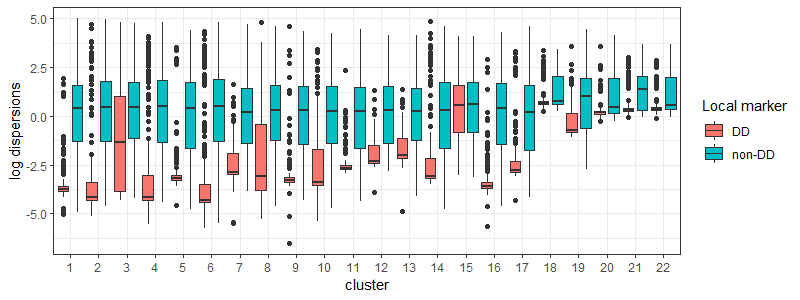}
\caption{Distribution of posterior estimates of dispersions for local DD and non-DD genes for each cluster.}
\label{fig:local_marker_phi_dist}
\end{figure}

We present the heatmaps to compare estimated unique parameters of local marker genes for each cluster are shown in Web Figure \ref{fig:local_marker_summary_2} and \ref{fig:local_marker_summary_3}. Further, we present the relationship between the mean expressions and dispersions and highlight the local marker genes for each cluster in Web Figure \ref{fig:local_marker_genes_position}.
No clear pattern is observed between the local features of the genes and the relationship between the unique parameters. In addition, we compare the estimated unique parameters between local marker and non-marker genes for each cluster; differences in the posterior estimated unique parameters are evident for local marker genes in $18$ out of $22$ clusters. (Web Figure \ref{fig:local_marker_mu_dist} and \ref{fig:local_marker_phi_dist}).

\begin{table}[!t]
	\begin{center}
		\caption{Top 5 local DE genes for each cluster.}
		\scalebox{0.8}{\begin{tabular}{c  c  c  c  c  c  c  c}
			\hline
			Cluster 1 & Cluster 2 & Cluster 3 & Cluster 4 & Cluster 5 & Cluster 6 & Cluster 7 & Cluster 8 \\
			\hline
			Arhgap29 & Rit2 & Smc6 & Gm40421 & Gsx2 & Ifitm3 & Cnih2 & Mapk3 \\
			Ifi27 & Prc1 & Neto2 & 0610010F05Rik & Hecw1 & Gm36660 & Rfc3 & Ttk \\
			1700093K21Rik & Tenm2 & Smim18 & Cnih2 & Srrm4 & Gm28209 & Hist1h1e & Rab3a \\
			Zfp935 & 1700093K21Rik & Timeless & Nlgn1 & Mcm4 & Msra & Nav3 & Smc6 \\
			Tmem150c & Cdh4 & Kcnk13 & Brd8 & Fzd1 & Spock1 & Mycbp2 & Ina \\
			\hline
                Cluster 9 & Cluster 10 & Cluster 11 & Cluster 12 & Cluster 13 & Cluster 14 & Cluster 15 & Cluster 16 \\ 
			\hline
                Clic1 & Comt & Prokr2 & Nup85 & Clic1 & Lrrc7 & Agtr1a & Txlnb \\
                Nup62 & Fbln1 & Gm10457 & 1110017D15Rik & Serp2 & Nr2f1 & Kif4 & Dcaf17 \\
                Schip1 & Mapk3 & Smc4 & Xlr3a & Sobp & Loxl2 & Dct & Snhg12 \\
                Mcm5 & Tfap2c & 1500035N22Rik & Calb1 & Zic4 & Mcm7 & Lamp5 & Dkc1 \\
                Gm28196 & Dsn1 & Dll1 & Nefm & Reep2 & Spag5 & Syp & Manf \\
                \hline
                Cluster 17 & Cluster 18 & Cluster 19 & Cluster 20 & Cluster 21 & Cluster 22 &  &  \\
                \hline
                Tagln2 & Olfr655 & Comt & Fst & Pcgf5 & Pcgf5 & & \\
                Dusp6 & Gm16152 & Pcgf5 & Igfbpl1 & Syn1 & Dusp4 & & \\
                Shb & Gm29771 & Timeless & Cbfa2t3 & Synpr & Nek6 & & \\
                Efnb2 & BC030500 & Wnt7b & Mt3 & Spats2l & Scg3 & & \\
                Foxm1 & Khdrbs2 & Gpsm2 & Wnt10a & Kcnh4 & Shpk & & \\ \hline
		\end{tabular}}
	\end{center}
\end{table}

\begin{table}[!t]
	\begin{center}
		\caption{Top 5 local DD genes for each cluster.}
		\scalebox{0.8}{\begin{tabular}{c  c  c  c  c  c  c  c}
			\hline
			Cluster 1 & Cluster 2 & Cluster 3 & Cluster 4 & Cluster 5 & Cluster 6 & Cluster 7 & Cluster 8 \\
			\hline
			Ntm & Tenm2 & Doc2b & Gm40421 & Fzd1 & Gm28209 & Tssc4 & Glra1 \\
			1700093K21Rik & 1700093K21Rik & Numbl & Gm13425 & Gsx2 & Ifitm3 & Mgat4c & Vit \\
			Ifi27 & Cnrip1 & Fbln1 & Nlgn1 & Kcnb2 & Msra & Sobp & 4430402I18Rik \\
			Arhgap29 & Gm48283 & A830011K09Rik & Cnih2 & Mcm4 & Spock1 & Phf24 & Calb1 \\
			Zfp935 & Rgs8 & Bnip3 & Gabra2 & Ppp2r2c & Htra1 & Lsm3 & Rspo3 \\
			\hline
                Cluster 9 & Cluster 10 & Cluster 11 & Cluster 12 & Cluster 13 & Cluster 14 & Cluster 15 & Cluster 16 \\ 
			\hline
                Gm17322 & Fbln1 & 1500035N22Rik & Nup85 & Prim2 & Tgfb2 & Pcdhb6 & Snhg12 \\
                Tenm1 & Zcchc18 & Prokr2 & Ncapg2 & Kif4 & Lrrc7 & Gfra4 & Txlnb \\
                Rhbdl2 & Efnb1 & Sfxn3 & Atcay & Cck & Nr2f1 & Suz12 & Nlgn1 \\
                Zfpm2 & Ccno & Acss1 & 9330159F19Rik & Sobp & Mcm7 & Hspa8 & H19 \\
                Clic1 & Ackr3 & Zbtb20 & Xlr3a & Reep2 & Gm13425 & Chd5 & Rec114 \\
                \hline
                Cluster 17 & Cluster 18 & Cluster 19 & Cluster 20 & Cluster 21 & Cluster 22 &  &  \\
                \hline
                Tagln2 & Pcgf5 & Slc25a5 & Igfbpl1 & Srrm4 & Traf4 & & \\
                Cdh10 & Khdrbs2 & Wnt7b & Acrbp & Syn1 & Dusp4 & & \\
                Efnb2 & Rbms1 & Etv4 & Wnt10a & Tbx5 & Rest & & \\
                Dusp6 & Tubb3 & Nrcam & Myl6b & Bmp3 & Spc25 & & \\
                Arhgap29 & Gm16152 & Mirt1 & Bcl11a & Abhd11 & Arhgef25 & & \\ \hline
		\end{tabular}}
	\end{center}
\end{table}

\clearpage

\subsection{Posterior Estimated Latent Counts} \label{sec:posterior_latent_counts}

\begin{figure}[!t]
	\centering
	\subfigure{\includegraphics[width=0.9\textwidth]{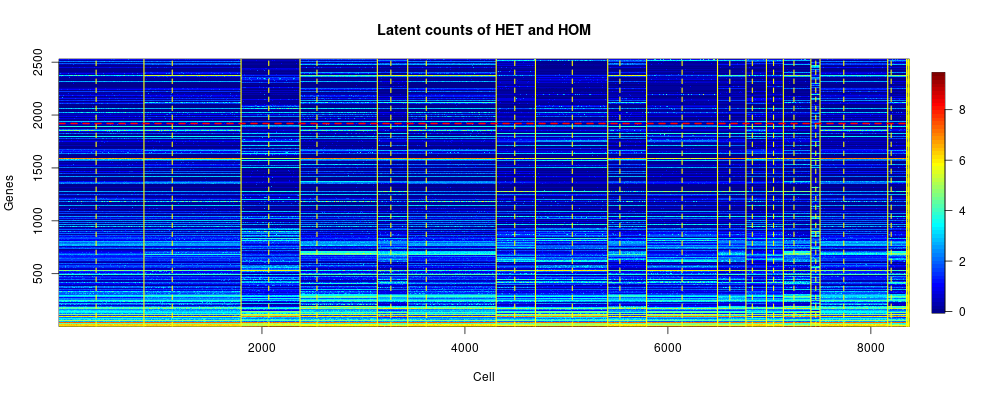}}
	\caption{Heat-map of posterior estimated latent gene-counts for HET and HOM. Genes are reordered by global DD tail probabilities, genes above the red horizontal line are global DD, and vice versa. Cells for reordered by the point estimate of posterior allocations. Cells from different clusters are separated by solid lines and cells from different datasets are separated by dashed lines.}
	\label{fig:latent_counts_heatmap_DD}
\end{figure}

We compute the posterior estimated latent counts for all cells and compare between different clusters. Web Figure \ref{fig:latent_counts_heatmap_DD} provides a heat-map of the estimated latent counts; cells are ordered by the clustering estimate, with solid vertical lines separating cells from different clusters and dashed vertical lines separating HET and HOM within cluster. Genes are reordered by global DD tail probabilities, with global DD genes above the horizontal line. Corresponding figures for the observed counts are shown in Web Appendix \ref{sec:global_marker_genes}. 

For each gene, posterior estimated latent counts and observed counts for cells within each clusters are similar, and clear differences are observed across cells from different clusters. In addition, we use t-SNE (a commonly used dimensional reduction method for visualising gene expressions) to visualize similarities between cells within each cluster and differences across clusters. Applying t-SNE to the posterior estimated latent counts for genes which are global DE and DD shows a clear separation between clusters.

\clearpage

\subsection{Posterior Predictive Checks} \label{sec:ppc_summary}

\begin{figure}[!t]
	\centering
	\includegraphics[width=0.80\textwidth]{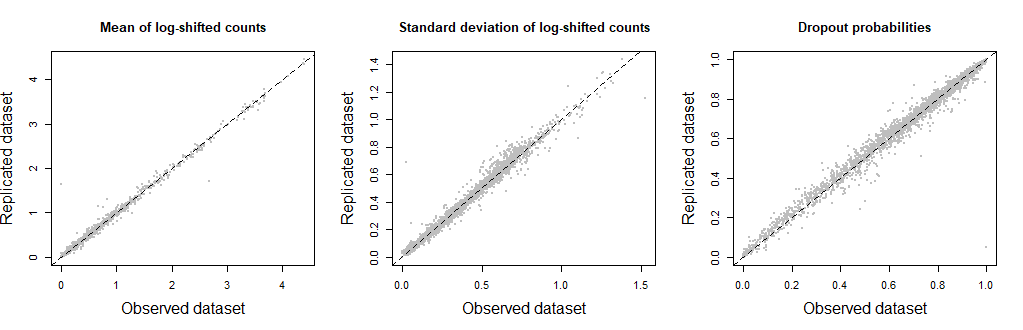}
	\caption{Gene-wise comparison of the mean of the log shifted counts, standard deviation of the log shifted counts and dropout probabilities for posterior predictive checks with single replicate.}
	\label{fig:ppc_single_2}
\end{figure}

By comparing the observed and replicated statistics match gene-wise in Web Figure \ref{fig:ppc_single_2}, we show the observed and replicated statistics are similar which further supporting the model fit.

\clearpage 

\section{\textit{Important} Genes} \label{Web Appendix D}

\begin{figure}[!t]
	\centering
	\includegraphics[width=\textwidth]{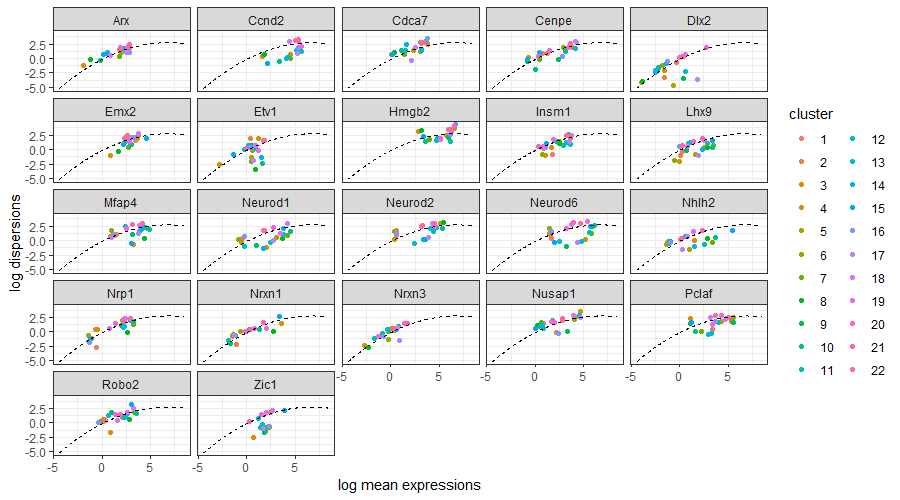}
	\caption{Posterior estimated mean expressions and dispersions (on the log-scale) for the important genes which are classified as both DE and DD. The dashed line shows the posterior estimated relationship between the mean expressions and dispersions.}
	\label{fig:important_genes_DEDD}
\end{figure}

\begin{figure}[!t]
	\centering
	\includegraphics[width=\textwidth]{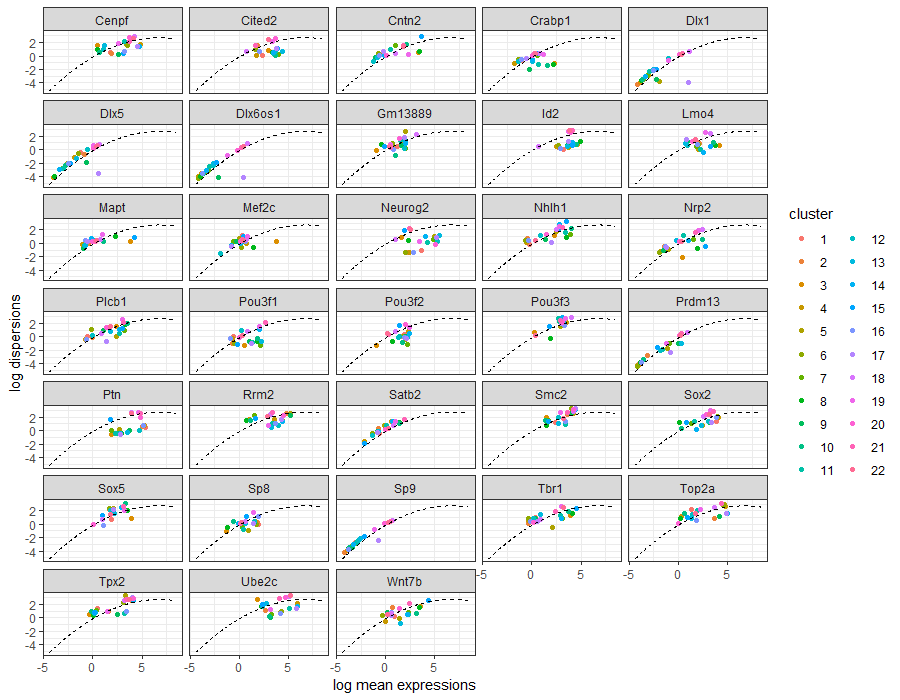}
	\caption{Posterior estimated mean expressions and dispersions (on the log-scale) for the important genes which are classified as  DE, but not DD. The dashed line shows the posterior estimated relationship between the mean expressions and dispersions.}
	\label{fig:important_genes_DEnonDD}
\end{figure}

\begin{figure}[!t]
	\centering
	\includegraphics[width=6cm]{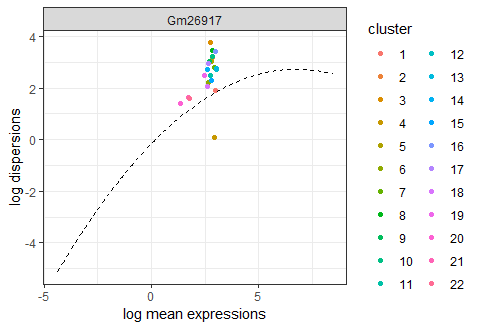}
	\caption{Posterior estimated mean expressions and dispersions (on the log-scale) for the important genes which are classified as DD, but not DE. The dashed line shows the posterior estimated  relationship between the mean expressions and dispersions.}
	\label{fig:important_genes_DDnonDE}
\end{figure}

\begin{figure}[!t]
	\centering
	\includegraphics[width=\textwidth]{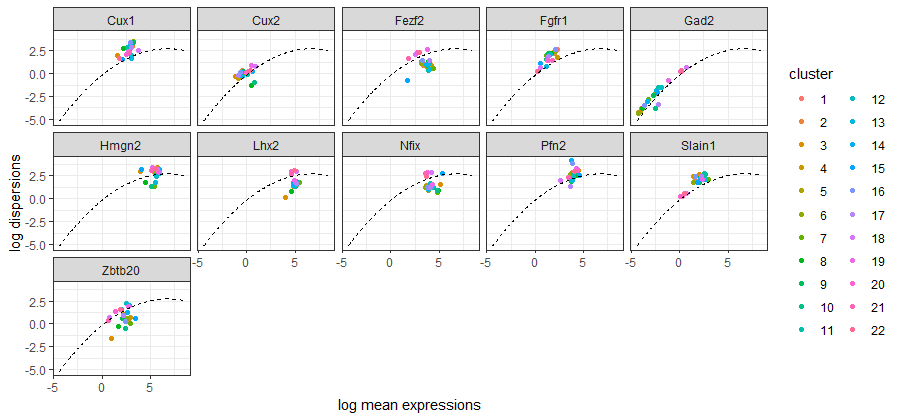}
	\caption{Posterior estimated mean expressions and dispersions (on the log-scale) for the important genes which are classified as both not DE and DD. The dashed line shows the posterior estimated relationship between the mean expressions and dispersions.}
	\label{fig:important_genes_nonDEnonDD}
\end{figure}

We were provided with a list of 70 \textit{important} genes, that are of particular interest for this experimental data. Information and summary statistics including the posterior estimated mean expressions and dispersions and  global features of these \textit{important} genes are shown below.
In the following, we present the list of \textit{important} genes (Web Table \ref{tab:important_genes}), and those which are classified as global marker genes are shown in Web Table \ref{tab:important_genes_DE2} and \ref{tab:important_genes_DD2}. 
Further, we present the relationship between posterior estimated mean expressions and dispersions with different local features in Web Figure \ref{fig:important_genes_DEDD}, \ref{fig:important_genes_DEnonDD}, \ref{fig:important_genes_DDnonDE} and \ref{fig:important_genes_nonDEnonDD}. Lastly, we show the heatmap of estimated unique parameters for the important genes in Web Figure \ref{fig:local_marker_genes_position}.

\begin{table}[!t]
	\centering
	\caption{List of important genes.}
	\scalebox{0.9}{\begin{tabular}{llllllllll}
			\hline
			Dlx6os1 & Sp9     & Nrxn3   & Dlx1    & Ccnd2   & Arx     & Dlx5   & Top2a & Rrm2   & Pclaf \\
			Hmgb2   & Cdca7   & Gm13889 & Etv1    & Cenpf   & Gm26917 & Sp8    & Gad2  & Hmgn2  & Cenpe \\
			Insm1   & Nusap1  & Tpx2    & Neurod6 & Cntn2   & Mef2c   & Mapt   & Tbr1  & Nrp1   & Wnt7b \\
			Id2     & Neurod1 & Nrxn1   & Satb2   & Neurog2 & Crabp1  & Lhx2   & Zic1  & Mfap4  & Nrp2  \\
			Ccnd2   & Nhlh1   & Plcb1   & Nhlh2   & Lhx9    & Lmo4    & Prdm13 & Emx2  & Cited2 & Insm1 \\
			Ptn     & Cux2    & Wnt7b   & Pou3f3  & Cux1    & Pou3f1  & Zbtb20 & Nfix  & Pfn2   & Ube2c \\
			Fezf2   & Sox2    & Neurod2 & Sox5    & Slain1  & Fgfr1   & Pou3f2 & Robo2 & Dlx2   & Smc2  \\ \hline
	\end{tabular}}
	\label{tab:important_genes}
\end{table}

\begin{table}[h]
\centering
\caption{List of important genes which are classified as DE.}
\scalebox{0.9}{\begin{tabular}{l l l l l l l l l l}
\hline
Pou3f3 & Satb2 & Nrp2 & Cntn2 & Lhx9 & Nhlh1 & Cenpf & Tbr1 & Dlx1 & Dlx2  \\
Cdca7 & Sp9 & Neurod1 & Gm13889 & Nusap1 & Plcb1 & Insm1 & Tpx2 & Ube2c & Arx  \\ 
Sox2 & Nhlh2 & Neurog2 & Cenpe & Lmo4 & Prdm13 & Pou3f2 & Smc2 & Pou3f1 & Dlx6os1 \\ 
Dlx5 & Ptn & Neurod6 & Ccnd2 & Sox5 & Cited2 & Hmgb2 & Nrp1 & Crabp1 & Pclaf \\ 
Zic1 & Mfap4 & Neurod2 & Top2a & Mapt & Mef2c & Rrm2 & Id2 & Etv1 & Nrxn3 \\
Sp8 & Wnt7b & Robo2 & Nrxn1 & Emx2 &  &  &  &  &   \\ \hline
\end{tabular}}
\label{tab:important_genes_DE2}
\end{table}

\begin{table}[h]
\centering
\caption{List of important genes which are classified as DD.}
\scalebox{0.9}{\begin{tabular}{llllllllll}
\hline
Lhx9 & Dlx2 & Cdca7 & Neurod1 & Nusap1 & Insm1 & Arx & Nhlh2 & Cenpe & Neurod6 \\
Ccnd2 & Hmgb2 & Nrp1 & Pclaf & Zic1 & Mfap4 & Neurod2 & Etv1 & Nrxn3 & Robo2 \\
Gm26917 & Nrxn1 & Emx2 &  &  &  &  &  &  &   \\ \hline
\end{tabular}}
\label{tab:important_genes_DD2}
\end{table}

\begin{figure}[!t]
	\centering
	\subfigure{\includegraphics[width=0.45\textwidth]{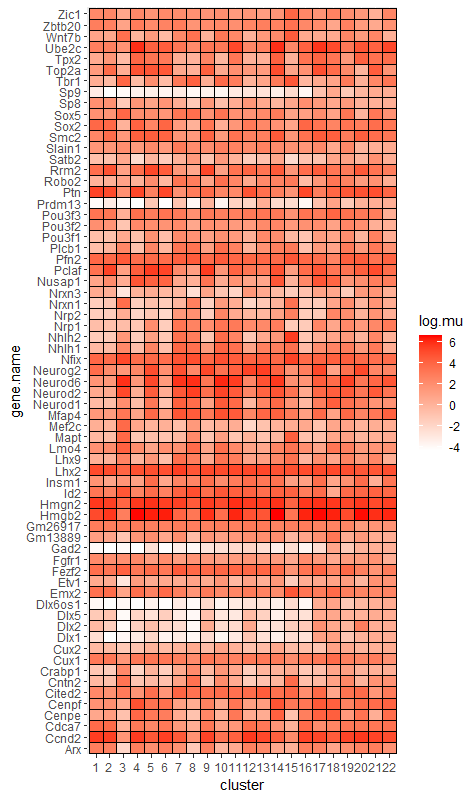}}
        \subfigure{\includegraphics[width=0.45\textwidth]{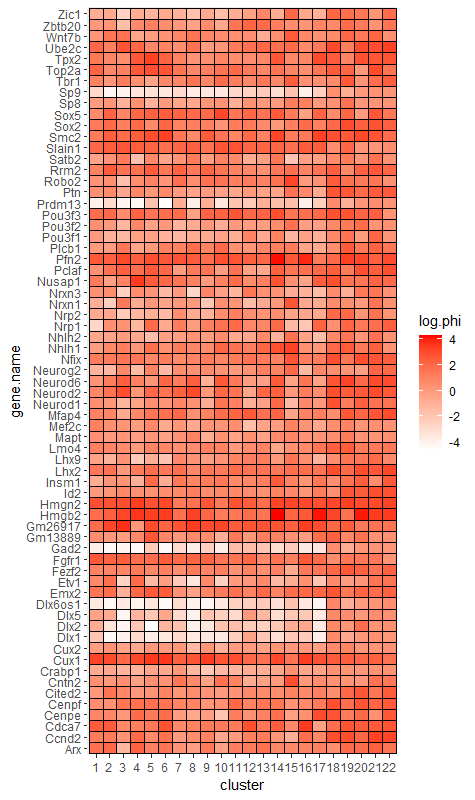}}
	\caption{Posterior means of the unique parameters of important genes for all clusters for mean expression (left) and dispersion (right).}
	\label{Fig:important_genes_unique}
\end{figure}

\clearpage

\bibliographystyle{plainnat} 
\bibliography{main}